\documentclass[longauth]{aaEC}

\usepackage{graphicx}
\usepackage{natbib}
\usepackage{scalerel}

\usepackage[table]{xcolor}

\usepackage{txfonts}
\usepackage[pdfencoding=auto,psdextra]{hyperref}
\hypersetup{
    colorlinks=true,
    linkcolor=blue,
    filecolor=magenta,      
    urlcolor=blue,
    citecolor=blue
}
\usepackage{orcidlink}

\makeatletter
\renewcommand*\aa@pageof{, page \thepage{} of \pageref*{LastPage}}
\makeatother

\usepackage[utf8]{inputenc}

\usepackage[switch, modulo]{lineno}

\usepackage{euclid}

\begin{document}
%
%

\title{Euclid Quick Data Release (Q1)}
\subtitle{The Strong Lensing Discovery Engine A -- System overview and lens catalogue}    

\newcommand{\orcid}[1]{\href{https://orcid.org/#1}{\orcidlink{#1}}}   

\author{Euclid Collaboration: M.~Walmsley\orcid{0000-0002-6408-4181}\thanks{\email{m.walmsley@utoronto.ca}}\inst{\ref{aff1},\ref{aff2}}
\and P.~Holloway\orcid{0009-0002-8896-6100}\inst{\ref{aff3}}
\and N.~E.~P.~Lines\orcid{0009-0004-7751-1914}\inst{\ref{aff4}}
\and K.~Rojas\orcid{0000-0003-1391-6854}\inst{\ref{aff5},\ref{aff4}}
\and T.~E.~Collett\orcid{0000-0001-5564-3140}\inst{\ref{aff4}}
\and A.~Verma\orcid{0000-0002-0730-0781}\inst{\ref{aff3}}
\and T.~Li\orcid{0009-0005-5008-0381}\inst{\ref{aff4}}
\and J.~W.~Nightingale\orcid{0000-0002-8987-7401}\inst{\ref{aff6}}
\and G.~Despali\orcid{0000-0001-6150-4112}\inst{\ref{aff7},\ref{aff8},\ref{aff9}}
\and S.~Schuldt\orcid{0000-0003-2497-6334}\inst{\ref{aff10},\ref{aff11}}
\and R.~Gavazzi\orcid{0000-0002-5540-6935}\inst{\ref{aff12},\ref{aff13}}
\and A.~Melo\orcid{0000-0002-6449-3970}\inst{\ref{aff14},\ref{aff15}}
\and R.~B.~Metcalf\orcid{0000-0003-3167-2574}\inst{\ref{aff7},\ref{aff8}}
\and I.~T.~Andika\orcid{0000-0001-6102-9526}\inst{\ref{aff15},\ref{aff14}}
\and L.~Leuzzi\orcid{0009-0006-4479-7017}\inst{\ref{aff7},\ref{aff8}}
\and A.~Manj\'on-Garc\'ia\orcid{0000-0002-7413-8825}\inst{\ref{aff16}}
\and R.~Pearce-Casey\inst{\ref{aff17}}
\and S.~H.~Vincken\inst{\ref{aff5}}
\and J.~Wilde\orcid{0000-0002-4460-7379}\inst{\ref{aff18}}
\and V.~Busillo\orcid{0009-0000-6049-1073}\inst{\ref{aff19},\ref{aff20},\ref{aff21}}
\and C.~Tortora\orcid{0000-0001-7958-6531}\inst{\ref{aff19}}
\and J.~A.~Acevedo~Barroso\orcid{0000-0002-9654-1711}\inst{\ref{aff22}}
\and H.~Dole\orcid{0000-0002-9767-3839}\inst{\ref{aff23}}
\and L.~R.~Ecker\orcid{0009-0005-3508-2469}\inst{\ref{aff24},\ref{aff25}}
\and J.~Pearson\orcid{0000-0001-8555-8561}\inst{\ref{aff17}}
\and P.~J.~Marshall\orcid{0000-0002-0113-5770}\inst{\ref{aff26},\ref{aff27}}
\and A.~More\inst{\ref{aff28}}
\and T.~Saifollahi\orcid{0000-0002-9554-7660}\inst{\ref{aff29}}
\and J.~Gracia-Carpio\inst{\ref{aff25}}
\and E.~Baeten\inst{\ref{aff30}}
\and C.~Cornen\orcid{0000-0002-7786-2798}\inst{\ref{aff30}}
\and L.~C.~Johnson\orcid{0000-0001-6421-0953}\inst{\ref{aff31}}
\and C.~Macmillan\inst{\ref{aff3}}
\and S.~Kruk\orcid{0000-0001-8010-8879}\inst{\ref{aff32}}
\and K.~A.~Remmelgas\inst{\ref{aff32}}
\and B.~Cl\'ement\orcid{0000-0002-7966-3661}\inst{\ref{aff22},\ref{aff33}}
\and H.~Degaudenzi\orcid{0000-0002-5887-6799}\inst{\ref{aff34}}
\and F.~Courbin\orcid{0000-0003-0758-6510}\inst{\ref{aff18},\ref{aff35}}
\and J.~Bovy\orcid{0000-0001-6855-442X}\inst{\ref{aff1}}
\and S.~Casas\orcid{0000-0002-4751-5138}\inst{\ref{aff36}}
\and H.~Dannerbauer\orcid{0000-0001-7147-3575}\inst{\ref{aff37}}
\and J.~M.~Diego\orcid{0000-0001-9065-3926}\inst{\ref{aff38}}
\and K.~Finner\orcid{0000-0002-4462-0709}\inst{\ref{aff39}}
\and A.~Galan\orcid{0000-0003-2547-9815}\inst{\ref{aff15},\ref{aff14}}
\and C.~Giocoli\orcid{0000-0002-9590-7961}\inst{\ref{aff8},\ref{aff9}}
\and N.~B.~Hogg\orcid{0000-0001-9346-4477}\inst{\ref{aff40}}
\and K.~Jahnke\orcid{0000-0003-3804-2137}\inst{\ref{aff41}}
\and J.~Katona\orcid{0009-0001-5371-8935}\inst{\ref{aff42},\ref{aff43}}
\and A.~Kov\'acs\orcid{0000-0002-5825-579X}\inst{\ref{aff43},\ref{aff44}}
\and C.~De~Leo\inst{\ref{aff45}}
\and G.~Mahler\orcid{0000-0003-3266-2001}\inst{\ref{aff46},\ref{aff47},\ref{aff48}}
\and M.~Millon\orcid{0000-0001-7051-497X}\inst{\ref{aff49}}
\and B.~C.~Nagam\orcid{0000-0002-3724-7694}\inst{\ref{aff50},\ref{aff51}}
\and P.~Nugent\orcid{0000-0002-3389-0586}\inst{\ref{aff52}}
\and A.~Sainz~de~Murieta\inst{\ref{aff4}}
\and C.~M.~O'Riordan\orcid{0000-0003-2227-1998}\inst{\ref{aff14}}
\and D.~Sluse\orcid{0000-0001-6116-2095}\inst{\ref{aff46}}
\and A.~Sonnenfeld\orcid{0000-0002-6061-5977}\inst{\ref{aff53}}
\and C.~Spiniello\orcid{0000-0002-3909-6359}\inst{\ref{aff3}}
\and S.~Serjeant\orcid{0000-0002-0517-7943}\inst{\ref{aff17}}
\and T.~T.~Thai\orcid{0000-0002-8408-4816}\inst{\ref{aff54}}
\and L.~Ulivi\orcid{0009-0001-3291-5382}\inst{\ref{aff55},\ref{aff56},\ref{aff57}}
\and G.~L.~Walth\orcid{0000-0002-6313-6808}\inst{\ref{aff39}}
\and L.~Weisenbach\orcid{0000-0003-1175-8004}\inst{\ref{aff4}}
\and M.~Zumalacarregui\orcid{0000-0002-9943-6490}\inst{\ref{aff58}}
\and N.~Aghanim\orcid{0000-0002-6688-8992}\inst{\ref{aff23}}
\and B.~Altieri\orcid{0000-0003-3936-0284}\inst{\ref{aff32}}
\and A.~Amara\inst{\ref{aff59}}
\and S.~Andreon\orcid{0000-0002-2041-8784}\inst{\ref{aff60}}
\and N.~Auricchio\orcid{0000-0003-4444-8651}\inst{\ref{aff8}}
\and H.~Aussel\orcid{0000-0002-1371-5705}\inst{\ref{aff61}}
\and C.~Baccigalupi\orcid{0000-0002-8211-1630}\inst{\ref{aff62},\ref{aff63},\ref{aff64},\ref{aff65}}
\and M.~Baldi\orcid{0000-0003-4145-1943}\inst{\ref{aff66},\ref{aff8},\ref{aff9}}
\and A.~Balestra\orcid{0000-0002-6967-261X}\inst{\ref{aff67}}
\and S.~Bardelli\orcid{0000-0002-8900-0298}\inst{\ref{aff8}}
\and P.~Battaglia\orcid{0000-0002-7337-5909}\inst{\ref{aff8}}
\and F.~Bernardeau\inst{\ref{aff68},\ref{aff13}}
\and A.~Biviano\orcid{0000-0002-0857-0732}\inst{\ref{aff63},\ref{aff62}}
\and A.~Bonchi\orcid{0000-0002-2667-5482}\inst{\ref{aff69}}
\and D.~Bonino\orcid{0000-0002-3336-9977}\inst{\ref{aff70}}
\and E.~Branchini\orcid{0000-0002-0808-6908}\inst{\ref{aff71},\ref{aff72},\ref{aff60}}
\and M.~Brescia\orcid{0000-0001-9506-5680}\inst{\ref{aff20},\ref{aff19}}
\and J.~Brinchmann\orcid{0000-0003-4359-8797}\inst{\ref{aff73},\ref{aff74}}
\and S.~Camera\orcid{0000-0003-3399-3574}\inst{\ref{aff75},\ref{aff76},\ref{aff70}}
\and G.~Ca\~nas-Herrera\orcid{0000-0003-2796-2149}\inst{\ref{aff77},\ref{aff78},\ref{aff79}}
\and V.~Capobianco\orcid{0000-0002-3309-7692}\inst{\ref{aff70}}
\and C.~Carbone\orcid{0000-0003-0125-3563}\inst{\ref{aff11}}
\and V.~F.~Cardone\inst{\ref{aff80},\ref{aff81}}
\and J.~Carretero\orcid{0000-0002-3130-0204}\inst{\ref{aff82},\ref{aff83}}
\and F.~J.~Castander\orcid{0000-0001-7316-4573}\inst{\ref{aff84},\ref{aff85}}
\and M.~Castellano\orcid{0000-0001-9875-8263}\inst{\ref{aff80}}
\and G.~Castignani\orcid{0000-0001-6831-0687}\inst{\ref{aff8}}
\and S.~Cavuoti\orcid{0000-0002-3787-4196}\inst{\ref{aff19},\ref{aff21}}
\and K.~C.~Chambers\orcid{0000-0001-6965-7789}\inst{\ref{aff86}}
\and A.~Cimatti\inst{\ref{aff87}}
\and C.~Colodro-Conde\inst{\ref{aff88}}
\and G.~Congedo\orcid{0000-0003-2508-0046}\inst{\ref{aff89}}
\and C.~J.~Conselice\orcid{0000-0003-1949-7638}\inst{\ref{aff2}}
\and L.~Conversi\orcid{0000-0002-6710-8476}\inst{\ref{aff90},\ref{aff32}}
\and Y.~Copin\orcid{0000-0002-5317-7518}\inst{\ref{aff91}}
\and L.~Corcione\orcid{0000-0002-6497-5881}\inst{\ref{aff70}}
\and H.~M.~Courtois\orcid{0000-0003-0509-1776}\inst{\ref{aff92}}
\and M.~Cropper\orcid{0000-0003-4571-9468}\inst{\ref{aff93}}
\and A.~Da~Silva\orcid{0000-0002-6385-1609}\inst{\ref{aff94},\ref{aff95}}
\and G.~De~Lucia\orcid{0000-0002-6220-9104}\inst{\ref{aff63}}
\and A.~M.~Di~Giorgio\orcid{0000-0002-4767-2360}\inst{\ref{aff96}}
\and C.~Dolding\orcid{0009-0003-7199-6108}\inst{\ref{aff93}}
\and F.~Dubath\orcid{0000-0002-6533-2810}\inst{\ref{aff34}}
\and C.~A.~J.~Duncan\orcid{0009-0003-3573-0791}\inst{\ref{aff2}}
\and X.~Dupac\inst{\ref{aff32}}
\and A.~Ealet\orcid{0000-0003-3070-014X}\inst{\ref{aff91}}
\and S.~Escoffier\orcid{0000-0002-2847-7498}\inst{\ref{aff97}}
\and M.~Fabricius\orcid{0000-0002-7025-6058}\inst{\ref{aff25},\ref{aff24}}
\and M.~Farina\orcid{0000-0002-3089-7846}\inst{\ref{aff96}}
\and R.~Farinelli\inst{\ref{aff8}}
\and F.~Faustini\orcid{0000-0001-6274-5145}\inst{\ref{aff69},\ref{aff80}}
\and F.~Finelli\orcid{0000-0002-6694-3269}\inst{\ref{aff8},\ref{aff98}}
\and S.~Fotopoulou\orcid{0000-0002-9686-254X}\inst{\ref{aff99}}
\and M.~Frailis\orcid{0000-0002-7400-2135}\inst{\ref{aff63}}
\and E.~Franceschi\orcid{0000-0002-0585-6591}\inst{\ref{aff8}}
\and M.~Fumana\orcid{0000-0001-6787-5950}\inst{\ref{aff11}}
\and S.~Galeotta\orcid{0000-0002-3748-5115}\inst{\ref{aff63}}
\and K.~George\orcid{0000-0002-1734-8455}\inst{\ref{aff24}}
\and W.~Gillard\orcid{0000-0003-4744-9748}\inst{\ref{aff97}}
\and B.~Gillis\orcid{0000-0002-4478-1270}\inst{\ref{aff89}}
\and P.~G\'omez-Alvarez\orcid{0000-0002-8594-5358}\inst{\ref{aff100},\ref{aff32}}
\and B.~R.~Granett\orcid{0000-0003-2694-9284}\inst{\ref{aff60}}
\and A.~Grazian\orcid{0000-0002-5688-0663}\inst{\ref{aff67}}
\and F.~Grupp\inst{\ref{aff25},\ref{aff24}}
\and L.~Guzzo\orcid{0000-0001-8264-5192}\inst{\ref{aff10},\ref{aff60},\ref{aff101}}
\and S.~Gwyn\orcid{0000-0001-8221-8406}\inst{\ref{aff102}}
\and S.~V.~H.~Haugan\orcid{0000-0001-9648-7260}\inst{\ref{aff103}}
\and H.~Hoekstra\orcid{0000-0002-0641-3231}\inst{\ref{aff79}}
\and W.~Holmes\inst{\ref{aff104}}
\and I.~M.~Hook\orcid{0000-0002-2960-978X}\inst{\ref{aff105}}
\and F.~Hormuth\inst{\ref{aff106}}
\and A.~Hornstrup\orcid{0000-0002-3363-0936}\inst{\ref{aff107},\ref{aff108}}
\and P.~Hudelot\inst{\ref{aff13}}
\and M.~Jhabvala\inst{\ref{aff109}}
\and B.~Joachimi\orcid{0000-0001-7494-1303}\inst{\ref{aff110}}
\and E.~Keih\"anen\orcid{0000-0003-1804-7715}\inst{\ref{aff111}}
\and S.~Kermiche\orcid{0000-0002-0302-5735}\inst{\ref{aff97}}
\and A.~Kiessling\orcid{0000-0002-2590-1273}\inst{\ref{aff104}}
\and B.~Kubik\orcid{0009-0006-5823-4880}\inst{\ref{aff91}}
\and M.~K\"ummel\orcid{0000-0003-2791-2117}\inst{\ref{aff24}}
\and M.~Kunz\orcid{0000-0002-3052-7394}\inst{\ref{aff112}}
\and H.~Kurki-Suonio\orcid{0000-0002-4618-3063}\inst{\ref{aff113},\ref{aff114}}
\and O.~Lahav\orcid{0000-0002-1134-9035}\inst{\ref{aff110}}
\and Q.~Le~Boulc'h\inst{\ref{aff115}}
\and A.~M.~C.~Le~Brun\orcid{0000-0002-0936-4594}\inst{\ref{aff116}}
\and D.~Le~Mignant\orcid{0000-0002-5339-5515}\inst{\ref{aff12}}
\and S.~Ligori\orcid{0000-0003-4172-4606}\inst{\ref{aff70}}
\and P.~B.~Lilje\orcid{0000-0003-4324-7794}\inst{\ref{aff103}}
\and V.~Lindholm\orcid{0000-0003-2317-5471}\inst{\ref{aff113},\ref{aff114}}
\and I.~Lloro\orcid{0000-0001-5966-1434}\inst{\ref{aff117}}
\and G.~Mainetti\orcid{0000-0003-2384-2377}\inst{\ref{aff115}}
\and D.~Maino\inst{\ref{aff10},\ref{aff11},\ref{aff101}}
\and E.~Maiorano\orcid{0000-0003-2593-4355}\inst{\ref{aff8}}
\and O.~Mansutti\orcid{0000-0001-5758-4658}\inst{\ref{aff63}}
\and S.~Marcin\inst{\ref{aff118}}
\and O.~Marggraf\orcid{0000-0001-7242-3852}\inst{\ref{aff119}}
\and M.~Martinelli\orcid{0000-0002-6943-7732}\inst{\ref{aff80},\ref{aff81}}
\and N.~Martinet\orcid{0000-0003-2786-7790}\inst{\ref{aff12}}
\and F.~Marulli\orcid{0000-0002-8850-0303}\inst{\ref{aff7},\ref{aff8},\ref{aff9}}
\and R.~Massey\orcid{0000-0002-6085-3780}\inst{\ref{aff48}}
\and S.~Maurogordato\inst{\ref{aff120}}
\and H.~J.~McCracken\orcid{0000-0002-9489-7765}\inst{\ref{aff13}}
\and E.~Medinaceli\orcid{0000-0002-4040-7783}\inst{\ref{aff8}}
\and S.~Mei\orcid{0000-0002-2849-559X}\inst{\ref{aff121},\ref{aff122}}
\and Y.~Mellier\inst{\ref{aff123},\ref{aff13}}
\and M.~Meneghetti\orcid{0000-0003-1225-7084}\inst{\ref{aff8},\ref{aff9}}
\and E.~Merlin\orcid{0000-0001-6870-8900}\inst{\ref{aff80}}
\and G.~Meylan\inst{\ref{aff22}}
\and A.~Mora\orcid{0000-0002-1922-8529}\inst{\ref{aff124}}
\and M.~Moresco\orcid{0000-0002-7616-7136}\inst{\ref{aff7},\ref{aff8}}
\and L.~Moscardini\orcid{0000-0002-3473-6716}\inst{\ref{aff7},\ref{aff8},\ref{aff9}}
\and R.~Nakajima\orcid{0009-0009-1213-7040}\inst{\ref{aff119}}
\and C.~Neissner\orcid{0000-0001-8524-4968}\inst{\ref{aff125},\ref{aff83}}
\and R.~C.~Nichol\orcid{0000-0003-0939-6518}\inst{\ref{aff59}}
\and S.-M.~Niemi\inst{\ref{aff77}}
\and C.~Padilla\orcid{0000-0001-7951-0166}\inst{\ref{aff125}}
\and S.~Paltani\orcid{0000-0002-8108-9179}\inst{\ref{aff34}}
\and F.~Pasian\orcid{0000-0002-4869-3227}\inst{\ref{aff63}}
\and K.~Pedersen\inst{\ref{aff126}}
\and W.~J.~Percival\orcid{0000-0002-0644-5727}\inst{\ref{aff127},\ref{aff128},\ref{aff129}}
\and V.~Pettorino\inst{\ref{aff77}}
\and S.~Pires\orcid{0000-0002-0249-2104}\inst{\ref{aff61}}
\and G.~Polenta\orcid{0000-0003-4067-9196}\inst{\ref{aff69}}
\and M.~Poncet\inst{\ref{aff130}}
\and L.~A.~Popa\inst{\ref{aff131}}
\and L.~Pozzetti\orcid{0000-0001-7085-0412}\inst{\ref{aff8}}
\and F.~Raison\orcid{0000-0002-7819-6918}\inst{\ref{aff25}}
\and R.~Rebolo\orcid{0000-0003-3767-7085}\inst{\ref{aff88},\ref{aff132},\ref{aff133}}
\and A.~Renzi\orcid{0000-0001-9856-1970}\inst{\ref{aff134},\ref{aff135}}
\and J.~Rhodes\orcid{0000-0002-4485-8549}\inst{\ref{aff104}}
\and G.~Riccio\inst{\ref{aff19}}
\and E.~Romelli\orcid{0000-0003-3069-9222}\inst{\ref{aff63}}
\and M.~Roncarelli\orcid{0000-0001-9587-7822}\inst{\ref{aff8}}
\and R.~Saglia\orcid{0000-0003-0378-7032}\inst{\ref{aff24},\ref{aff25}}
\and Z.~Sakr\orcid{0000-0002-4823-3757}\inst{\ref{aff136},\ref{aff137},\ref{aff138}}
\and A.~G.~S\'anchez\orcid{0000-0003-1198-831X}\inst{\ref{aff25}}
\and D.~Sapone\orcid{0000-0001-7089-4503}\inst{\ref{aff139}}
\and B.~Sartoris\orcid{0000-0003-1337-5269}\inst{\ref{aff24},\ref{aff63}}
\and J.~A.~Schewtschenko\orcid{0000-0002-4913-6393}\inst{\ref{aff89}}
\and M.~Schirmer\orcid{0000-0003-2568-9994}\inst{\ref{aff41}}
\and P.~Schneider\orcid{0000-0001-8561-2679}\inst{\ref{aff119}}
\and T.~Schrabback\orcid{0000-0002-6987-7834}\inst{\ref{aff140}}
\and A.~Secroun\orcid{0000-0003-0505-3710}\inst{\ref{aff97}}
\and G.~Seidel\orcid{0000-0003-2907-353X}\inst{\ref{aff41}}
\and M.~Seiffert\orcid{0000-0002-7536-9393}\inst{\ref{aff104}}
\and S.~Serrano\orcid{0000-0002-0211-2861}\inst{\ref{aff85},\ref{aff141},\ref{aff84}}
\and P.~Simon\inst{\ref{aff119}}
\and C.~Sirignano\orcid{0000-0002-0995-7146}\inst{\ref{aff134},\ref{aff135}}
\and G.~Sirri\orcid{0000-0003-2626-2853}\inst{\ref{aff9}}
\and A.~Spurio~Mancini\orcid{0000-0001-5698-0990}\inst{\ref{aff142}}
\and L.~Stanco\orcid{0000-0002-9706-5104}\inst{\ref{aff135}}
\and J.~Steinwagner\orcid{0000-0001-7443-1047}\inst{\ref{aff25}}
\and P.~Tallada-Cresp\'{i}\orcid{0000-0002-1336-8328}\inst{\ref{aff82},\ref{aff83}}
\and A.~N.~Taylor\inst{\ref{aff89}}
\and I.~Tereno\inst{\ref{aff94},\ref{aff143}}
\and N.~Tessore\orcid{0000-0002-9696-7931}\inst{\ref{aff110}}
\and S.~Toft\orcid{0000-0003-3631-7176}\inst{\ref{aff144},\ref{aff145}}
\and R.~Toledo-Moreo\orcid{0000-0002-2997-4859}\inst{\ref{aff146}}
\and F.~Torradeflot\orcid{0000-0003-1160-1517}\inst{\ref{aff83},\ref{aff82}}
\and I.~Tutusaus\orcid{0000-0002-3199-0399}\inst{\ref{aff137}}
\and E.~A.~Valentijn\inst{\ref{aff51}}
\and L.~Valenziano\orcid{0000-0002-1170-0104}\inst{\ref{aff8},\ref{aff98}}
\and J.~Valiviita\orcid{0000-0001-6225-3693}\inst{\ref{aff113},\ref{aff114}}
\and T.~Vassallo\orcid{0000-0001-6512-6358}\inst{\ref{aff24},\ref{aff63}}
\and G.~Verdoes~Kleijn\orcid{0000-0001-5803-2580}\inst{\ref{aff51}}
\and A.~Veropalumbo\orcid{0000-0003-2387-1194}\inst{\ref{aff60},\ref{aff72},\ref{aff71}}
\and Y.~Wang\orcid{0000-0002-4749-2984}\inst{\ref{aff147}}
\and J.~Weller\orcid{0000-0002-8282-2010}\inst{\ref{aff24},\ref{aff25}}
\and A.~Zacchei\orcid{0000-0003-0396-1192}\inst{\ref{aff63},\ref{aff62}}
\and G.~Zamorani\orcid{0000-0002-2318-301X}\inst{\ref{aff8}}
\and F.~M.~Zerbi\inst{\ref{aff60}}
\and E.~Zucca\orcid{0000-0002-5845-8132}\inst{\ref{aff8}}
\and V.~Allevato\orcid{0000-0001-7232-5152}\inst{\ref{aff19}}
\and M.~Ballardini\orcid{0000-0003-4481-3559}\inst{\ref{aff148},\ref{aff149},\ref{aff8}}
\and M.~Bolzonella\orcid{0000-0003-3278-4607}\inst{\ref{aff8}}
\and E.~Bozzo\orcid{0000-0002-8201-1525}\inst{\ref{aff34}}
\and C.~Burigana\orcid{0000-0002-3005-5796}\inst{\ref{aff150},\ref{aff98}}
\and R.~Cabanac\orcid{0000-0001-6679-2600}\inst{\ref{aff137}}
\and A.~Cappi\inst{\ref{aff8},\ref{aff120}}
\and D.~Di~Ferdinando\inst{\ref{aff9}}
\and J.~A.~Escartin~Vigo\inst{\ref{aff25}}
\and L.~Gabarra\orcid{0000-0002-8486-8856}\inst{\ref{aff3}}
\and M.~Huertas-Company\orcid{0000-0002-1416-8483}\inst{\ref{aff88},\ref{aff37},\ref{aff151},\ref{aff152}}
\and J.~Mart\'{i}n-Fleitas\orcid{0000-0002-8594-569X}\inst{\ref{aff124}}
\and S.~Matthew\orcid{0000-0001-8448-1697}\inst{\ref{aff89}}
\and N.~Mauri\orcid{0000-0001-8196-1548}\inst{\ref{aff87},\ref{aff9}}
\and A.~Pezzotta\orcid{0000-0003-0726-2268}\inst{\ref{aff153},\ref{aff25}}
\and M.~P\"ontinen\orcid{0000-0001-5442-2530}\inst{\ref{aff113}}
\and C.~Porciani\orcid{0000-0002-7797-2508}\inst{\ref{aff119}}
\and I.~Risso\orcid{0000-0003-2525-7761}\inst{\ref{aff154}}
\and V.~Scottez\inst{\ref{aff123},\ref{aff155}}
\and M.~Sereno\orcid{0000-0003-0302-0325}\inst{\ref{aff8},\ref{aff9}}
\and M.~Tenti\orcid{0000-0002-4254-5901}\inst{\ref{aff9}}
\and M.~Viel\orcid{0000-0002-2642-5707}\inst{\ref{aff62},\ref{aff63},\ref{aff65},\ref{aff64},\ref{aff156}}
\and M.~Wiesmann\orcid{0009-0000-8199-5860}\inst{\ref{aff103}}
\and Y.~Akrami\orcid{0000-0002-2407-7956}\inst{\ref{aff157},\ref{aff158}}
\and S.~Anselmi\orcid{0000-0002-3579-9583}\inst{\ref{aff135},\ref{aff134},\ref{aff159}}
\and M.~Archidiacono\orcid{0000-0003-4952-9012}\inst{\ref{aff10},\ref{aff101}}
\and F.~Atrio-Barandela\orcid{0000-0002-2130-2513}\inst{\ref{aff160}}
\and C.~Benoist\inst{\ref{aff120}}
\and K.~Benson\inst{\ref{aff93}}
\and P.~Bergamini\orcid{0000-0003-1383-9414}\inst{\ref{aff10},\ref{aff8}}
\and D.~Bertacca\orcid{0000-0002-2490-7139}\inst{\ref{aff134},\ref{aff67},\ref{aff135}}
\and M.~Bethermin\orcid{0000-0002-3915-2015}\inst{\ref{aff29}}
\and L.~Blot\orcid{0000-0002-9622-7167}\inst{\ref{aff161},\ref{aff159}}
\and S.~Borgani\orcid{0000-0001-6151-6439}\inst{\ref{aff162},\ref{aff62},\ref{aff63},\ref{aff64},\ref{aff156}}
\and M.~L.~Brown\orcid{0000-0002-0370-8077}\inst{\ref{aff2}}
\and S.~Bruton\orcid{0000-0002-6503-5218}\inst{\ref{aff163}}
\and A.~Calabro\orcid{0000-0003-2536-1614}\inst{\ref{aff80}}
\and B.~Camacho~Quevedo\orcid{0000-0002-8789-4232}\inst{\ref{aff85},\ref{aff84}}
\and F.~Caro\inst{\ref{aff80}}
\and C.~S.~Carvalho\inst{\ref{aff143}}
\and T.~Castro\orcid{0000-0002-6292-3228}\inst{\ref{aff63},\ref{aff64},\ref{aff62},\ref{aff156}}
\and Y.~Charles\inst{\ref{aff12}}
\and F.~Cogato\orcid{0000-0003-4632-6113}\inst{\ref{aff7},\ref{aff8}}
\and A.~R.~Cooray\orcid{0000-0002-3892-0190}\inst{\ref{aff164}}
\and O.~Cucciati\orcid{0000-0002-9336-7551}\inst{\ref{aff8}}
\and S.~Davini\orcid{0000-0003-3269-1718}\inst{\ref{aff72}}
\and F.~De~Paolis\orcid{0000-0001-6460-7563}\inst{\ref{aff165},\ref{aff166},\ref{aff167}}
\and G.~Desprez\orcid{0000-0001-8325-1742}\inst{\ref{aff51}}
\and A.~D\'iaz-S\'anchez\orcid{0000-0003-0748-4768}\inst{\ref{aff16}}
\and J.~J.~Diaz\inst{\ref{aff88}}
\and S.~Di~Domizio\orcid{0000-0003-2863-5895}\inst{\ref{aff71},\ref{aff72}}
\and A.~Enia\orcid{0000-0002-0200-2857}\inst{\ref{aff66},\ref{aff8}}
\and Y.~Fang\inst{\ref{aff24}}
\and A.~G.~Ferrari\orcid{0009-0005-5266-4110}\inst{\ref{aff9}}
\and A.~Finoguenov\orcid{0000-0002-4606-5403}\inst{\ref{aff113}}
\and A.~Fontana\orcid{0000-0003-3820-2823}\inst{\ref{aff80}}
\and A.~Franco\orcid{0000-0002-4761-366X}\inst{\ref{aff166},\ref{aff165},\ref{aff167}}
\and K.~Ganga\orcid{0000-0001-8159-8208}\inst{\ref{aff121}}
\and J.~Garc\'ia-Bellido\orcid{0000-0002-9370-8360}\inst{\ref{aff157}}
\and T.~Gasparetto\orcid{0000-0002-7913-4866}\inst{\ref{aff63}}
\and V.~Gautard\inst{\ref{aff168}}
\and E.~Gaztanaga\orcid{0000-0001-9632-0815}\inst{\ref{aff84},\ref{aff85},\ref{aff4}}
\and F.~Giacomini\orcid{0000-0002-3129-2814}\inst{\ref{aff9}}
\and G.~Gozaliasl\orcid{0000-0002-0236-919X}\inst{\ref{aff169},\ref{aff113}}
\and M.~Guidi\orcid{0000-0001-9408-1101}\inst{\ref{aff66},\ref{aff8}}
\and C.~M.~Gutierrez\orcid{0000-0001-7854-783X}\inst{\ref{aff170}}
\and A.~Hall\orcid{0000-0002-3139-8651}\inst{\ref{aff89}}
\and W.~G.~Hartley\inst{\ref{aff34}}
\and S.~Hemmati\orcid{0000-0003-2226-5395}\inst{\ref{aff39}}
\and C.~Hern\'andez-Monteagudo\orcid{0000-0001-5471-9166}\inst{\ref{aff133},\ref{aff88}}
\and H.~Hildebrandt\orcid{0000-0002-9814-3338}\inst{\ref{aff171}}
\and J.~Hjorth\orcid{0000-0002-4571-2306}\inst{\ref{aff126}}
\and J.~J.~E.~Kajava\orcid{0000-0002-3010-8333}\inst{\ref{aff172},\ref{aff173}}
\and Y.~Kang\orcid{0009-0000-8588-7250}\inst{\ref{aff34}}
\and V.~Kansal\orcid{0000-0002-4008-6078}\inst{\ref{aff174},\ref{aff175}}
\and D.~Karagiannis\orcid{0000-0002-4927-0816}\inst{\ref{aff148},\ref{aff176}}
\and K.~Kiiveri\inst{\ref{aff111}}
\and C.~C.~Kirkpatrick\inst{\ref{aff111}}
\and J.~Le~Graet\orcid{0000-0001-6523-7971}\inst{\ref{aff97}}
\and L.~Legrand\orcid{0000-0003-0610-5252}\inst{\ref{aff177},\ref{aff178}}
\and M.~Lembo\orcid{0000-0002-5271-5070}\inst{\ref{aff148},\ref{aff149}}
\and F.~Lepori\orcid{0009-0000-5061-7138}\inst{\ref{aff179}}
\and G.~Leroy\orcid{0009-0004-2523-4425}\inst{\ref{aff47},\ref{aff48}}
\and G.~F.~Lesci\orcid{0000-0002-4607-2830}\inst{\ref{aff7},\ref{aff8}}
\and J.~Lesgourgues\orcid{0000-0001-7627-353X}\inst{\ref{aff36}}
\and T.~I.~Liaudat\orcid{0000-0002-9104-314X}\inst{\ref{aff180}}
\and A.~Loureiro\orcid{0000-0002-4371-0876}\inst{\ref{aff181},\ref{aff182}}
\and J.~Macias-Perez\orcid{0000-0002-5385-2763}\inst{\ref{aff183}}
\and G.~Maggio\orcid{0000-0003-4020-4836}\inst{\ref{aff63}}
\and M.~Magliocchetti\orcid{0000-0001-9158-4838}\inst{\ref{aff96}}
\and E.~A.~Magnier\orcid{0000-0002-7965-2815}\inst{\ref{aff86}}
\and F.~Mannucci\orcid{0000-0002-4803-2381}\inst{\ref{aff57}}
\and R.~Maoli\orcid{0000-0002-6065-3025}\inst{\ref{aff45},\ref{aff80}}
\and C.~J.~A.~P.~Martins\orcid{0000-0002-4886-9261}\inst{\ref{aff184},\ref{aff73}}
\and L.~Maurin\orcid{0000-0002-8406-0857}\inst{\ref{aff23}}
\and M.~Miluzio\inst{\ref{aff32},\ref{aff185}}
\and P.~Monaco\orcid{0000-0003-2083-7564}\inst{\ref{aff162},\ref{aff63},\ref{aff64},\ref{aff62}}
\and C.~Moretti\orcid{0000-0003-3314-8936}\inst{\ref{aff65},\ref{aff156},\ref{aff63},\ref{aff62},\ref{aff64}}
\and G.~Morgante\inst{\ref{aff8}}
\and C.~Murray\inst{\ref{aff121}}
\and S.~Nadathur\orcid{0000-0001-9070-3102}\inst{\ref{aff4}}
\and K.~Naidoo\orcid{0000-0002-9182-1802}\inst{\ref{aff4}}
\and A.~Navarro-Alsina\orcid{0000-0002-3173-2592}\inst{\ref{aff119}}
\and S.~Nesseris\orcid{0000-0002-0567-0324}\inst{\ref{aff157}}
\and F.~Passalacqua\orcid{0000-0002-8606-4093}\inst{\ref{aff134},\ref{aff135}}
\and K.~Paterson\orcid{0000-0001-8340-3486}\inst{\ref{aff41}}
\and L.~Patrizii\inst{\ref{aff9}}
\and A.~Pisani\orcid{0000-0002-6146-4437}\inst{\ref{aff97},\ref{aff186}}
\and D.~Potter\orcid{0000-0002-0757-5195}\inst{\ref{aff179}}
\and S.~Quai\orcid{0000-0002-0449-8163}\inst{\ref{aff7},\ref{aff8}}
\and M.~Radovich\orcid{0000-0002-3585-866X}\inst{\ref{aff67}}
\and P.-F.~Rocci\inst{\ref{aff23}}
\and S.~Sacquegna\orcid{0000-0002-8433-6630}\inst{\ref{aff165},\ref{aff166},\ref{aff167}}
\and M.~Sahl\'en\orcid{0000-0003-0973-4804}\inst{\ref{aff187}}
\and D.~B.~Sanders\orcid{0000-0002-1233-9998}\inst{\ref{aff86}}
\and E.~Sarpa\orcid{0000-0002-1256-655X}\inst{\ref{aff65},\ref{aff156},\ref{aff64}}
\and C.~Scarlata\orcid{0000-0002-9136-8876}\inst{\ref{aff50}}
\and J.~Schaye\orcid{0000-0002-0668-5560}\inst{\ref{aff79}}
\and A.~Schneider\orcid{0000-0001-7055-8104}\inst{\ref{aff179}}
\and D.~Sciotti\orcid{0009-0008-4519-2620}\inst{\ref{aff80},\ref{aff81}}
\and E.~Sellentin\inst{\ref{aff188},\ref{aff79}}
\and F.~Shankar\orcid{0000-0001-8973-5051}\inst{\ref{aff189}}
\and L.~C.~Smith\orcid{0000-0002-3259-2771}\inst{\ref{aff190}}
\and K.~Tanidis\orcid{0000-0001-9843-5130}\inst{\ref{aff3}}
\and G.~Testera\inst{\ref{aff72}}
\and R.~Teyssier\orcid{0000-0001-7689-0933}\inst{\ref{aff186}}
\and S.~Tosi\orcid{0000-0002-7275-9193}\inst{\ref{aff71},\ref{aff154}}
\and A.~Troja\orcid{0000-0003-0239-4595}\inst{\ref{aff134},\ref{aff135}}
\and M.~Tucci\inst{\ref{aff34}}
\and C.~Valieri\inst{\ref{aff9}}
\and A.~Venhola\orcid{0000-0001-6071-4564}\inst{\ref{aff191}}
\and D.~Vergani\orcid{0000-0003-0898-2216}\inst{\ref{aff8}}
\and G.~Vernardos\orcid{0000-0001-8554-7248}\inst{\ref{aff192},\ref{aff193}}
\and G.~Verza\orcid{0000-0002-1886-8348}\inst{\ref{aff194}}
\and P.~Vielzeuf\orcid{0000-0003-2035-9339}\inst{\ref{aff97}}
\and N.~A.~Walton\orcid{0000-0003-3983-8778}\inst{\ref{aff190}}
\and D.~Scott\orcid{0000-0002-6878-9840}\inst{\ref{aff195}}}
										   
\institute{David A. Dunlap Department of Astronomy \& Astrophysics, University of Toronto, 50 St George Street, Toronto, Ontario M5S 3H4, Canada\label{aff1}
\and
Jodrell Bank Centre for Astrophysics, Department of Physics and Astronomy, University of Manchester, Oxford Road, Manchester M13 9PL, UK\label{aff2}
\and
Department of Physics, Oxford University, Keble Road, Oxford OX1 3RH, UK\label{aff3}
\and
Institute of Cosmology and Gravitation, University of Portsmouth, Portsmouth PO1 3FX, UK\label{aff4}
\and
University of Applied Sciences and Arts of Northwestern Switzerland, School of Engineering, 5210 Windisch, Switzerland\label{aff5}
\and
School of Mathematics, Statistics and Physics, Newcastle University, Herschel Building, Newcastle-upon-Tyne, NE1 7RU, UK\label{aff6}
\and
Dipartimento di Fisica e Astronomia "Augusto Righi" - Alma Mater Studiorum Universit\`a di Bologna, via Piero Gobetti 93/2, 40129 Bologna, Italy\label{aff7}
\and
INAF-Osservatorio di Astrofisica e Scienza dello Spazio di Bologna, Via Piero Gobetti 93/3, 40129 Bologna, Italy\label{aff8}
\and
INFN-Sezione di Bologna, Viale Berti Pichat 6/2, 40127 Bologna, Italy\label{aff9}
\and
Dipartimento di Fisica "Aldo Pontremoli", Universit\`a degli Studi di Milano, Via Celoria 16, 20133 Milano, Italy\label{aff10}
\and
INAF-IASF Milano, Via Alfonso Corti 12, 20133 Milano, Italy\label{aff11}
\and
Aix-Marseille Universit\'e, CNRS, CNES, LAM, Marseille, France\label{aff12}
\and
Institut d'Astrophysique de Paris, UMR 7095, CNRS, and Sorbonne Universit\'e, 98 bis boulevard Arago, 75014 Paris, France\label{aff13}
\and
Max-Planck-Institut f\"ur Astrophysik, Karl-Schwarzschild-Str.~1, 85748 Garching, Germany\label{aff14}
\and
Technical University of Munich, TUM School of Natural Sciences, Physics Department, James-Franck-Str.~1, 85748 Garching, Germany\label{aff15}
\and
Departamento F\'isica Aplicada, Universidad Polit\'ecnica de Cartagena, Campus Muralla del Mar, 30202 Cartagena, Murcia, Spain\label{aff16}
\and
School of Physical Sciences, The Open University, Milton Keynes, MK7 6AA, UK\label{aff17}
\and
Institut de Ci\`{e}ncies del Cosmos (ICCUB), Universitat de Barcelona (IEEC-UB), Mart\'{i} i Franqu\`{e}s 1, 08028 Barcelona, Spain\label{aff18}
\and
INAF-Osservatorio Astronomico di Capodimonte, Via Moiariello 16, 80131 Napoli, Italy\label{aff19}
\and
Department of Physics "E. Pancini", University Federico II, Via Cinthia 6, 80126, Napoli, Italy\label{aff20}
\and
INFN section of Naples, Via Cinthia 6, 80126, Napoli, Italy\label{aff21}
\and
Institute of Physics, Laboratory of Astrophysics, Ecole Polytechnique F\'ed\'erale de Lausanne (EPFL), Observatoire de Sauverny, 1290 Versoix, Switzerland\label{aff22}
\and
Universit\'e Paris-Saclay, CNRS, Institut d'astrophysique spatiale, 91405, Orsay, France\label{aff23}
\and
Universit\"ats-Sternwarte M\"unchen, Fakult\"at f\"ur Physik, Ludwig-Maximilians-Universit\"at M\"unchen, Scheinerstrasse 1, 81679 M\"unchen, Germany\label{aff24}
\and
Max Planck Institute for Extraterrestrial Physics, Giessenbachstr. 1, 85748 Garching, Germany\label{aff25}
\and
Kavli Institute for Particle Astrophysics \& Cosmology (KIPAC), Stanford University, Stanford, CA 94305, USA\label{aff26}
\and
SLAC National Accelerator Laboratory, 2575 Sand Hill Road, Menlo Park, CA 94025, USA\label{aff27}
\and
The Inter-University Centre for Astronomy and Astrophysics, Post Bag 4, Ganeshkhind, Pune 411007, India\label{aff28}
\and
Universit\'e de Strasbourg, CNRS, Observatoire astronomique de Strasbourg, UMR 7550, 67000 Strasbourg, France\label{aff29}
\and
Citizen Scientist, Zooniverse c/o University of Oxford,  Keble Road, Oxford OX1 3RH, UK\label{aff30}
\and
Center for Interdisciplinary Exploration and Research in Astrophysics (CIERA) and Department of Physics and Astronomy, Northwestern University, 1800 Sherman Ave., Evanston, IL 60201, USA\label{aff31}
\and
ESAC/ESA, Camino Bajo del Castillo, s/n., Urb. Villafranca del Castillo, 28692 Villanueva de la Ca\~nada, Madrid, Spain\label{aff32}
\and
SCITAS, Ecole Polytechnique F\'ed\'erale de Lausanne (EPFL), 1015 Lausanne, Switzerland\label{aff33}
\and
Department of Astronomy, University of Geneva, ch. d'Ecogia 16, 1290 Versoix, Switzerland\label{aff34}
\and
Instituci\'o Catalana de Recerca i Estudis Avan\c{c}ats (ICREA), Passeig de Llu\'{\i}s Companys 23, 08010 Barcelona, Spain\label{aff35}
\and
Institute for Theoretical Particle Physics and Cosmology (TTK), RWTH Aachen University, 52056 Aachen, Germany\label{aff36}
\and
Instituto de Astrof\'isica de Canarias (IAC); Departamento de Astrof\'isica, Universidad de La Laguna (ULL), 38200, La Laguna, Tenerife, Spain\label{aff37}
\and
Instituto de F\'isica de Cantabria, Edificio Juan Jord\'a, Avenida de los Castros, 39005 Santander, Spain\label{aff38}
\and
Caltech/IPAC, 1200 E. California Blvd., Pasadena, CA 91125, USA\label{aff39}
\and
Laboratoire univers et particules de Montpellier, Universit\'e de Montpellier, CNRS, 34090 Montpellier, France\label{aff40}
\and
Max-Planck-Institut f\"ur Astronomie, K\"onigstuhl 17, 69117 Heidelberg, Germany\label{aff41}
\and
ELTE E\"otv\"os Lor\'and University, Institute of Physics and Astronomy, P\'azm\'any P. st. 1/A, H-1171 Budapest, Hungary\label{aff42}
\and
MTA-CSFK Lend\"ulet Large-Scale Structure Research Group, Konkoly-Thege Mikl\'os \'ut 15-17, H-1121 Budapest, Hungary\label{aff43}
\and
Konkoly Observatory, HUN-REN CSFK, MTA Centre of Excellence, Budapest, Konkoly Thege Mikl\'os {\'u}t 15-17. H-1121, Hungary\label{aff44}
\and
Dipartimento di Fisica, Sapienza Universit\`a di Roma, Piazzale Aldo Moro 2, 00185 Roma, Italy\label{aff45}
\and
STAR Institute, University of Li{\`e}ge, Quartier Agora, All\'ee du six Ao\^ut 19c, 4000 Li\`ege, Belgium\label{aff46}
\and
Department of Physics, Centre for Extragalactic Astronomy, Durham University, South Road, Durham, DH1 3LE, UK\label{aff47}
\and
Department of Physics, Institute for Computational Cosmology, Durham University, South Road, Durham, DH1 3LE, UK\label{aff48}
\and
Institute for Particle Physics and Astrophysics, Dept. of Physics, ETH Zurich, Wolfgang-Pauli-Strasse 27, 8093 Zurich, Switzerland\label{aff49}
\and
Minnesota Institute for Astrophysics, University of Minnesota, 116 Church St SE, Minneapolis, MN 55455, USA\label{aff50}
\and
Kapteyn Astronomical Institute, University of Groningen, PO Box 800, 9700 AV Groningen, The Netherlands\label{aff51}
\and
Lawrence Berkeley National Laboratory, One Cyclotron Road, Berkeley, CA 94720, USA\label{aff52}
\and
Department of Astronomy, School of Physics and Astronomy, Shanghai Jiao Tong University, Shanghai 200240, China\label{aff53}
\and
National Astronomical Observatory of Japan, 2-21-1 Osawa, Mitaka, Tokyo 181-8588, Japan\label{aff54}
\and
University of Trento, Via Sommarive 14, I-38123 Trento, Italy\label{aff55}
\and
Dipartimento di Fisica e Astronomia, Universit\`{a} di Firenze, via G. Sansone 1, 50019 Sesto Fiorentino, Firenze, Italy\label{aff56}
\and
INAF-Osservatorio Astrofisico di Arcetri, Largo E. Fermi 5, 50125, Firenze, Italy\label{aff57}
\and
Max Planck Institute for Gravitational Physics (Albert Einstein Institute), Am Muhlenberg 1, D-14476 Potsdam-Golm, Germany\label{aff58}
\and
School of Mathematics and Physics, University of Surrey, Guildford, Surrey, GU2 7XH, UK\label{aff59}
\and
INAF-Osservatorio Astronomico di Brera, Via Brera 28, 20122 Milano, Italy\label{aff60}
\and
Universit\'e Paris-Saclay, Universit\'e Paris Cit\'e, CEA, CNRS, AIM, 91191, Gif-sur-Yvette, France\label{aff61}
\and
IFPU, Institute for Fundamental Physics of the Universe, via Beirut 2, 34151 Trieste, Italy\label{aff62}
\and
INAF-Osservatorio Astronomico di Trieste, Via G. B. Tiepolo 11, 34143 Trieste, Italy\label{aff63}
\and
INFN, Sezione di Trieste, Via Valerio 2, 34127 Trieste TS, Italy\label{aff64}
\and
SISSA, International School for Advanced Studies, Via Bonomea 265, 34136 Trieste TS, Italy\label{aff65}
\and
Dipartimento di Fisica e Astronomia, Universit\`a di Bologna, Via Gobetti 93/2, 40129 Bologna, Italy\label{aff66}
\and
INAF-Osservatorio Astronomico di Padova, Via dell'Osservatorio 5, 35122 Padova, Italy\label{aff67}
\and
Institut de Physique Th\'eorique, CEA, CNRS, Universit\'e Paris-Saclay 91191 Gif-sur-Yvette Cedex, France\label{aff68}
\and
Space Science Data Center, Italian Space Agency, via del Politecnico snc, 00133 Roma, Italy\label{aff69}
\and
INAF-Osservatorio Astrofisico di Torino, Via Osservatorio 20, 10025 Pino Torinese (TO), Italy\label{aff70}
\and
Dipartimento di Fisica, Universit\`a di Genova, Via Dodecaneso 33, 16146, Genova, Italy\label{aff71}
\and
INFN-Sezione di Genova, Via Dodecaneso 33, 16146, Genova, Italy\label{aff72}
\and
Instituto de Astrof\'isica e Ci\^encias do Espa\c{c}o, Universidade do Porto, CAUP, Rua das Estrelas, PT4150-762 Porto, Portugal\label{aff73}
\and
Faculdade de Ci\^encias da Universidade do Porto, Rua do Campo de Alegre, 4150-007 Porto, Portugal\label{aff74}
\and
Dipartimento di Fisica, Universit\`a degli Studi di Torino, Via P. Giuria 1, 10125 Torino, Italy\label{aff75}
\and
INFN-Sezione di Torino, Via P. Giuria 1, 10125 Torino, Italy\label{aff76}
\and
European Space Agency/ESTEC, Keplerlaan 1, 2201 AZ Noordwijk, The Netherlands\label{aff77}
\and
Institute Lorentz, Leiden University, Niels Bohrweg 2, 2333 CA Leiden, The Netherlands\label{aff78}
\and
Leiden Observatory, Leiden University, Einsteinweg 55, 2333 CC Leiden, The Netherlands\label{aff79}
\and
INAF-Osservatorio Astronomico di Roma, Via Frascati 33, 00078 Monteporzio Catone, Italy\label{aff80}
\and
INFN-Sezione di Roma, Piazzale Aldo Moro, 2 - c/o Dipartimento di Fisica, Edificio G. Marconi, 00185 Roma, Italy\label{aff81}
\and
Centro de Investigaciones Energ\'eticas, Medioambientales y Tecnol\'ogicas (CIEMAT), Avenida Complutense 40, 28040 Madrid, Spain\label{aff82}
\and
Port d'Informaci\'{o} Cient\'{i}fica, Campus UAB, C. Albareda s/n, 08193 Bellaterra (Barcelona), Spain\label{aff83}
\and
Institute of Space Sciences (ICE, CSIC), Campus UAB, Carrer de Can Magrans, s/n, 08193 Barcelona, Spain\label{aff84}
\and
Institut d'Estudis Espacials de Catalunya (IEEC),  Edifici RDIT, Campus UPC, 08860 Castelldefels, Barcelona, Spain\label{aff85}
\and
Institute for Astronomy, University of Hawaii, 2680 Woodlawn Drive, Honolulu, HI 96822, USA\label{aff86}
\and
Dipartimento di Fisica e Astronomia "Augusto Righi" - Alma Mater Studiorum Universit\`a di Bologna, Viale Berti Pichat 6/2, 40127 Bologna, Italy\label{aff87}
\and
Instituto de Astrof\'{\i}sica de Canarias, V\'{\i}a L\'actea, 38205 La Laguna, Tenerife, Spain\label{aff88}
\and
Institute for Astronomy, University of Edinburgh, Royal Observatory, Blackford Hill, Edinburgh EH9 3HJ, UK\label{aff89}
\and
European Space Agency/ESRIN, Largo Galileo Galilei 1, 00044 Frascati, Roma, Italy\label{aff90}
\and
Universit\'e Claude Bernard Lyon 1, CNRS/IN2P3, IP2I Lyon, UMR 5822, Villeurbanne, F-69100, France\label{aff91}
\and
UCB Lyon 1, CNRS/IN2P3, IUF, IP2I Lyon, 4 rue Enrico Fermi, 69622 Villeurbanne, France\label{aff92}
\and
Mullard Space Science Laboratory, University College London, Holmbury St Mary, Dorking, Surrey RH5 6NT, UK\label{aff93}
\and
Departamento de F\'isica, Faculdade de Ci\^encias, Universidade de Lisboa, Edif\'icio C8, Campo Grande, PT1749-016 Lisboa, Portugal\label{aff94}
\and
Instituto de Astrof\'isica e Ci\^encias do Espa\c{c}o, Faculdade de Ci\^encias, Universidade de Lisboa, Campo Grande, 1749-016 Lisboa, Portugal\label{aff95}
\and
INAF-Istituto di Astrofisica e Planetologia Spaziali, via del Fosso del Cavaliere, 100, 00100 Roma, Italy\label{aff96}
\and
Aix-Marseille Universit\'e, CNRS/IN2P3, CPPM, Marseille, France\label{aff97}
\and
INFN-Bologna, Via Irnerio 46, 40126 Bologna, Italy\label{aff98}
\and
School of Physics, HH Wills Physics Laboratory, University of Bristol, Tyndall Avenue, Bristol, BS8 1TL, UK\label{aff99}
\and
FRACTAL S.L.N.E., calle Tulip\'an 2, Portal 13 1A, 28231, Las Rozas de Madrid, Spain\label{aff100}
\and
INFN-Sezione di Milano, Via Celoria 16, 20133 Milano, Italy\label{aff101}
\and
NRC Herzberg, 5071 West Saanich Rd, Victoria, BC V9E 2E7, Canada\label{aff102}
\and
Institute of Theoretical Astrophysics, University of Oslo, P.O. Box 1029 Blindern, 0315 Oslo, Norway\label{aff103}
\and
Jet Propulsion Laboratory, California Institute of Technology, 4800 Oak Grove Drive, Pasadena, CA, 91109, USA\label{aff104}
\and
Department of Physics, Lancaster University, Lancaster, LA1 4YB, UK\label{aff105}
\and
Felix Hormuth Engineering, Goethestr. 17, 69181 Leimen, Germany\label{aff106}
\and
Technical University of Denmark, Elektrovej 327, 2800 Kgs. Lyngby, Denmark\label{aff107}
\and
Cosmic Dawn Center (DAWN), Denmark\label{aff108}
\and
NASA Goddard Space Flight Center, Greenbelt, MD 20771, USA\label{aff109}
\and
Department of Physics and Astronomy, University College London, Gower Street, London WC1E 6BT, UK\label{aff110}
\and
Department of Physics and Helsinki Institute of Physics, Gustaf H\"allstr\"omin katu 2, 00014 University of Helsinki, Finland\label{aff111}
\and
Universit\'e de Gen\`eve, D\'epartement de Physique Th\'eorique and Centre for Astroparticle Physics, 24 quai Ernest-Ansermet, CH-1211 Gen\`eve 4, Switzerland\label{aff112}
\and
Department of Physics, P.O. Box 64, 00014 University of Helsinki, Finland\label{aff113}
\and
Helsinki Institute of Physics, Gustaf H{\"a}llstr{\"o}min katu 2, University of Helsinki, Helsinki, Finland\label{aff114}
\and
Centre de Calcul de l'IN2P3/CNRS, 21 avenue Pierre de Coubertin 69627 Villeurbanne Cedex, France\label{aff115}
\and
Laboratoire d'etude de l'Univers et des phenomenes eXtremes, Observatoire de Paris, Universit\'e PSL, Sorbonne Universit\'e, CNRS, 92190 Meudon, France\label{aff116}
\and
SKA Observatory, Jodrell Bank, Lower Withington, Macclesfield, Cheshire SK11 9FT, UK\label{aff117}
\and
University of Applied Sciences and Arts of Northwestern Switzerland, School of Computer Science, 5210 Windisch, Switzerland\label{aff118}
\and
Universit\"at Bonn, Argelander-Institut f\"ur Astronomie, Auf dem H\"ugel 71, 53121 Bonn, Germany\label{aff119}
\and
Universit\'e C\^{o}te d'Azur, Observatoire de la C\^{o}te d'Azur, CNRS, Laboratoire Lagrange, Bd de l'Observatoire, CS 34229, 06304 Nice cedex 4, France\label{aff120}
\and
Universit\'e Paris Cit\'e, CNRS, Astroparticule et Cosmologie, 75013 Paris, France\label{aff121}
\and
CNRS-UCB International Research Laboratory, Centre Pierre Binetruy, IRL2007, CPB-IN2P3, Berkeley, USA\label{aff122}
\and
Institut d'Astrophysique de Paris, 98bis Boulevard Arago, 75014, Paris, France\label{aff123}
\and
Aurora Technology for European Space Agency (ESA), Camino bajo del Castillo, s/n, Urbanizacion Villafranca del Castillo, Villanueva de la Ca\~nada, 28692 Madrid, Spain\label{aff124}
\and
Institut de F\'{i}sica d'Altes Energies (IFAE), The Barcelona Institute of Science and Technology, Campus UAB, 08193 Bellaterra (Barcelona), Spain\label{aff125}
\and
DARK, Niels Bohr Institute, University of Copenhagen, Jagtvej 155, 2200 Copenhagen, Denmark\label{aff126}
\and
Waterloo Centre for Astrophysics, University of Waterloo, Waterloo, Ontario N2L 3G1, Canada\label{aff127}
\and
Department of Physics and Astronomy, University of Waterloo, Waterloo, Ontario N2L 3G1, Canada\label{aff128}
\and
Perimeter Institute for Theoretical Physics, Waterloo, Ontario N2L 2Y5, Canada\label{aff129}
\and
Centre National d'Etudes Spatiales -- Centre spatial de Toulouse, 18 avenue Edouard Belin, 31401 Toulouse Cedex 9, France\label{aff130}
\and
Institute of Space Science, Str. Atomistilor, nr. 409 M\u{a}gurele, Ilfov, 077125, Romania\label{aff131}
\and
Consejo Superior de Investigaciones Cientificas, Calle Serrano 117, 28006 Madrid, Spain\label{aff132}
\and
Universidad de La Laguna, Departamento de Astrof\'{\i}sica, 38206 La Laguna, Tenerife, Spain\label{aff133}
\and
Dipartimento di Fisica e Astronomia "G. Galilei", Universit\`a di Padova, Via Marzolo 8, 35131 Padova, Italy\label{aff134}
\and
INFN-Padova, Via Marzolo 8, 35131 Padova, Italy\label{aff135}
\and
Institut f\"ur Theoretische Physik, University of Heidelberg, Philosophenweg 16, 69120 Heidelberg, Germany\label{aff136}
\and
Institut de Recherche en Astrophysique et Plan\'etologie (IRAP), Universit\'e de Toulouse, CNRS, UPS, CNES, 14 Av. Edouard Belin, 31400 Toulouse, France\label{aff137}
\and
Universit\'e St Joseph; Faculty of Sciences, Beirut, Lebanon\label{aff138}
\and
Departamento de F\'isica, FCFM, Universidad de Chile, Blanco Encalada 2008, Santiago, Chile\label{aff139}
\and
Universit\"at Innsbruck, Institut f\"ur Astro- und Teilchenphysik, Technikerstr. 25/8, 6020 Innsbruck, Austria\label{aff140}
\and
Satlantis, University Science Park, Sede Bld 48940, Leioa-Bilbao, Spain\label{aff141}
\and
Department of Physics, Royal Holloway, University of London, TW20 0EX, UK\label{aff142}
\and
Instituto de Astrof\'isica e Ci\^encias do Espa\c{c}o, Faculdade de Ci\^encias, Universidade de Lisboa, Tapada da Ajuda, 1349-018 Lisboa, Portugal\label{aff143}
\and
Cosmic Dawn Center (DAWN)\label{aff144}
\and
Niels Bohr Institute, University of Copenhagen, Jagtvej 128, 2200 Copenhagen, Denmark\label{aff145}
\and
Universidad Polit\'ecnica de Cartagena, Departamento de Electr\'onica y Tecnolog\'ia de Computadoras,  Plaza del Hospital 1, 30202 Cartagena, Spain\label{aff146}
\and
Infrared Processing and Analysis Center, California Institute of Technology, Pasadena, CA 91125, USA\label{aff147}
\and
Dipartimento di Fisica e Scienze della Terra, Universit\`a degli Studi di Ferrara, Via Giuseppe Saragat 1, 44122 Ferrara, Italy\label{aff148}
\and
Istituto Nazionale di Fisica Nucleare, Sezione di Ferrara, Via Giuseppe Saragat 1, 44122 Ferrara, Italy\label{aff149}
\and
INAF, Istituto di Radioastronomia, Via Piero Gobetti 101, 40129 Bologna, Italy\label{aff150}
\and
Universit\'e PSL, Observatoire de Paris, Sorbonne Universit\'e, CNRS, LERMA, 75014, Paris, France\label{aff151}
\and
Universit\'e Paris-Cit\'e, 5 Rue Thomas Mann, 75013, Paris, France\label{aff152}
\and
INAF - Osservatorio Astronomico di Brera, via Emilio Bianchi 46, 23807 Merate, Italy\label{aff153}
\and
INAF-Osservatorio Astronomico di Brera, Via Brera 28, 20122 Milano, Italy, and INFN-Sezione di Genova, Via Dodecaneso 33, 16146, Genova, Italy\label{aff154}
\and
ICL, Junia, Universit\'e Catholique de Lille, LITL, 59000 Lille, France\label{aff155}
\and
ICSC - Centro Nazionale di Ricerca in High Performance Computing, Big Data e Quantum Computing, Via Magnanelli 2, Bologna, Italy\label{aff156}
\and
Instituto de F\'isica Te\'orica UAM-CSIC, Campus de Cantoblanco, 28049 Madrid, Spain\label{aff157}
\and
CERCA/ISO, Department of Physics, Case Western Reserve University, 10900 Euclid Avenue, Cleveland, OH 44106, USA\label{aff158}
\and
Laboratoire Univers et Th\'eorie, Observatoire de Paris, Universit\'e PSL, Universit\'e Paris Cit\'e, CNRS, 92190 Meudon, France\label{aff159}
\and
Departamento de F{\'\i}sica Fundamental. Universidad de Salamanca. Plaza de la Merced s/n. 37008 Salamanca, Spain\label{aff160}
\and
Center for Data-Driven Discovery, Kavli IPMU (WPI), UTIAS, The University of Tokyo, Kashiwa, Chiba 277-8583, Japan\label{aff161}
\and
Dipartimento di Fisica - Sezione di Astronomia, Universit\`a di Trieste, Via Tiepolo 11, 34131 Trieste, Italy\label{aff162}
\and
California Institute of Technology, 1200 E California Blvd, Pasadena, CA 91125, USA\label{aff163}
\and
Department of Physics \& Astronomy, University of California Irvine, Irvine CA 92697, USA\label{aff164}
\and
Department of Mathematics and Physics E. De Giorgi, University of Salento, Via per Arnesano, CP-I93, 73100, Lecce, Italy\label{aff165}
\and
INFN, Sezione di Lecce, Via per Arnesano, CP-193, 73100, Lecce, Italy\label{aff166}
\and
INAF-Sezione di Lecce, c/o Dipartimento Matematica e Fisica, Via per Arnesano, 73100, Lecce, Italy\label{aff167}
\and
CEA Saclay, DFR/IRFU, Service d'Astrophysique, Bat. 709, 91191 Gif-sur-Yvette, France\label{aff168}
\and
Department of Computer Science, Aalto University, PO Box 15400, Espoo, FI-00 076, Finland\label{aff169}
\and
Instituto de Astrof\'\i sica de Canarias, c/ Via Lactea s/n, La Laguna 38200, Spain. Departamento de Astrof\'\i sica de la Universidad de La Laguna, Avda. Francisco Sanchez, La Laguna, 38200, Spain\label{aff170}
\and
Ruhr University Bochum, Faculty of Physics and Astronomy, Astronomical Institute (AIRUB), German Centre for Cosmological Lensing (GCCL), 44780 Bochum, Germany\label{aff171}
\and
Department of Physics and Astronomy, Vesilinnantie 5, 20014 University of Turku, Finland\label{aff172}
\and
Serco for European Space Agency (ESA), Camino bajo del Castillo, s/n, Urbanizacion Villafranca del Castillo, Villanueva de la Ca\~nada, 28692 Madrid, Spain\label{aff173}
\and
ARC Centre of Excellence for Dark Matter Particle Physics, Melbourne, Australia\label{aff174}
\and
Centre for Astrophysics \& Supercomputing, Swinburne University of Technology,  Hawthorn, Victoria 3122, Australia\label{aff175}
\and
Department of Physics and Astronomy, University of the Western Cape, Bellville, Cape Town, 7535, South Africa\label{aff176}
\and
DAMTP, Centre for Mathematical Sciences, Wilberforce Road, Cambridge CB3 0WA, UK\label{aff177}
\and
Kavli Institute for Cosmology Cambridge, Madingley Road, Cambridge, CB3 0HA, UK\label{aff178}
\and
Department of Astrophysics, University of Zurich, Winterthurerstrasse 190, 8057 Zurich, Switzerland\label{aff179}
\and
IRFU, CEA, Universit\'e Paris-Saclay 91191 Gif-sur-Yvette Cedex, France\label{aff180}
\and
Oskar Klein Centre for Cosmoparticle Physics, Department of Physics, Stockholm University, Stockholm, SE-106 91, Sweden\label{aff181}
\and
Astrophysics Group, Blackett Laboratory, Imperial College London, London SW7 2AZ, UK\label{aff182}
\and
Univ. Grenoble Alpes, CNRS, Grenoble INP, LPSC-IN2P3, 53, Avenue des Martyrs, 38000, Grenoble, France\label{aff183}
\and
Centro de Astrof\'{\i}sica da Universidade do Porto, Rua das Estrelas, 4150-762 Porto, Portugal\label{aff184}
\and
HE Space for European Space Agency (ESA), Camino bajo del Castillo, s/n, Urbanizacion Villafranca del Castillo, Villanueva de la Ca\~nada, 28692 Madrid, Spain\label{aff185}
\and
Department of Astrophysical Sciences, Peyton Hall, Princeton University, Princeton, NJ 08544, USA\label{aff186}
\and
Theoretical astrophysics, Department of Physics and Astronomy, Uppsala University, Box 515, 751 20 Uppsala, Sweden\label{aff187}
\and
Mathematical Institute, University of Leiden, Einsteinweg 55, 2333 CA Leiden, The Netherlands\label{aff188}
\and
School of Physics \& Astronomy, University of Southampton, Highfield Campus, Southampton SO17 1BJ, UK\label{aff189}
\and
Institute of Astronomy, University of Cambridge, Madingley Road, Cambridge CB3 0HA, UK\label{aff190}
\and
Space physics and astronomy research unit, University of Oulu, Pentti Kaiteran katu 1, FI-90014 Oulu, Finland\label{aff191}
\and
Department of Physics and Astronomy, Lehman College of the CUNY, Bronx, NY 10468, USA\label{aff192}
\and
American Museum of Natural History, Department of Astrophysics, New York, NY 10024, USA\label{aff193}
\and
Center for Computational Astrophysics, Flatiron Institute, 162 5th Avenue, 10010, New York, NY, USA\label{aff194}
\and
Department of Physics and Astronomy, University of British Columbia, Vancouver, BC V6T 1Z1, Canada\label{aff195}}    

\abstract{
We present a catalogue of 497 galaxy-galaxy strong lenses in the Euclid Quick Release 1 data (63 deg$^2$). In the initial 0.45\% of \Euclid's surveys, we double the total number of known lens candidates with space-based imaging. Our catalogue includes 250 grade A candidates,  the vast majority of which (243) were previously unpublished. \Euclid's resolution reveals rare lens configurations of scientific value including double-source-plane lenses, edge-on lenses, complete Einstein rings, and quadruply-imaged lenses. We resolve lenses with small Einstein radii ($\theta_{\rm E} < \ang{;;1}$) in large numbers for the first time. These lenses are found through an initial sweep by deep learning models, followed by Space Warps citizen scientist inspection, expert vetting, and system-by-system modelling. Our search approach scales straightforwardly to Euclid Data Release 1 and, without changes, would yield approximately \num{7000} high-confidence (grade A or B) lens candidates by late 2026. Further extrapolating to the complete Euclid Wide Survey implies a likely yield of over \num{100000} high-confidence candidates, transforming strong lensing science. 
}
%
%
%

   \titlerunning{Euclid Q1: The Strong Lensing Discovery Engine -- System overview}
   \authorrunning{Euclid Collaboration: M.~Walmsley et. al.}
   
   \maketitle
%
%
%
\section{\label{sc:Intro}Introduction}

\Euclid will resolve 1.5 billion galaxies over the next ten years \citep{EuclidSkyOverview}.
Within those 1.5 billion galaxies will be of order \num{100000} galaxy-galaxy strong lenses \citep{collettPopulationGalaxyGalaxyStrong2015} -- around 100 times more than currently known. These lenses will enable model-independent tests of cosmology \citep{sharmaTestingCosmologyDouble2023}, break degeneracies in dark energy parameters
\citep{liCosmologyLargePopulations2024}, test dark matter models at small scales through the detection of low-mass dark subhalos \citep{oriordanSensitivityStrongLensing2023}, measure the density profile of massive galaxies to determine how they grow, and more (see \citealt{shajib_strong_2024} for a review). In this series of papers, we develop and apply a method (our `discovery engine') to find those lenses.\\

Most strong lenses are expected to have Einstein radii smaller than \ang{;;1.0}, below the resolution limits of ground-based surveys \citep{collettPopulationGalaxyGalaxyStrong2015, sonnenfeldStrongLensingSelection2023}. \Euclid's space-based point spread function (PSF) can resolve lenses down to an Einstein radii of around \ang{;;0.6} (Sect. \ref{sec:modelling}) -- close to the peak of the Einstein radii distribution. The combination of resolving smaller Einstein radii with a wide 0.5 deg$^2$ field-of-view \citep{EuclidSkyVIS, EuclidSkyNISP} makes \Euclid the most efficient instrument for finding strong lenses ever built. 

\begin{figure}
    \centering
    \includegraphics[width=\columnwidth]{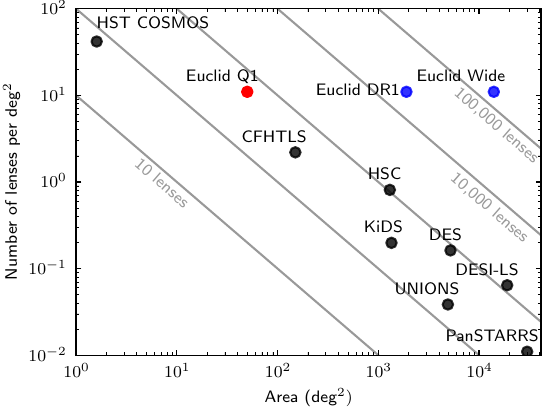}
    \caption{Lens-finding capabilities of \Euclid compared to recent surveys. We show the number of A/B grade lenses detected per deg$^2$ vs. total survey area in deg$^2$, with contours of constant total lenses detected. The `Euclid Q1' datapoint uses lenses detected in this series of works. Estimates for Euclid DR1 and Euclid Wide multiply our rate of detected lenses with the appropriate area; see Sect. \ref{sc:conclusion_and_outlook} for detailed forecasting scenarios. All numbers are approximate due to the subjective nature of grading and the varying details of each search approach, and from searches focusing on galaxy-galaxy lenses in alignment with this series of works. References: HST COSMOS \citep{faureFirstCatalogStrong2008}, CFHTLS \citep{Gavazzi2014}, HSC \citep{sonnenfeldSurveyGravitationallylensedObjects2018,sonnenfeldSurveyGravitationallylensedObjects2020a,Canameras2021,shu22_HOLISMOKES8, wongSurveyGravitationallyLensed2022,jaelani24,    schuldt_holismokes_2025, schuldtEtAl25b}, DES \citep{jacobsExtendedCatalogGalaxyGalaxy2019,rojasImpactHumanExpert2023,gonzalez_discovering_2025}, DESI Legacy Surveys \citep{huang_finding_2020,huang_discovering_2021,storferNewStrongGravitational2024}, UNIONS \citep{savaryStrongLensingUNIONS2022,acevedoBarrosoEtAl25}, 
    KiDS \citep{liHighqualityStrongLens2021},
    PanSTARSS \citep{Canameras2020}.}
    \label{fig:lens_surveys}
\end{figure}

The rate of detectable lenses in \Euclid is  10 to 100 times higher than in previous wide-area surveys, as shown in Fig.~\ref{fig:lens_surveys} (citations in caption). Visual searches of DES, DESI Legacy Survey, UNIONS, KiDS, and PanSTARSS find around 0.1 lenses per deg$^2$. Searches of HSC, which has LSST-like depth, identify around 1 lens per deg$^2$. Preliminary searches of the Euclid Early Release Observations \citep{AcevedoBarroso24,nagam_euclid_2025}
identify 10 lenses per deg$^2$. This work introduces a visual search through the first images from \Euclid's main surveys -- Quick Data Release 1 (Q1, \citealt{Q1-TP001}). 

Previous visual searches for galaxy-galaxy strong lenses typically include at least one of three common components (see \citealt{lemon_searching_2024} Sect. 4 for a recent review). Expert visual inspection recruits professional astronomers to look through candidate images. Experts typically search 1 to \num{10000} images due to the time required to look through each image, restricting application to specific fields (e.g., 
\citealt{barkanaPossibleGravitationalLens1999,fassnachtStrongGravitationalLens2004,faureFirstCatalogStrong2008}) or to specific sources prioritised using catalogue-level measurements (e.g., 
\citealt{myersCosmicLensAllSky2003,agnelloQuasarLensesSouth2019}). Citizen science efforts instead distribute the inspection task among a `crowd' of volunteers -- members of the public who freely contribute their time to find lenses.  This approach was pioneered by Space Warps 
\citep{Marshall2016,moreSpaceWarpsII2016} and has also been applied by Hubble Asteroid Hunters 
\citep{garvin_hubble_2022}. The Space Warps Hyper Suprime-Cam (HSC) projects searched approximately \num{150000} galaxies per month \citep{sonnenfeldSurveyGravitationallylensedObjects2020a}, far faster than expert inspection and of comparable scale to current surveys (prior to \Euclid). Finally, automated image searches augment human visual searches by using algorithms to prioritise the most promising galaxies for inspection. 
Recent automated searches universally use deep learning models. Supervised models trained with labelled lens and non-lens examples are the most common \citep[e.g.,][]{Petrillo2017, Petrillo2019a, Petrillo2019b,Lanusse2017,Schaefer2017,Pourrahmani+18,Davies2019,metcalf_strong_2019,Canameras2020,huang_finding_2020,Gentile+21,Li+20,liHighqualityStrongLens2021,rezaei_machine_2022, rojas_search_2022,savaryStrongLensingUNIONS2022,canameras_holismokes_2024,Pearce-Casey24, huang_desi_2025}. New directions include self-supervised learning \citep{Stein2022} and ensembles \citep[e.g.,][]{andikaStreamlinedLensedQuasar2023,nagamDenseLensUsingDenseNet2023,nagam+24,ishida_combining_2025}.
Most known strong lenses were identified by a deep learning sweep followed by expert inspection. 

In this work, we combine the strengths of AI, citizen scientists, and experts into a single `discovery engine' that efficiently searches through Euclid Quick Release 1 \citep{Q1cite} and, ultimately, through all \Euclid's surveys. We use an ensemble of deep learning models to select galaxies to show to citizen science volunteers. The volunteers then select promising candidates for validation through expert vetting and detailed lens modelling. 
Our discoveries vindicate the science value of AI foundation models in astronomy; the best-performing model in our ensemble, Zoobot, is a generalist model not specifically designed for strong lensing.\\

\begin{figure*}[ht!]
    \centering
    \includegraphics[width=0.49\textwidth]{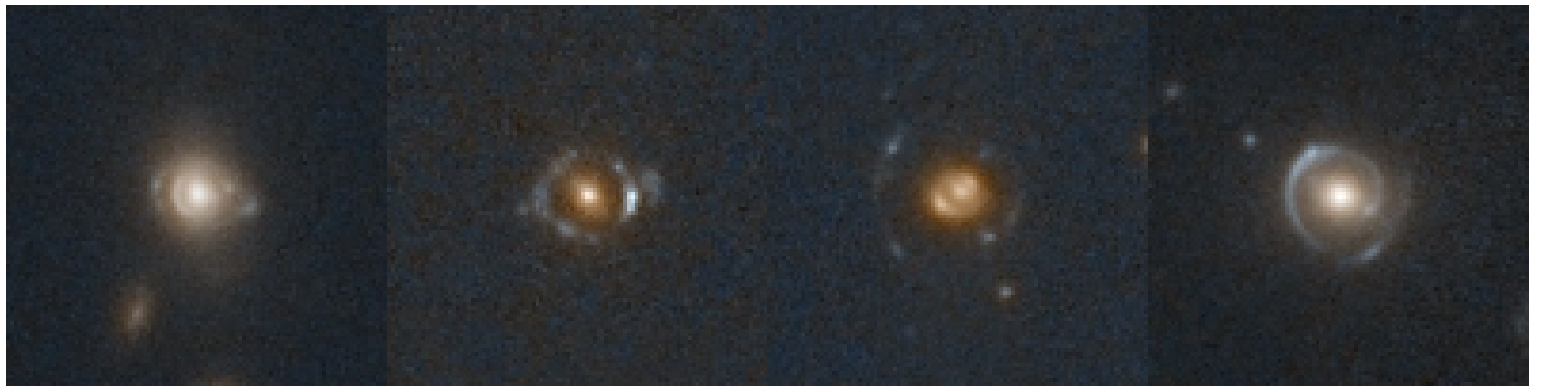}
    \hfill
    \includegraphics[width=0.49\textwidth]{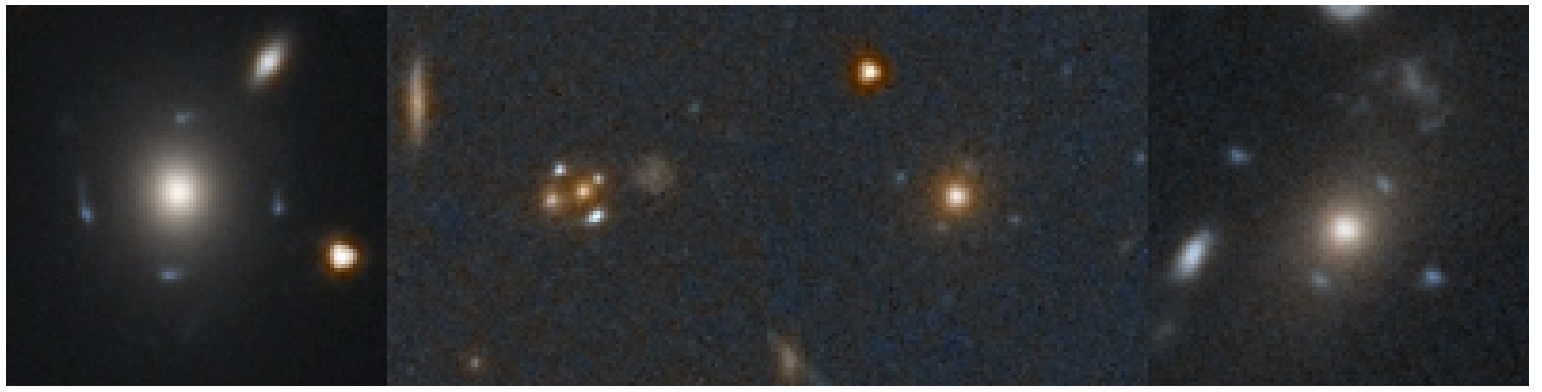}

    \centerline{\footnotesize{(a) Double source plane (left) and quadruple image (right) lens candidates}}

    \includegraphics[width=\textwidth]{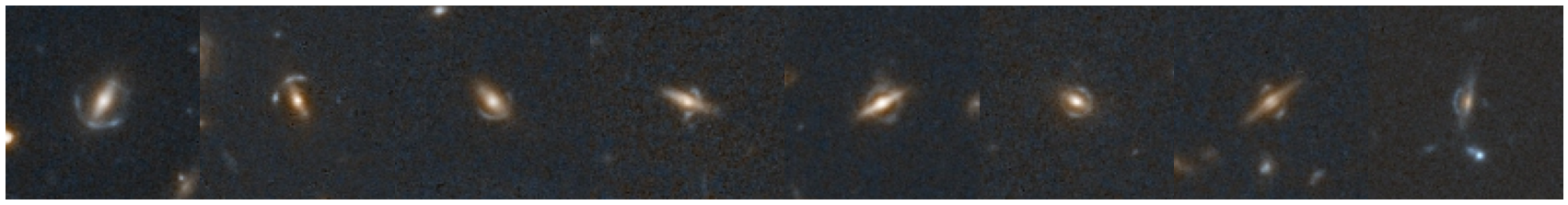}
    \centerline{\footnotesize{(b) Top-ranked edge-on lens candidates}}

    \includegraphics[width=\textwidth]{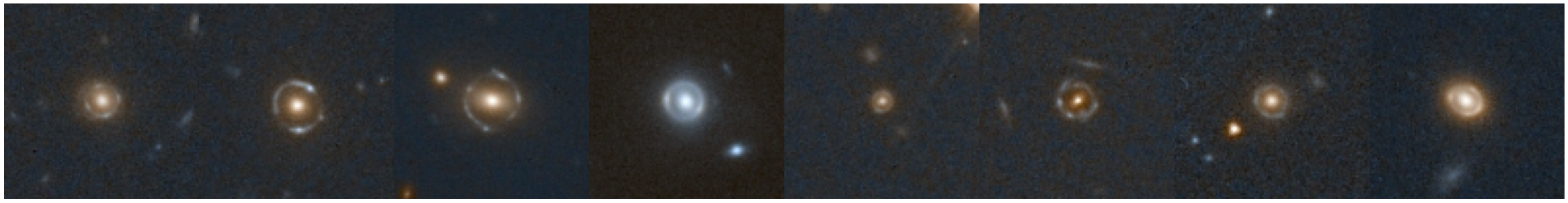}
    \centerline{\footnotesize{(c) Top-ranked Einstein ring lens candidates}}

    \includegraphics[width=\textwidth]{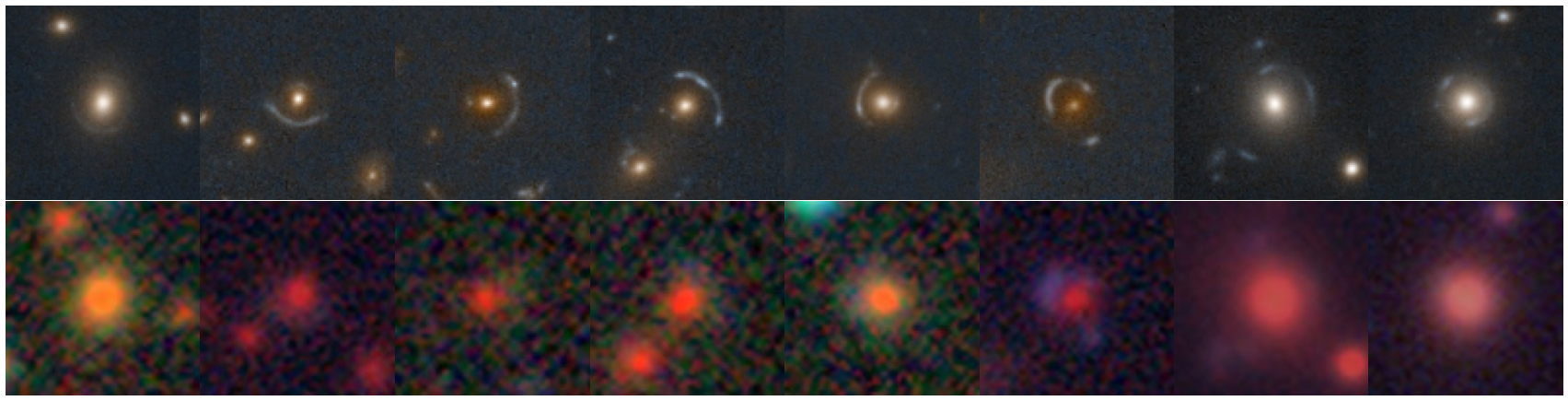}
    \centerline{\footnotesize{(d) Top-ranked lens candidates (excluding those above) shown with \Euclid (upper row) vs. Legacy Survey (lower row) images}}

    \includegraphics[width=\textwidth]{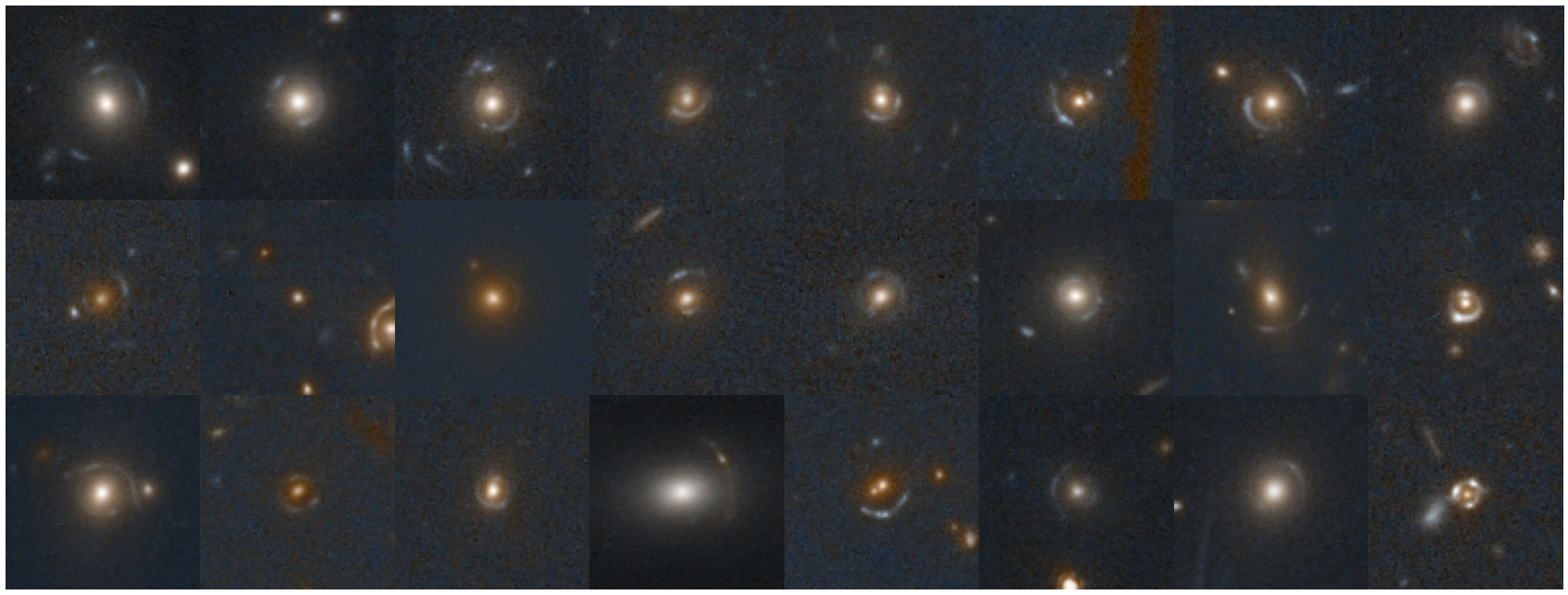}
    \centerline{\footnotesize{(e) Top-ranked lens candidates, continued (see Fig. \ref{fig:lens_subset_by_score} for extended gallery)}}
    
    \caption{Notable and highest-scoring strong lens candidates found in Q1. Includes Einstein rings, double source plane lenses, quads, and edge-on lenses. Each cutout has a $\ang{;;10}$ field-of-view. Figure \ref{fig:lens_subset_by_score} shows strong lens candidates as a function of expert visual inspection score.}
    \label{fig:intro_grade_a}
\end{figure*}

We find 497 strong lens candidates in \Euclid's initial Q1 release -- a similar count to recent searches of wide-area surveys \citep{jacobsExtendedCatalogGalaxyGalaxy2019, Stein2022, Canameras2021, schuldt_holismokes_2025, schuldtEtAl25b, gonzalez_discovering_2025} despite covering only 50 deg$^2$. 
We double the total number of known lens candidates with space-based imaging \citep{boltonSloanLensACS2008, huang_finding_2020}.
Our discoveries include rare galaxy-galaxy configurations, most notably double source plane lenses. Double source plane lenses are considered valuable complementary probes to constrain cosmological parameters and dark energy \citep{gavazziSloanLensACS2007, collett_cosmological_2014}. Such systems are more likely to be found in high-spatial resolution imaging rather than ground-based seeing-limited surveys. We explore \Euclid's promise in detecting these systems \citep{Q1-SP054}. We also identify approximately 20 candidate edge-on lens galaxies; these may help break the degeneracy between baryon and dark matter mass slopes \citep{treuSWELLSSurveyLarge2011}. We show that \Euclid can find lenses around low-mass galaxies, with a majority (56\%) of our successfully-modelled candidates having Einstein radii between $\ang{;;0.5}$ and $\ang{;;1.0}$. Looking ahead, our Q1 search is a practical test of lens finding in \Euclid and allows us a chance to optimise our approach ahead of the much larger Euclid Data Release 1 (DR1; \num{1900} deg$^2$ in late 2026). 

This overview paper opens a series of papers detailing our Q1 search.
\citet{Q1-SP052} presents early discoveries using spectroscopically-identified high-velocity-dispersion galaxies, which also form part of the initial training set for our machine learning models. \citet{Q1-SP053} presents those machine learning models and compares their performance on real \Euclid images. \citet{Q1-SP054} presents the double source plane lenses discovered in Q1, with preliminary modelling for forecasting cosmological parameters. \citet{Q1-SP059} presents a Bayesian ensemble method combining lens classifiers to further optimise lens discovery within our discovery engine for DR1. 

\section{\label{sc:design_motivation}  Design motivation}

Designing a search system is challenging because \Euclid has both an extremely large number of galaxies to search and an extremely high rate of detectable strong lenses.
\Euclid's surveys will ultimately find 300 million galaxies brighter than $\IE < 22.5$ mag (our primary selection cut, see Sect. \ref{sec:data}). This is on par with the largest strong lens searches to date \citep{Canameras2021,schuldt_holismokes_2025,gonzalez_discovering_2025}.
Larger samples require more accurate automated searches; otherwise, the number of false positives (mistakenly flagged non-lenses) grows to overwhelm any capacity for manual inspection. To illustrate: applying a 99.9\% accurate machine learning model to one million galaxies would generate one thousand false positives, which is easily feasible for manual inspection, while applying the same model to 100 million galaxies would generate 100 thousand false positives.
Further, \Euclid's order-of-magnitude higher rate of detectable lenses (Fig. \ref{fig:lens_surveys}) requires a high search efficiency throughout our system. Astronomical searches typically first apply a cheap (in time, compute, etc.) filter to everything, and then sequentially apply more expensive filters as the remaining sample shrinks. When lenses are more common, we expect to keep more lens candidates at every stage, which `costs' more than applying the same series of filters to a survey of equal size but containing fewer lenses.
Finally, in order to share our catalogue alongside the Q1 public data release, we needed to carry out our search in around six weeks. 

Our approach employs three different lens finding methods: a broad deep learning search, refinement by citizen scientists, and finally confirmation by expert visual inspection. We briefly review each one below.

Deep learning models can search for strong lenses in arbitrarily large numbers of images but the resulting samples are generally impure, i.e., deep learning-identified samples often contain a high rate of non-lenses. \cite{sonnenfeldSurveyGravitationallylensedObjects2020a} notes that experts validated 46 of 1480 lens candidates found by YATTALENS in HSC \citep{sonnenfeldSurveyGravitationallylensedObjects2018}, 89 of \num{3500} found in KiDS by \cite{Petrillo2019b}, etc. 
Previous work suggests that training on simulations leads to models that are excellent at finding simulated lenses but struggle when presented with real data, reflecting a broader `sim to real' generalisation gap well-known in computer science \citep{zhaoSimtoRealTransferDeep2020} and astronomy \citep{margalef-bentabolGalaxyMergerChallenge2024}.  Painting simulated lenses onto real images mitigates this issue \citep{Petrillo2017, jacobs_finding_2017, rojas_search_2022, Canameras2020, Canameras2021}. 
Using foundation models, pre-trained on real images with labels collected for other purposes, also mitigates this issue \citep{Pearce-Casey24}. We apply both strategies (Sect. \ref{sec:machine_learning}).

Citizen scientists can search large, but not limitless, numbers of images. They are adept at finding the unexpected and have shown that they can extrapolate beyond their training sample set \citep{lintottGalaxyZooHannys2009,Geach2015}.
Previous citizen science searches for strong lenses have mixed in known images (simulations or previously-discovered lenses) to measure how individual volunteers respond, and then combined their responses on new images via Bayes' rule (see Sect. \ref{sec:citizen_science}). 

Professional astronomers (`experts' hereafter) are generally assumed to have high accuracy and low variance in identifying lenses and recognising rare configurations - but can only search small (of order 10k) sets of images. This is particularly true when showing each image to several experts, as required for robust and consistent grading \citep{rojasImpactHumanExpert2023}. 

We combine all three search approaches, with each one playing to their strengths. Deep learning models make an initial prioritisation, to discard the bulk of galaxies that do not exhibit any features consistent with lensing. The machine learning (ML) models are also likely to incorrectly flag unusual images (e.g., instrumental artifacts) and lens-like galaxies (e.g., ring galaxies, galaxies with faint spiral arms or tidal tails, etc.). This large-but-impure sample is sent to citizen scientists to refine the sample, prioritising genuine lenses and rejecting obvious-to-a-human artifacts. The best candidates are then sent to professional astronomers who are experts in strong lens identification (members of the Euclid Strong Lensing Working Group) for a final grade on visual appearance.  Finally, we pass the promising candidates to a modelling pipeline (\texttt{PyAutoLens}, \citealt{nightingalePyAutoLensOpenSourceStrong2021}) to verify if the image can be matched to a physical source and lens configuration. 

This combination of search approaches is a natural and long-foreseen step in lens finding, but is only now being realised. Recently-published work by \cite{gonzalez_discovering_2025} independently uses a combination of ML model/volunteer/expert combination for searching 230 million (DES) images and successfully identifies 665 `probable' candidates (147 new). We ultimately find more new lenses and at a far higher rate, but this likely reflects our different target survey more than any differences in approach -- DES has already been extensively searched and \Euclid's sensitivity and spatial resolution (FWHM$_{\text{VIS}}^{\text{Euc}}\sim$\ \ang{;;0.16} for \Euclid, FWHM$_{g}^{\text{DES}}\sim$\ang{;;0.9} for DES) can better resolve small Einstein radii and fainter lensed images than ground-based surveys. We both use a training set combining hand-selected non-lenses with known and painted lenses, we both use the Space Warps citizen science project for volunteer annotation, and we both show a substantial increase in machine-selected lens purity compared to previous searches. The key differences are that this work uses lens modelling for additional validation, and that Gonzalez et.\ al.\ focused on using a vision transformer while this work uses five diverse deep learning models (including a vision transformer). 
Our best-performing model is an astronomy foundation model not specifically designed for lens finding (Sect. \ref{sec:machine_learning}).

\section{\label{sec:data}  Data}


\begin{table}
\caption{Approximate PSF (or seeing, when ground-based) and field-of-view for instruments used in previous strong lens searches, compared to \Euclid. DECam data for the Dark Energy Survey \citep{abbott_dark_2018}, Hyper Suprime-Cam (HSC) data for the HSC Wide survey \citep{Aihara2019}. \textit{Hubble} Space Telescope and \textit{James Webb} Space Telescope values from Space Telescope Science Institute documentation. }
\renewcommand{\arraystretch}{1.5}
\centering
\begin{tabular}{ccc}
\hline
\hline
\textbf{Instrument} & \textbf{PSF \ Seeing} & \textbf{Field-of-view} \\
\hline
DECam (DES, \SI{620}{\nano\meter}) & \ang{;;0.9} & \ang{;10800} \\
HSC (Wide, \SI{620}{\nano\meter}) & \ang{;;0.7} & \ang{;6480} \\
HST WFC3 (\SI{700}{\nano\meter}) & \ang{;;0.07} & \ang{;7.3} \\
JWST NIRCam (\SI{700}{\nano\meter}) & \ang{;;0.03} & \ang{;9.7} \\
\Euclid VIS (\SI{700}{\nano\meter}) & \ang{;;0.16} & \ang{;1900} \\
\hline
\end{tabular}
\label{tab:instrument_table}
\end{table}

\Euclid is uniquely capable at finding strong lenses because it combines a space-based PSF with a wide field-of-view. This reflects \Euclid's design as a survey space telescope intended to deliver high image quality over a wide area. Table \ref{tab:instrument_table} compares the PSF and field-of-view of \Euclid with other instruments used for previous lens searches. 










The Euclid Wide Survey (EWS) will ultimately image approximately \num{14400} deg$^2$.
Q1 \citep{Q1-TP001} is the first data release of EWS-like images, captured to the same depth and processed with the same pipeline. Q1 is therefore ideal for demonstrating the science that \Euclid will enable. We will show that -- despite covering only 63 deg$^2$, one-three-hundredth of the area of the final EWS -- Q1 includes as many detectable strong lenses as the largest lens searches to date.

We select the strong lens search sample from the Q1 MERge catalogue \citep{Q1-TP004} by applying the following cuts.
We require $\IE \leq 22.5$ mag (corresponding to $\texttt{FLUX\_DETECTION\_TOTAL}\geq 3.63078$) to restrict our search to reasonably bright sources. As with other searches, the lensing rate strongly depends on selection cuts -- here, preselection of galaxies that are bright in VIS. By cutting at $\IE < 22.5$ mag, we removed $\sim$90\% of galaxies in Q1, in exchange for throwing away $\sim 25\%$ of lenses. The lensing rate is lower for  galaxies that are faint in \IE since they are either: low mass, with smaller lensing cross sections; or, high redshift with fewer sources behind them.

Brightness and extent are strongly correlated and so our $\IE \leq 22.5$ mag cut also acts as a size cut; our 5th percentile has a \texttt{SEGMENTATION\_AREA} (pixel mask) of 259px.
We also require $\texttt{VIS\_DET}==1$ to select VIS-detected sources, no \texttt{GAIA\_CROSSMATCH} and $\texttt{MUMAX\_MINUS\_MAG}\geq -2.6$ to remove stars and point-like objects, $\texttt{MU\_MAX}\geq 15.0$ to remove saturated stars, and $\texttt{SPURIOUS\_PROB}< 0.05$ to reject artifacts. This gives a search sample of \num{1086556} sources (about \num{20000} sources per deg$^2$).

We use images from \Euclid's VIS and NISP instruments \citep{EuclidSkyVIS,EuclidSkyNISP} as processed for Q1 \citep{Q1-TP002,Q1-TP003, Q1-TP004}
We access the \Euclid images via the ESA Science Archive Service and the ESA Datalabs platform \citep{Datalabscite}.
We cut out each source with a field-of-view of \ang{;;15} (or 150$\times$150 MERge mosaic pixels, which are native instrument pixels resampled to \ang{;;0.1} per pixel). We then adjust the raw flux values in each cutout into  images suitable for human inspection, described below, and save the final images as colour JPEG for export and display.

The raw flux values recorded from astronomical sources is typically rescaled from an extremely high dynamic range to a range small enough for viewing. We create cutouts with one of two scalings: either arcsinh scaling, which `boosts the lows' with the aim of making fainter sources more visible, or midtone transfer function (MTF) scaling, which `boosts the mids' with the aim of making typical sources clearly visible against the dark background. Arcsinh scaling \citep{Lupton2004} uses $x^\prime = \sinh^{-1}(Qx)$, where $Q$ sets the aggressiveness of the scaling (we chose 500/1/0.5 for \IE/\YE/\JE). MTF scaling applies

\begin{equation}
x^\prime = \frac{(m-1) x}{(2 m-1) x-m},
\end{equation}
where $x$ is the original pixel value and $m$ is a stretch factor (we chose to set this automatically such that the central 100$\times$100 pixels had a mean of 0.2). We also apply a percentile clip of 99.85\% to remove the most extreme flux values before scaling.

Colour images require three channels (red/green/blue). We created cutouts with the following combinations: \YE/median\footnote{Intuitively, we define \YE as red, \IE as blue, and fill the green channel with the median value of each pixel across \YE and \IE}/\IE, \JE/median/\IE, and \JE/\YE/\IE, plus a greyscale VIS-only version. 
Naively combining the \IE and NISP bands into RGB images led to `blurrier' images because of the lower resolution NISP bands. Instead, we constructed our images using \IE for the luminosity and all selected bands for the hue and saturation (HSL processing).

\begin{figure}
    \centering
    \includegraphics[width=\columnwidth]{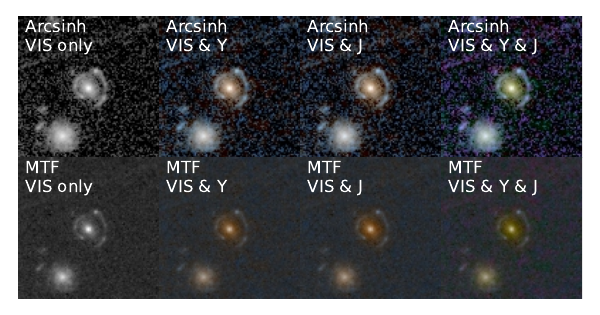}
    \caption{Illustration of image processing options. The same lens candidate is shown with each scaling (arcsinh or midtone transfer function) and colouring (greyscale \IE, \YE/median/\IE, \JE/median/\IE, \JE/\YE/\IE) option. To maximise resolution we set the luminance using the \IE image in all cases.} 
    \label{fig:colouring_options}
\end{figure}

Figure \ref{fig:colouring_options} shows an example cutout processed with each possible processing choice. There is no ideal choice for all galaxies; instead, we provided all image versions to the machine-learning teams (Sect. \ref{sec:machine_learning} and \citealt{Q1-SP053}) and we showed grids combining multiple image versions to volunteers (Sect. \ref{sec:citizen_science}) and experts (Sect. \ref{sec:expert}).

\section{\label{sec:machine_learning}  Machine learning overview}

Our strategy is informed by \citet{Pearce-Casey24}, hereafter PC25. PC25 invited Euclid Consortium members to submit automated scores for galaxies in the \Euclid ERO Perseus field. Participants were free to use any training dataset and machine learning approach. Accuracy was evaluated against expert visual inspection scores collected in parallel and reported in \cite{AcevedoBarroso24}. The results led PC25 to make two recommendations for training effective models. 

First, following earlier work on other surveys \citep[e.g.,][]{Petrillo2017, jacobs_finding_2017, rojas_search_2022, Canameras2020, Canameras2021, moreSpaceWarpsII2016}, PC25 suggest that `painting' lenses on real galaxies is more effective than training only on purely simulated images. All but one team trained on purely-simulated images based on the Euclid Flagship simulations \citep{EuclidSkyFlagship}. Those models did well at finding lenses inserted into those simulations, but performed far worse when applied to real images. This suggests that simulations may not capture the full diversity of real galaxies, real artifacts, and real lenses (see also \citealt{metcalf_strong_2019}). We therefore focused on creating painted lenses for our training data. 

Second, PC25 found that the Zoobot foundation models had the best performance. Foundation models \citep{Bommasani2021} are deep learning models pre-trained on diverse tasks with plentiful data.  Pre-training teaches the models to extract generally useful features that can then be used for new tasks where data are scarce. The Zoobot models are pre-trained to answer a broad set of galaxy morphology questions on real data \citep{Walmsley2023zoobot}. Strong lensing and galaxy morphology share similar features -- recognising galaxy shapes -- and so Zoobot can recognise strong lenses without any further training, and does so better than similar models trained from scratch (PC25). We use an improved `Zoobot 2.0' version \citep{walmsley_scaling_2024} in our Discovery Engine, as part of a diverse ensemble of models. 

\citet{Q1-SP053}, continuing this paper series, describes the real and simulated training data, outlines each machine learning approach, makes a detailed comparison of their results with one another when judged against professional astronomers, and considers the implications for searching future \Euclid data. 

In short, \citet{Q1-SP053} again finds that Zoobot performs best, recovering 143 likely lens candidates in the top one thousand predictions from one million sources. But all models do well and no one model recovers every lens found by every other model. Together, this suggests that our current models are already good enough to find transformative numbers of strong lenses -- Zoobot alone would detect \num{7500} to \num{11000} lens candidates in the top \num{20000} galaxies in Euclid DR1 -- and that an ensemble of models with different architectures and training data would likely improve this further. 

\section{\label{sec:citizen_science}  Citizen science overview}

\subsection{Approach}

We use the dedicated strong gravitational lens search platform Space Warps, powered by the Zooniverse platform \citep{Marshall2016, moreSpaceWarpsII2016}, to visually inspect the top scoring cutouts, harnessing the visual inspection power of citizen scientists. This system has previously been demonstrated to be efficient and versatile in finding lens candidates for expert inspection and sifting out the false positives that dominate initial lens samples (e.g., \citealp{Geach2015,sonnenfeldSurveyGravitationallylensedObjects2020a, gonzalez_discovering_2025}). 

Our lens candidate inspection strategy was based partly on what would be feasible in the available six weeks. Because of the short timeline and uncertain level of citizen participation, we adopted a phased strategy for classification. While having the broad goal of classifying images from the parent sample (referred to as `test subjects') with the top 10k scores of all classifiers, we initially prioritised classifications of the union of the top thousand scored test subjects from each of the ML classifiers, under the assumption that these would be the most likely lens candidates. This allowed us to gauge citizen engagement in the project and develop the continuing strategy for classification, as well as allowing us to compare the performance between the ML models. 

We made an initial estimate of the performance of each model by considering the recovery of known lenses and the number of high scoring subjects in the early stages of the search that were judged to be lenses by experts. Two networks (models 1 and 4, see \citealp{Q1-SP053,Q1-SP059}) were particularly successful in finding lens candidates and so we supplemented the highest-ranked 10k galaxies from each network with the next-highest-ranked 10k-20k galaxies for those two networks only.  

Randomly selected objects from the initial galaxy sample that was shown to all networks were also classified. This enabled the comparison of the performance of the citizens and ML models (see \citealp{Q1-SP059}) and additionally provided an indication of incompleteness to systems that may have been missed by the ML classifiers. 

The project launched on November 19, 2024 and collected classifications from approximately \num{1800} users. In total \num{115329} unique systems were classified by citizens, including \num{78214} high-scoring objects, with the remainder from the random sampling. Collectively, this crowd made \num{857278} classifications from which lens candidates and rejects were identified. The classifications were made through the Space Warps project \citep{Marshall2016,moreSpaceWarpsII2016} on the Zooniverse platform.\footnote{\href{https://www.zooniverse.org/projects/aprajita/space-warps-esa-euclid}{https://www.zooniverse.org/projects/aprajita/space-warps-esa-euclid}}

Four colour settings were shown to the citizen inspectors based on combinations of the \Euclid VIS+NISP filters with arcsinh stretch in \IE only, \IE-\YE, \IE-\JE and \IE-\YE-\JE in the MTF stretch (see Sect. \ref{sec:data} for more information). These settings were chosen based on known and simulated lenses to highlight typical lensing features. 
For images where possible lensing signatures are seen, the citizens were instructed to place at least one marker in the image. They then have the option to classify the next image or discuss the subject with other citizens and scientists/experts on the project's `Talk' forum. 

\subsection{Aggregation}

Training images were interleaved into the images shown to citizens at a rate of $\sim$1 in 10. These training images were taken from \cite{Q1-SP052} and included both positive examples (simulations) and negatives, e.g., luminous red galaxies without lensing systems, ring galaxies, mergers, etc. We included \num{11706} such `training subjects', comprising \num{7320} non-lenses, and \num{4386} simulated lenses. These images help to train the citizens actively (rather than passively via info pages) on a wide variety of labelled examples. 
When a citizen classifies a training image, they receive automated live feedback (e.g., `Congratulations!! You’ve spotted a simulated gravitational lens!') through a pop-up.

The training images were also used in the aggregation of classifications received on any given test subject to produce a crowd probability or `score' for that test subject being a lens or not.
This aggregation process is described in detail in \citet{Marshall2016} and summarised here.
The posterior probability $P_{k+1}(L)\equiv P(L|\{C_{U_0},...,C_{U_k}\})$ for a given training subject, having received $k+1$ classifications $\{C_{U_0},...,C_{U_k}\}$ from users $\{U_0,...,U_k\}$ is given by
\begin{equation}\label{Eq: SWAP_Posterior}
P_{k+1}(L)= \frac{P(C_{U_k}|L)\cdot P_k(L)}{P(C_{U_k}|L)\cdot P_k (L)+P(C_{U_k}|{\sim} L)\cdot P_k({\sim} L)}
\end{equation}
where the users' classifications are denoted as `Lens' (`$L$') or `Non-Lens' (`${\sim} L$’). The classifications on training subjects are used to continuously update the skill of a given user. The skill of a user is measured by their ability to correctly identify lenses as lenses, $P(C_{U_k}|L)$, and their ability to correctly identify non-lenses, $P(C_{U_k}|{\sim} L)$. We define these skills as
\begin{equation}
    P(C_{U_k}=`L’| L )\approx \frac{1+N_{`L’}}{2+N_L}
\end{equation}
and 
\begin{equation}
    P(C_{U_k}=`{\sim} L’| {\sim} L )\approx \frac{1+N_{`{\sim} L’}}{2+N_{{\sim} L}}
\end{equation}
where $N_{`L’}$ ($N_{`{\sim} L’}$ ) denotes the number of simulated lenses (non-lenses) correctly classified and $N_L$ ($N_{{\sim} L}$) is the number of these systems inspected by that user at that point. The skill therefore broadly reflects the proportion of images of each class that a user correctly classified but assigns new users a score of 0.5, such that their classifications don't alter subject scores until they have classified at least one training image. These instantaneous user skills are used in Eq. (\ref{Eq: SWAP_Posterior}), allowing higher-skill users to change the lens posterior probability by a greater extent.

Each subject starts with a score $P_0(L)=5\times10^{-4}$ (based on a ball-park assumption for the approximate frequency of strong lensing in the galaxy population) which is updated each time a user classifies an image. This update is made through the Space Warps Analysis Pipeline (SWAP, \citealt{Marshall2016}). This update in score or posterior probability is shown in Fig.~\ref{Fig:Trajectories}. Each line or trajectory shows the evolving score per test subject with the number of classifications made. Systems move to the left (non-lens) or right (lens) with the size of the shift at each classification reflecting the (instantaneous) skill of the citizen who classified it. 

We ran SWAP in an online mode \citep{Marshall2016} whereby the users skills were updated after each classification
and to improve efficiency we retired images in real time based on their score. We set the retirement threshold as $P(L)<1\times10^{-5}$ with a minimum of six classifications representing a crowd consensus that a test subject does not contain a lens (shaded region in Fig.~\ref{Fig:Trajectories}). As the bulk of images will contain no lensing, removing these quickly from the stream means the project is 20\%--30\% more efficient than previous projects without automated retirement. Potential lenses, test subjects with $P(L)>1\times10^{-5}$, were only retired from the classification stream after 30 classifications had been made.
In general, we did not retire any training images unless they had artifacts which made their correct classification ambiguous, or were failed simulations, such as a lens drawn around no obvious lensing galaxy or on a bright artifact.

\subsection{Results of Visual Inspection}\label{subsec:results_visual_inspection}

\begin{figure}
    \centering
    \includegraphics[width=0.45\textwidth]{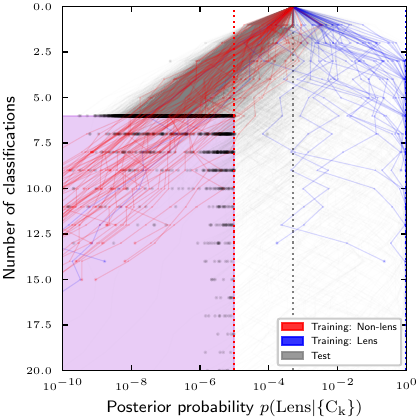}
    \caption{A random subset of score trajectories for test (grey) and training subjects (lenses: blue, non-lenses: red) in the Space Warps lens search. Subject which are voted as lenses (non-lenses) by citizens move towards the right (left), and users with higher skill can cause larger changes in score when they classify a subject. The vast majority of training subjects were correctly classified by the citizens. Subjects reaching the shaded region were removed from the platform to increase the efficiency of classification.}
    \label{Fig:Trajectories}
\end{figure}

\begin{figure}
    \centering
    \includegraphics[width=0.45\textwidth]{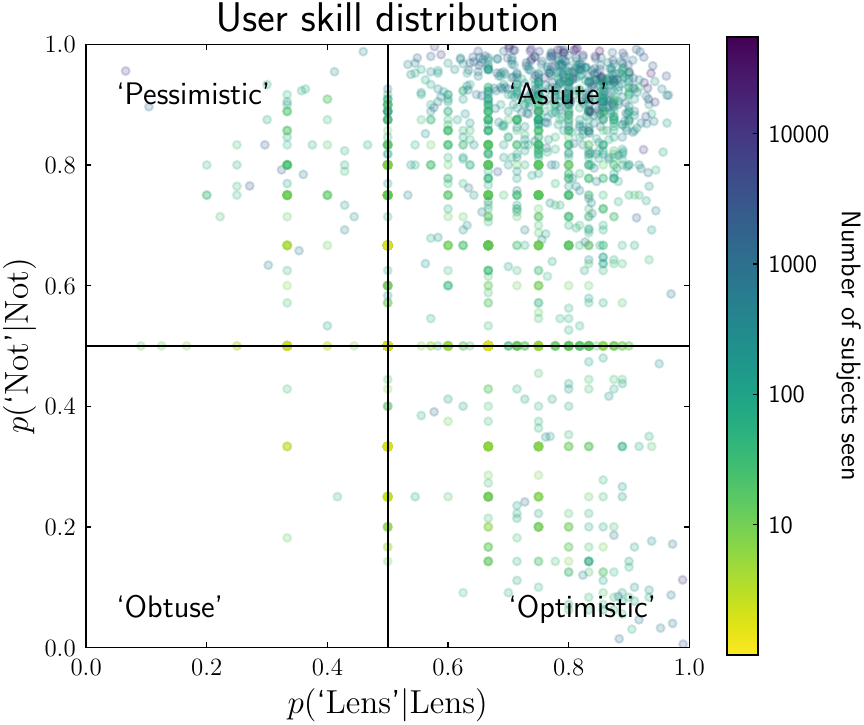}
    \caption{Distribution of user skills in the Space Warps lens search, based on their classifications of the training images for which the ground truth was known. The vast majority of users lie in the `Astute' quadrant, indicating they correctly classified the majority of lens and non-lens training images.}
    \label{Fig:User_skill_dist}
\end{figure}

Figure \ref{Fig:Trajectories} shows the score trajectories for a random subset of training and test subjects in the Space Warps lens search. The majority of training subjects were accurately classified: most non-lenses received low scores, while most lenses received high scores. The vast majority of test subjects received low scores, indicating they were not lenses, as expected. The crowd's skill distribution is shown in Fig.~\ref{Fig:User_skill_dist}. The vast majority of users are located in the `Astute' quadrant of user skill, indicating that they could correctly identify the majority of training subjects, both lenses and non-lenses. 
Out of the \num{114960} test subjects that were inspected by citizens, \num{2799} received a score of $p > 1\times10^{-5}$ and were subsequently inspected by experts. We treat citizen and ML scores independently in this paper;
but we refer the reader to \citet{holloway_bayesian_2024}, and \citet{Q1-SP059} for a detailed consideration of an ensemble of ML and citizen classifiers.

\section{\label{sec:expert}  Expert inspection}

The galaxies voted by Space Warps volunteers as most likely to be strong lenses were shown to experts -- i.e., professional astronomers -- for validation. The visual judgements of our experts determined which lenses were modelled (Sect. \ref{sec:modelling}) and ultimately reported in our lens catalogue (Sect. \ref{sc:catalogue}). 

\subsection{Expert inspection overview}

Our experts were professional astronomers who volunteered from the Euclid Consortium Strong Lensing Science Working Group.
61 volunteers made \num{80009} annotations of \num{8132} images\footnote{\num{7727} unique real candidates, after excluding simulated candidates and duplicates}.
Each expert was randomly assigned to annotate an image from the pool of candidates. 
Images were removed (`retired') once they received ten expert annotations.
This approach prioritises simplicity over efficiency. Random assignment is also ideal for managing our large and distributed group of experts.

The pool was initially filled with the top one thousand ML-identified candidates (see Sect. \ref{sec:machine_learning}) and continually updated with new candidates as identified by the Space Warps citizen scientists (see Sect. \ref{sec:citizen_science}). 
To understand which lenses our experts are likely to recover -- our expert selection function -- we also showed images painted with simulated strong lenses, selected to evenly cover a challenging range in Einstein radius and signal-to-noise ratio ($\ang{;;0.5}<\theta_{\rm E}<\ang{;;1.2}$ and $1<{\rm S/N}<200$).

Following \citet{AcevedoBarroso24} and in keeping with convention, we asked experts to assign each image as either Grade A (``confident lens: shows clear lensing features, no additional information is needed''), Grade B (``probable lens: shows lensing features but additional information is required to verify that it is a definite lens''),
Grade C (``possible lens: shows lensing features but they can be explained without resorting to gravitational lensing'') or not a lens (X).

We added two extensions to this conventional grading scheme. First, we introduced the new option Grade A+ (``Confident lens of individual scientific value, might be worth dedicated follow-up or a dedicated paper'') to flag rare configurations such as double source plane lenses \citep{Q1-SP054}. We expect these to be identified more frequently than in ground-based surveys thanks to \Euclid's PSF. Second, non-lenses may be marked as `otherwise interesting' in the hope of recording unusual galaxies seen while searching for lenses. We noted that some of our machine learning models \citep{Q1-SP053} confused these galaxies with lenses, likely because they were not represented in the training data for those models, and hence our search for lenses also functioned as an accidental search for unusual galaxies in general. 

The expert responses were used to calculate the lens score and lens grade reported in our catalogue (Sect. \ref{sc:catalogue}) after adjustments to account for the optimism of individual experts (Appendix \ref{app:calibrating_experts}) and to align our grades with previous work (Sect. \ref{sc:overall_results}). 

\subsection{Expert inspection results}


Experts inspected \num{7362} candidates identified by the Discovery Engine; the \num{4712} candidates rated in the top 1000 of each of the five deep learning models \citep{Q1-SP053}, and the \num{2650} candidates ranked most highly (SWAP score above $1\times10^{-5}$) by Space Warps volunteers. Experts also graded \num{365} ad hoc candidates identified outside our main search; see Sect. \ref{sc:additional_lens_candidates}.

We converted each expert response to a numerical score with a simple point system, following \citet{Canameras2020}:

\begin{equation}
\operatorname{\text{Score}}(\text{response})= \begin{cases}0 & \text { for response}=\text{X} \\ 1 & \text { for response}=\text{C} \\ 2 & \text { for response}=\text{B} \\ 3 & \text { for response}=\text{A or A+}  \end{cases}
\end{equation}

We then calculate a final score for each lens by aggregating these individual scores across the ten experts responding to each image. This aggregation starts with a simple mean, and then adds a correction to account for the typical optimism of each expert. See Appendix \ref{app:calibrating_experts} for more details. Where the same lens appears in overlapping cutouts, we manually select the correctly-centred cutout and assign the highest aggregate response from any of the overlapping cutouts.

We suggest that practitioners use the image gallery (Fig.~\ref{fig:lens_subset_by_score}) to choose a score cut that suits their desired purity. However, for convenience and by convention, we use score cuts to group lenses into `Grade A' and `Grade B' candidates. We define Grade A as candidates with lens score above 2 and Grade B as candidates with lens score above 1.5. We chose these cuts so that our `Grade A' and `Grade B' candidates have a similar visual confidence (roughly, accepting the inherent subjectivity) to the `Grade A' and `Grade B' candidates reported by previous works.

\section{\label{sec:modelling}  Modelling}

\subsection{Modelling overview}

We perform automated strong lens modelling of the highest expert-ranked candidates using the Euclid Strong Lens Modelling Pipeline\footnote{\href{https://github.com/Jammy2211/euclid\_strong\_lens\_modeling\_pipeline}{github.com/Jammy2211/euclid\_strong\_lens\_modeling\_pipeline}}, adaptated from the lens modelling software \texttt{PyAutoLens}\footnote{\href{https://github.com/Jammy2211/PyAutoLens}{github.com/Jammy2211/PyAutoLens}} \citep{nightingalePyAutoLensOpenSourceStrong2021}. 

The lens mass is modelled as an isothermal profile
\begin{equation}
\label{eqn:SPLEkap}
\kappa (\xi) = \frac{1}{1 + q^{\rm mass}} \bigg( \frac{\theta^{\rm mass}_{\rm E}}{\xi} \bigg)\;,
\end{equation}
where $\theta^{\rm mass}_{\rm E}$ is the Einstein radius, $q^{\rm mass}$ is the axis ratio, and $\xi$ is an elliptical radius measuring the distance to the centre in the image plane. Deflection angles are calculated using the method of \citet{kormann_isothermal_1994} in {\tt PyAutoLens}. External shear is included, parametrised as $(\gamma_1^{\rm ext}, \gamma_2^{\rm ext})$, with the shear magnitude and orientation given by
\begin{equation}
    \label{eq:shear}
    \gamma^{\rm ext} = \sqrt{\gamma_{\rm 1}^{\rm ext^{2}}+\gamma_{\rm 2}^{\rm ext^{2}}}, \, \,
    \tan{2\phi^{\rm ext}} = \frac{\gamma_{\rm 2}^{\rm ext}}{\gamma_{\rm 1}^{\rm ext}}\;.
\end{equation}
The deflection angles due to the external shear are computed analytically.

Table \ref{tab:pipeline_table} outlines our automated lens modelling pipeline. The pipeline models the lens galaxy's light using a multi-Gaussian expansion (MGE,  \citealt{he_unveiling_2024}), accounts for PSF blurring, and subtracts this model from the observed image. A mass model (isothermal distribution) ray-traces image pixels to the source plane, where a pixelized source reconstruction is performed using an adaptive Delaunay mesh. The pipeline iteratively fits various combinations of light, mass, and source models; the pipeline initially fits a simpler model using an MGE source for efficient and robust convergence towards accurate results, then subsequent stages employ the more complex Voronoi source reconstruction. The pipeline chains together five lens model fits in total. 

For further description of \texttt{PyAutoLens}, see Appendix \ref{app:modelling} and references therein, particularly \cite{he_unveiling_2024} and \cite{nightingaleScanningDarkMatter2024} for full details. A visual step-by-step guide to the {\tt PyAutoLens} likelihood function used in this work is  available via Jupyter notebooks.\footnote{\href{https://github.com/Jammy2211/autolens_likelihood_function}{github.com/Jammy2211/autolens\_likelihood\_function}}

\begin{table}
\caption{Pipeline composition used in the analysis built using \texttt{PyAutoLens}.}
\renewcommand{\arraystretch}{1.5}
\centering
\begin{tabular}{cccc}
\hline
\textbf{Pipeline} & \textbf{Component} & \textbf{Model} & \textbf{Prior info} \\ \hline
\textbf{Source}                             &  Lens light         & MGE            & -   \\ 
\textbf{Parametric}   & Lens mass          & SIE+Shear      & -                   \\ 
\textbf{(SP)}                                & Source light       & MGE            & -                   \\ \hline
\textbf{Source}                               & Lens light         & MGE            & \textbf{SP (fixed)}         \\
\textbf{Pixelization 1}  & Lens mass          & SIE+Shear      & \textbf{SP}         \\ 
  \textbf{(SPix1)}                               & Source light       & MPR            & -                   \\ \hline
 {\textbf{Source}}                & Lens light         & MGE            & \textbf{SP (fixed)}         \\ 
 {\textbf{Pixelization 2}} & Lens mass          & SIE+Shear      & -        \\
{\textbf{(SPix2)}}                  & Source light       & Voronoi            & -                   \\ \hline
 {\textbf{Mass}}                & Lens light         & MGE            & \textbf{SP1}         \\ 
 {\textbf{(M)}} & Lens mass          & SIE            & \textbf{SP1}        \\ 
               & Source light       & Delaunay            & \textbf{SP2}                   \\ \hline
\end{tabular}

\label{tab:pipeline_table}
\end{table}

\subsection{Modelling results}

The Euclid Strong Lens Modelling Pipeline was applied to
488
grade A or B lens candidates (i.e. with an expert vetting score greater than 2.0). The first step assessed whether the automated modelling was successful, based primarily on how well the model reproduced the observed lensed source emission. The critical curves of the mass model and the source plane were also evaluated. A successful lens model does not necessarily confirm the candidate as a strong lens but indicates that the model fit the galaxy image as expected. For instance, if the observed emission in the image-plane is singly imaged without a counter-image and the model reflects this, the fit is deemed successful, even though the candidate is not a strong lens. Overall, 374 out of 488 candidates ($77\%$) were successfully modelled.

Among the 374 successful fits, we evaluated whether the candidates were genuine strong lenses based on the models. Of these, 315
were judged as strong lenses, while 59 were judged as not to be. Notably, only seven of these 59 was initially graded as a Grade A lens during the first round of visual inspection. This result highlights the effectiveness of combining machine learning, citizen science, and expert validation in identifying genuine strong lenses.

\begin{figure*}
\centering
\includegraphics[width=0.15\textwidth]{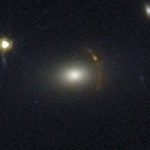}
\includegraphics[width=0.16\textwidth]{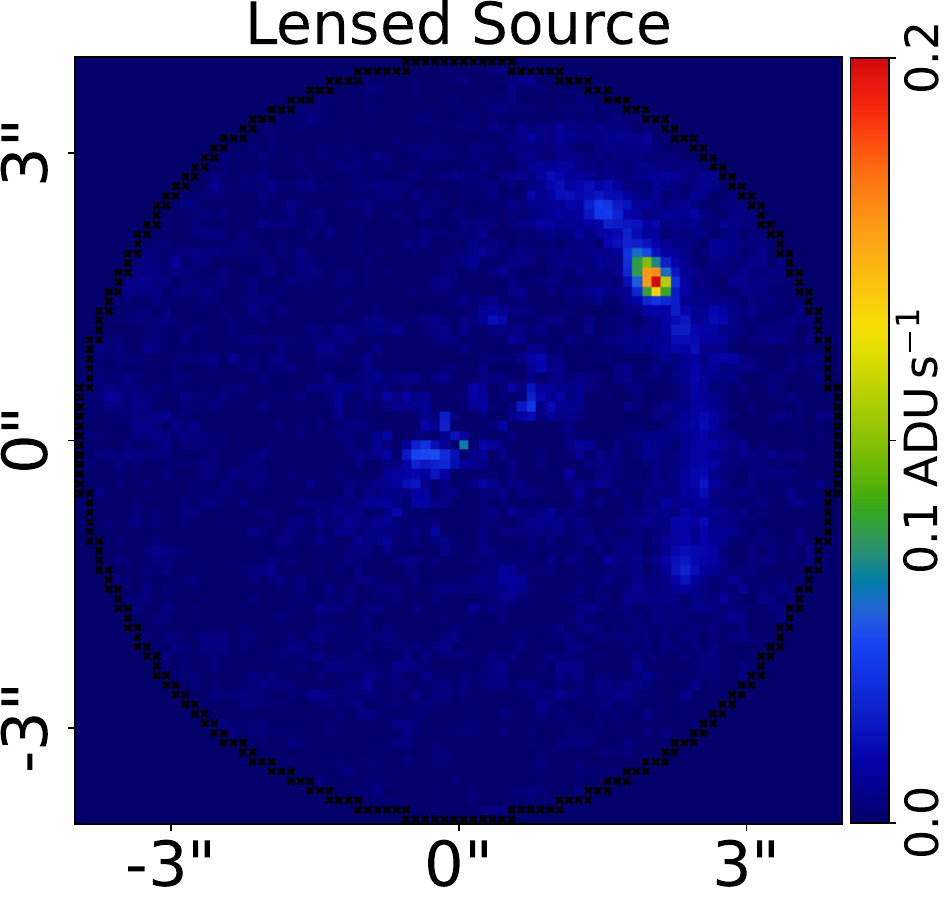}
\includegraphics[width=0.16\textwidth]{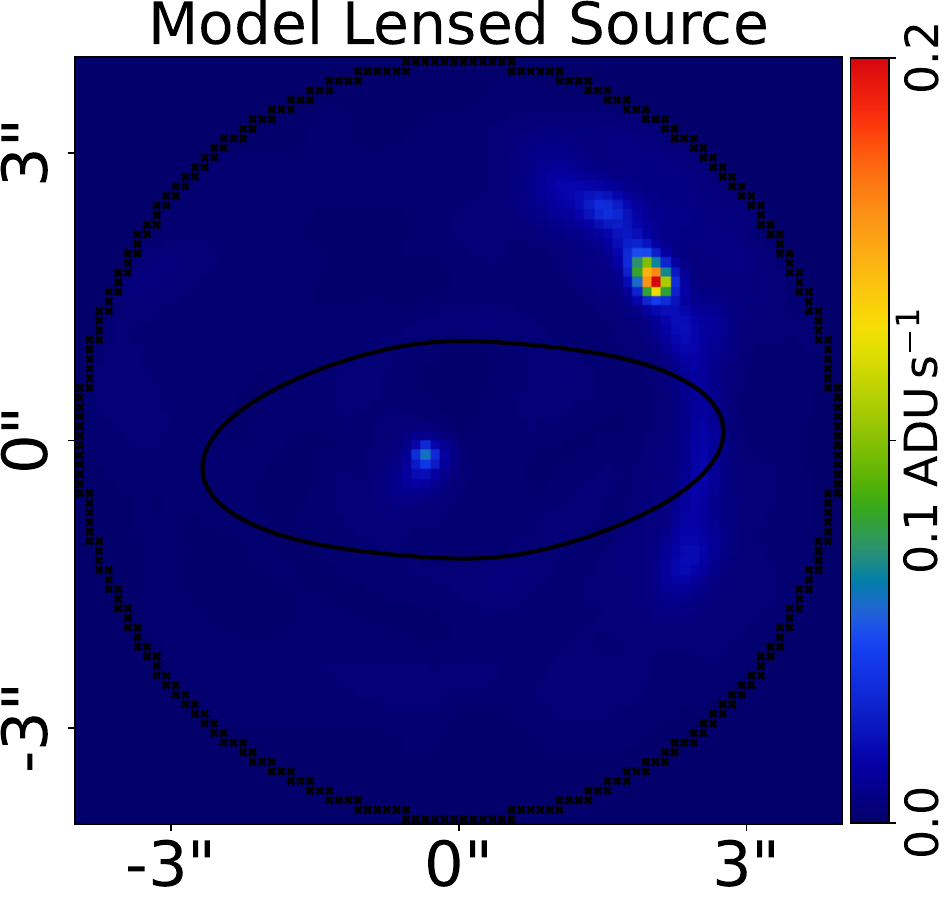}
\includegraphics[width=0.16\textwidth]{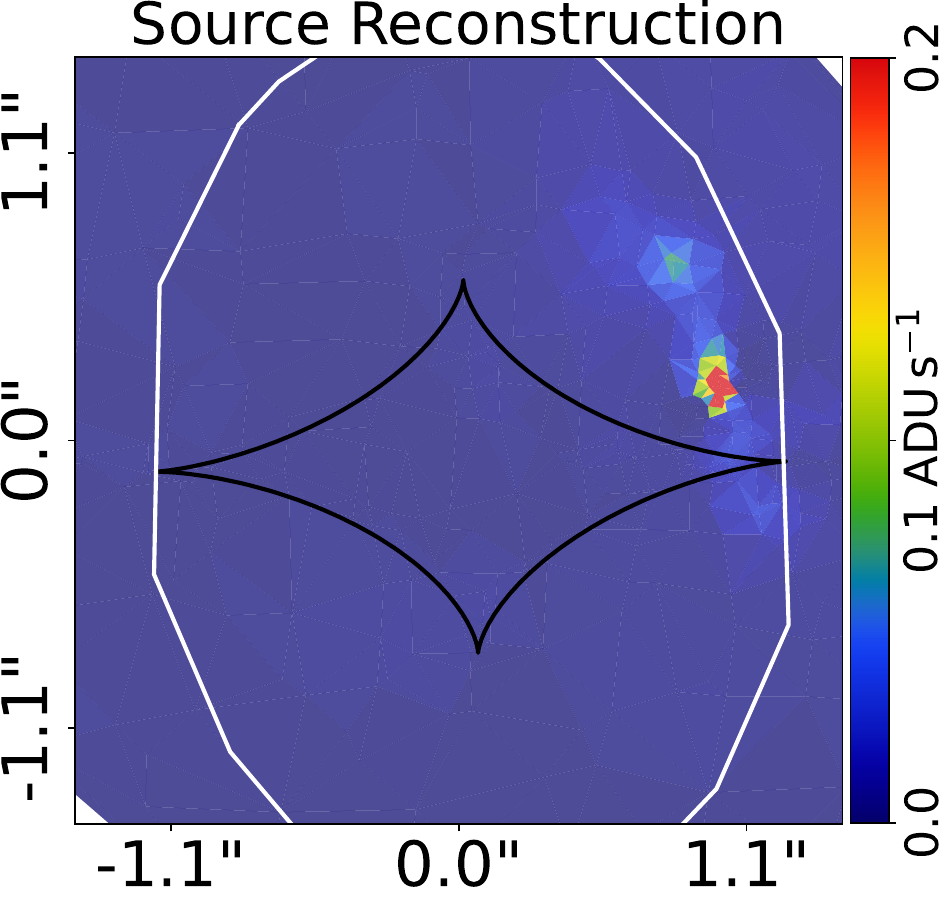}
\includegraphics[width=0.16\textwidth]{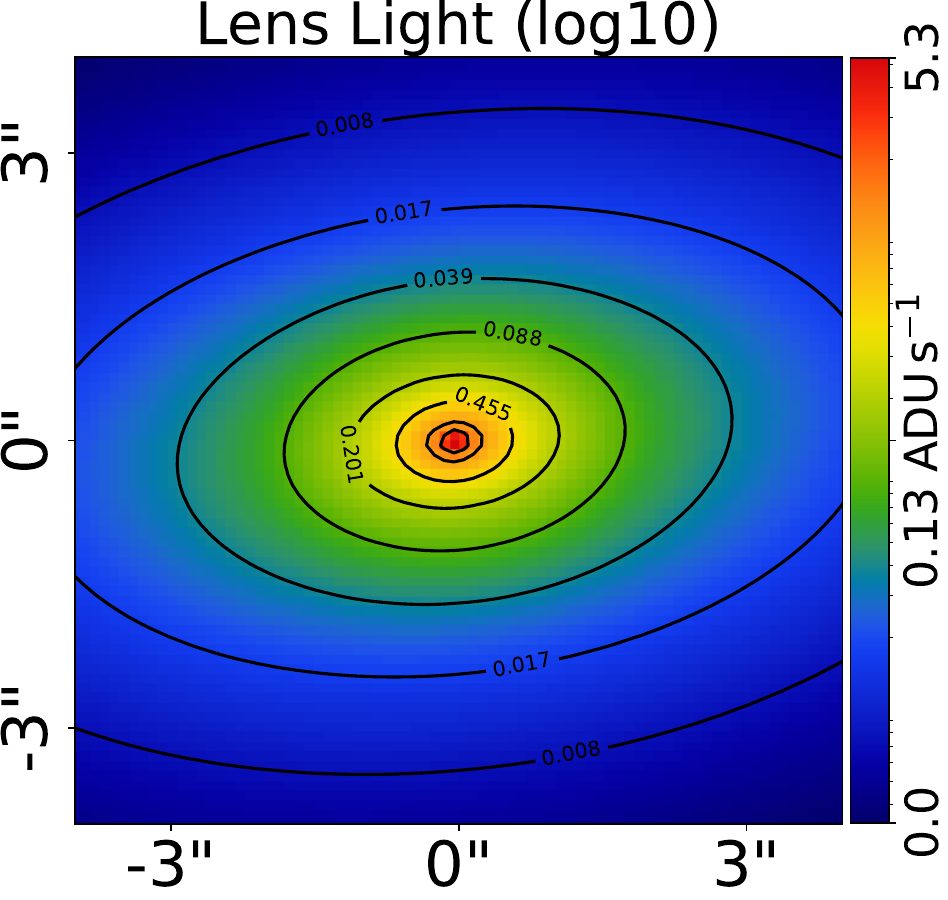}
\includegraphics[width=0.16\textwidth]{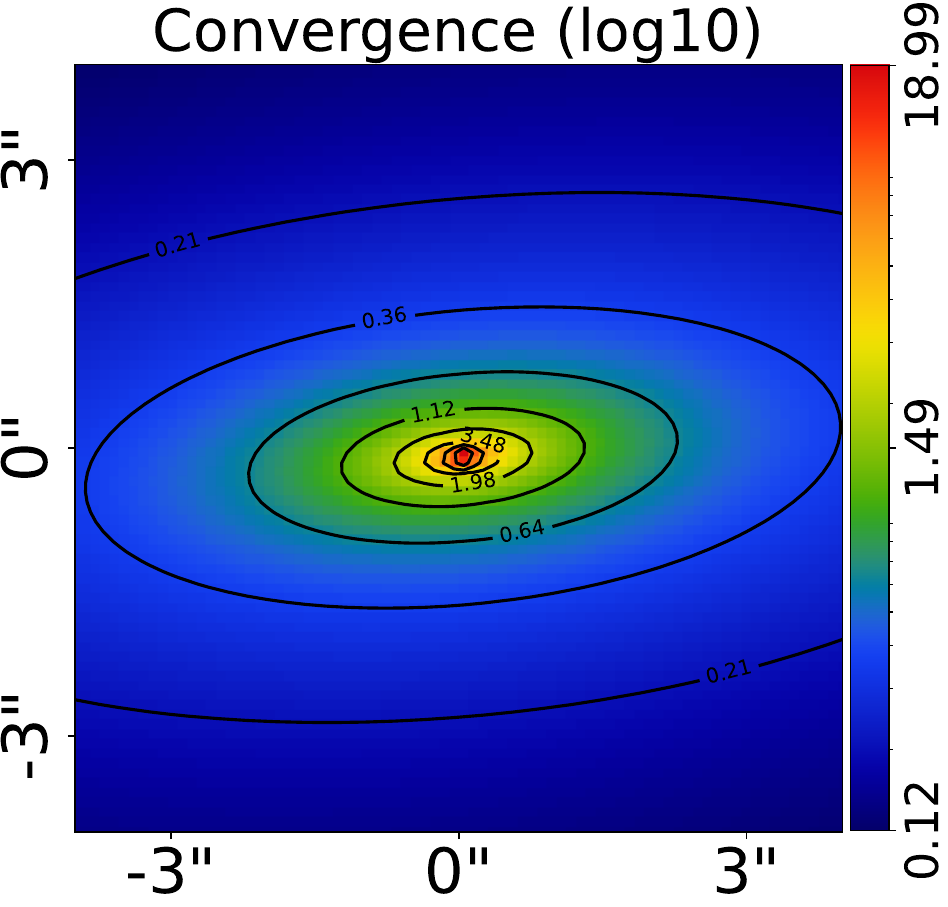}
\includegraphics[width=0.15\textwidth]{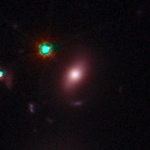}
\includegraphics[width=0.16\textwidth]{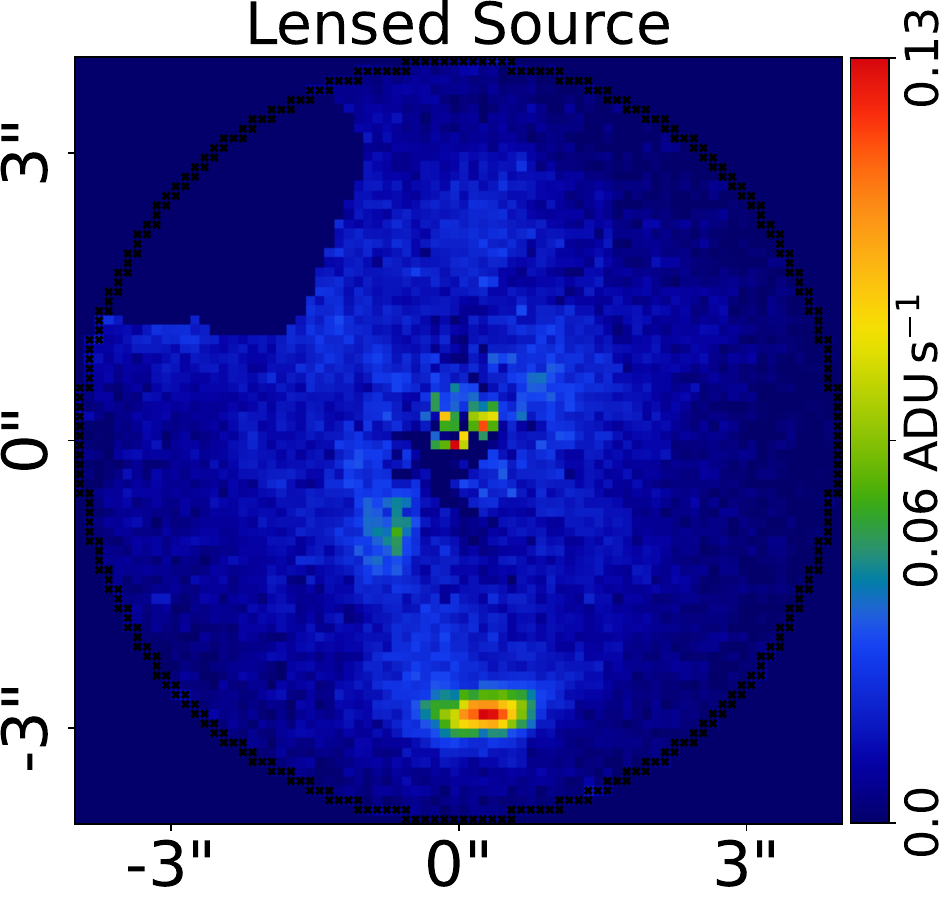}
\includegraphics[width=0.16\textwidth]{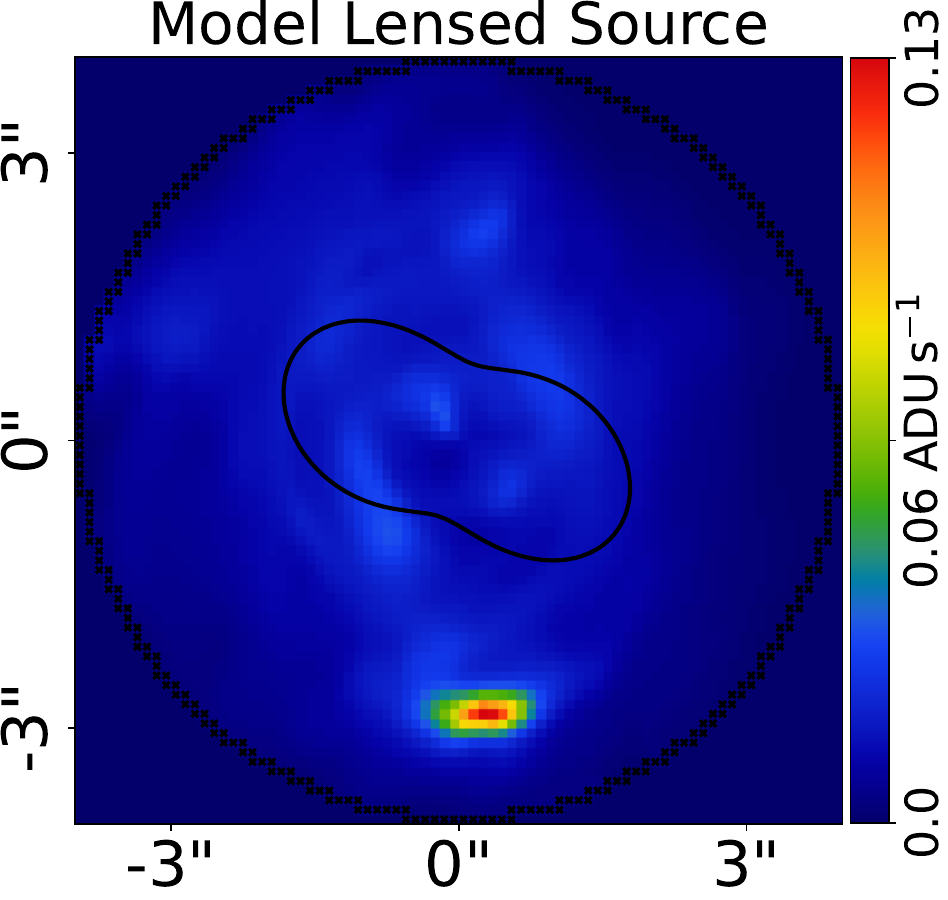}
\includegraphics[width=0.16\textwidth]{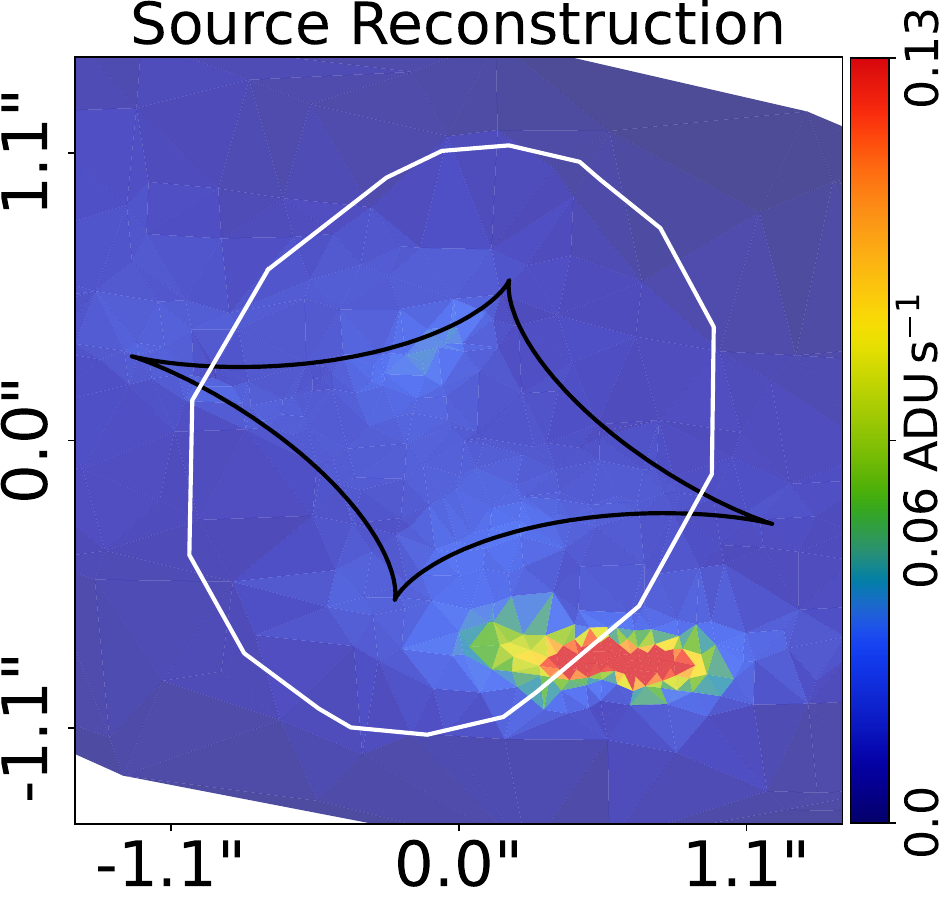}
\includegraphics[width=0.16\textwidth]{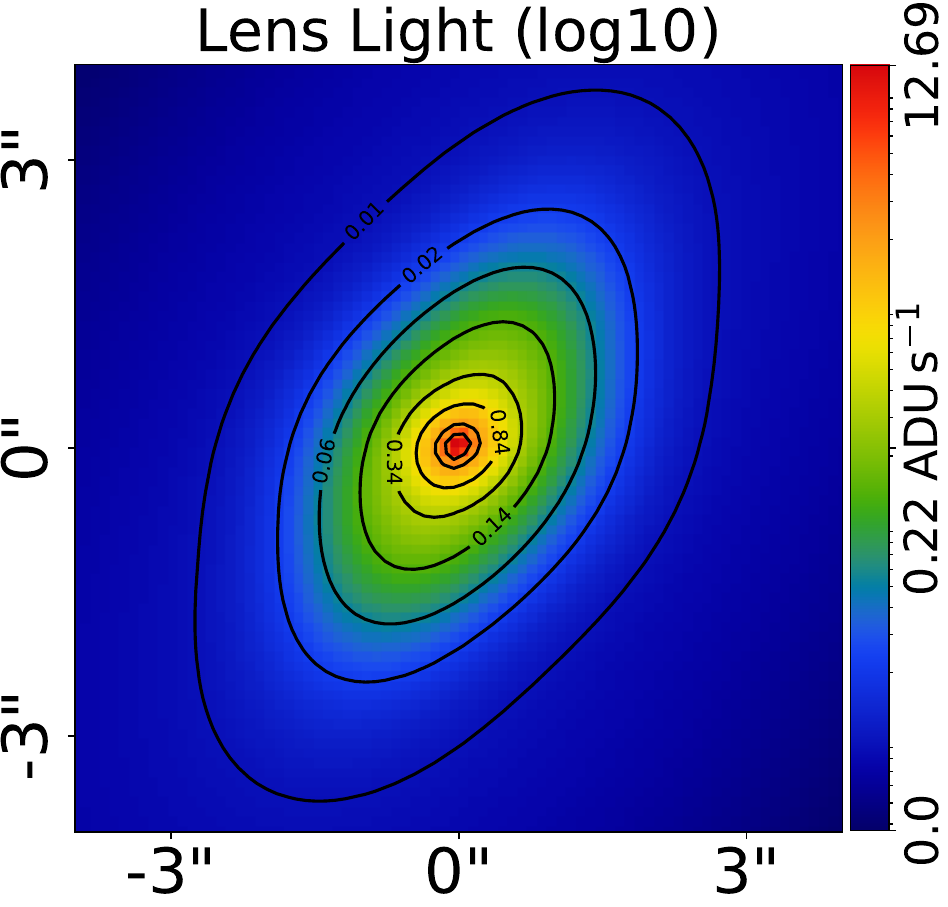}
\includegraphics[width=0.16\textwidth]{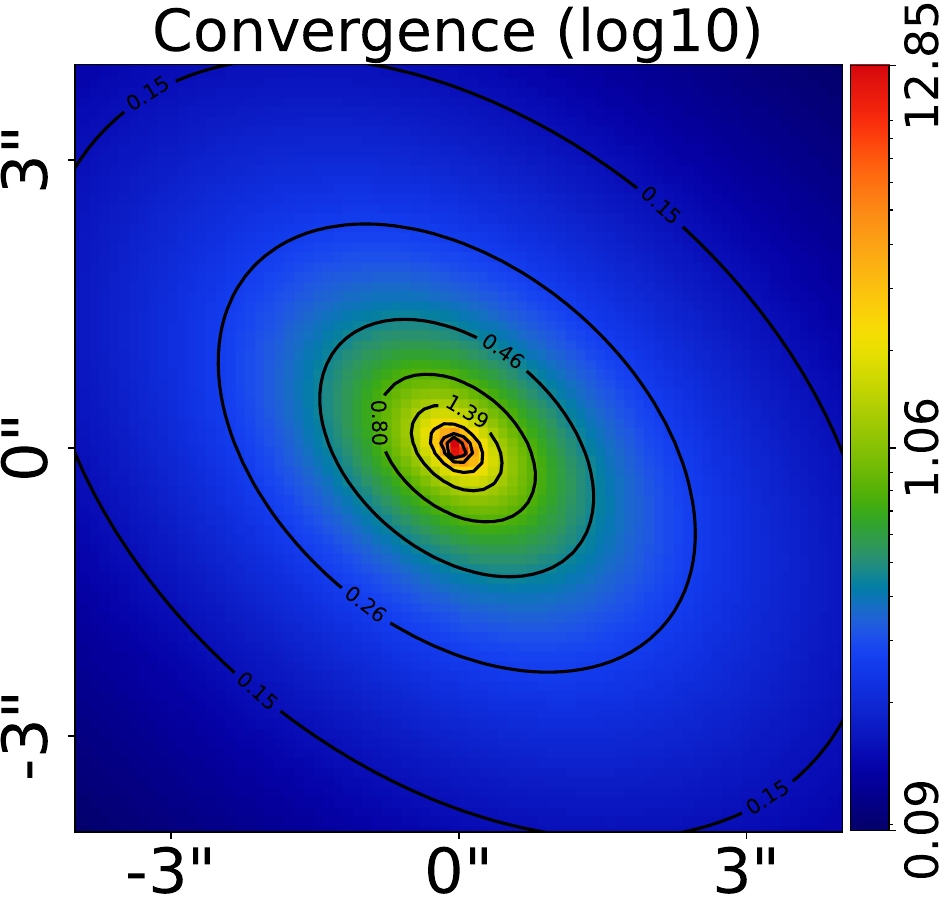}
\caption{
Lens models inferred by applying the Euclid Strong Lens Modelling Pipeline and \texttt{PyAutoLens} to our strong lens candidates. From left to right, the panels display: 
(i) the RGB image used during lens finding; 
(ii) the foreground lens-subtracted image; (iii) the model of the lensed source; (iv) the source-plane reconstruction on a Delaunay mesh; (v) the foreground lens light model on a $\log10$ colour scale; and (vi) the SIE mass model convergence on a $\log10$ colour scale. Tangential critical curves and caustics are marked with black lines, while radial critical curves and caustics are indicated with white lines. The first row illustrates how lens modelling confirms a strong lens by revealing a faint counter-image to a large source arc, which was undetectable through visual inspection of the observed \IE or RGB images. In contrast, the second row demonstrates a case where lens modelling rules out a lens candidate, as the model fails to uncover evidence of a counter-image in the data.
}
\label{fig:lens_models}
\end{figure*}

Figure \ref{fig:lens_models} presents example lens models fitted to candidates, demonstrating how lens modelling can confirm or rule out their status as strong lenses. The first row shows a case where the model uncovers a faint counter-image in the data, undetectable in the \IE data before lens subtraction or in the RGB images. This phenomenon occurs in 60 out of 374 candidates. The second row provides an example where the model fails to identify a counter-image for the candidate's lensed source emission, offering evidence that this candidate is not a strong lens (42 out of 59 candidates).

\begin{figure}
    \centering
    \includegraphics[width=\linewidth]{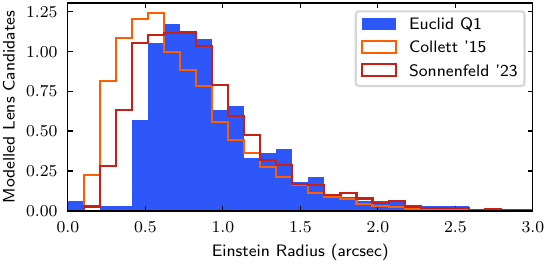}
    \caption{Distribution of Einstein radii for lens candidates with successful models (`Euclid Q1') compared to the expected Einstein radii of lenses in simulations of Euclid \citep{collettPopulationGalaxyGalaxyStrong2015,sonnenfeldStrongLensingSelection2023}. Normalised lens counts, with Q1 normalisation adjusted to match peaks.}
    \label{fig:lens_models_einstein_radii}
\end{figure}

Figure \ref{fig:lens_models_einstein_radii} shows the normalised distribution of Einstein radii recovered in our search. The distribution of radii is in excellent agreement with the expected distribution for lenses in the Universe as simulated by \citet{collettPopulationGalaxyGalaxyStrong2015} and \citet{sonnenfeldStrongLensingSelection2023}. This suggests that Einstein radius is not a strong selection effect for $\theta_{\rm E} = \ang{;;0.6}$, consistent with the selection function we estimate independently by injecting simulated lenses (Sect. \ref{sc:completeness}). Further, the agreement between forecasted and recovered radii provides evidence in favour of the accuracy of both the forecasts and our search approach.

The distribution comparison with simulated lens populations suggests a sharp drop in detectability below $\theta_{\rm E} = \ang{;;0.6}$ (see also \citealt{AcevedoBarroso24}, using five lenses discovered in the \Euclid ERO). Injections of simulated lens images, however, suggest lenses should still visible down to $\theta_{\rm E} = \ang{;;0.5}$. Identifying the exact lower limit of \Euclid's detection capability will require further work, and the limit itself may change as we refine\footnote{If the forecasts are correct in predicting a substantial population of low $\theta_{\rm E}$ lenses, a small improvement in the lower detection limit -- e.g., adding galaxy light subtraction -- would lead to a substantial increase in detected lenses.} our search approach. 

\subsection{Scaling lens modelling to DR1}

Traditional optimisation methods, such as \texttt{PyAutoLens}, are computationally intensive and often require significant manual input. While modelling the strong lenses presented in this paper took only a few days, applying the same approach to all lenses expected from DR1 and the full survey would be infeasible. 
One alternative is differentiable fitting codes \citep{guGIGALensFastBayesian2022, stoneCausticsPythonPackage2024}. These scale traditional lens modelling using gradient-based sampling on GPU accelerators.
Another alternative is neural network lens parameter estimation directly from the lens images, as in LEMON \citep{gentile_lemon_2023, Q1-SP063}.
LEMON estimates the lens mass and light parameters, such as the Einstein radius and foreground Sérsic index, and includes uncertainty estimates through its Bayesian framework. \citet{Q1-SP063} show that LEMON can accurately recover simulated lens parameters and provides estimates consistent with independent measurements on Euclidised, \Euclid ERO strong lenses, and the Q1 strong lenses presented here.  

\section{\label{sc:overall_results} Overall results}

\subsection{Discovery Engine search numbers}

Below, we summarise the counts from each stage of our search.

\begin{enumerate}
    \item Sources in Q1: \num{29767644}
    \item Sources passing selection cuts: \num{1086556}
    \item Candidates searched by ML: \num{1086554}
    \item Candidates shown to citizens
        \begin{enumerate}
        \item ML: \num{78214} 
        \item Random: \num{40000}
        \item Total (including overlap): \num{115329}
        \end{enumerate}
    \item Candidates shown to experts
        \begin{enumerate}
        \item ML only:  \num{4712}
        \item ML then Citizens: \num{2650}
        \item Total: \num{7362}
        \end{enumerate}
    \item Expert grades
        \begin{enumerate}
        \item Grade A: 250
        \item Grade B: 247
        \end{enumerate}
    \item Modelling of AB candidates
        \begin{enumerate}
        \item Successful fit, likely lens: 315 of 374
        \item Successful fit, unlikely lens: 59 of 374
        \item Unsuccessful fit, likely lens:  80 (ignoring group scale lenses, 61) 
        \end{enumerate}
\end{enumerate}

\subsection{Completeness}
\label{sc:completeness}

\begin{figure}
    \centering
    \includegraphics[width=\linewidth]{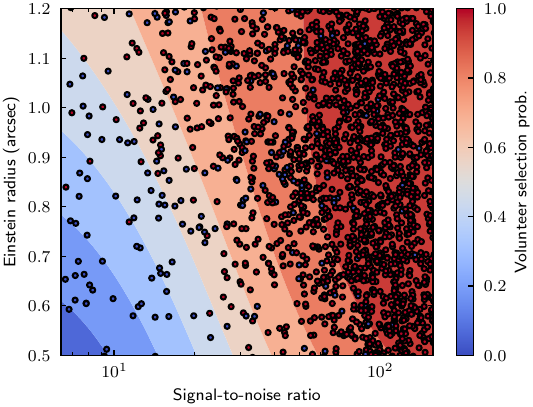}
    \caption{Selection function of volunteers with respect to Einstein radius and signal-to-noise estimated by injecting images with painted lenses from \citet{Q1-SP052}. Coloured points show volunteer responses to painted lenses (after SWAP aggregation). Contours show the estimated probability of a lens being selected. \Euclid's resolution allows us to recover lenses down to Einstein radii below \ang{;;0.5} (the lowest simulated). Signal-to-noise is the key limiting factor.}
    \label{fig:volunteer_selection_function}
\end{figure}

\begin{figure}
    \centering
    \includegraphics[width=\linewidth]{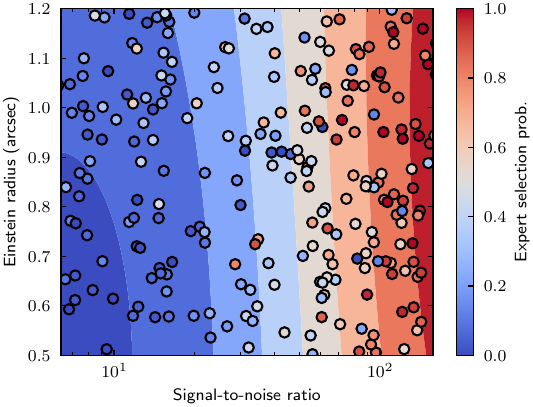}
    \caption{As in Fig.~\ref{fig:volunteer_selection_function}, for experts. Colours and contours show the actual and estimated scores given by experts to painted lenses. In this work we consider a score above 1.5 as Grade B and 2.0 as Grade A.}
    \label{fig:expert_selection_function}
\end{figure}

We used the responses of volunteers and experts to images with painted lenses \citep{Q1-SP053} to estimate completeness with respect to Einstein radii and signal-to-noise ratio. 
We find that both volunteers (Fig.~\ref{fig:volunteer_selection_function}) and experts (Fig.~\ref{fig:expert_selection_function}) can reliably recover bright lenses down to an Einstein radii of $\ang{;;0.5}$ (the lowest simulated), likely due to \Euclid's space-based resolution. Signal-to-noise ratio\footnote{Defined in Sect. 2.5 of \citealt{Q1-SP052}} is the key limiting factor; even for lenses of large Einstein radii, volunteer and expert responses rapidly drop as the signal-to-noise ratio falls below approximately \num{100} (as defined in \citealt{Q1-SP052}). Experts generally rate painted lenses of a signal-to-noise ratio below approximately 30 as Grade C (score below 1.5) or lower. Volunteers are more likely to correctly select these faint lenses (selection probability of approximately 50\%) but are also more optimistic overall, as we describe below. 
Fainter arcs become hard to recognise either because they become indistinguishable from background, or are `outshone' by a bright lensing galaxy. Combining images from deeper surveys (e.g., \citealt{melo_holismokes_2024}) may help recover arcs from background, while light subtraction or fully-automated methods may be necessary to recover arcs near bright lensing galaxies.

\begin{figure}
    \centering
    \includegraphics[width=\linewidth]{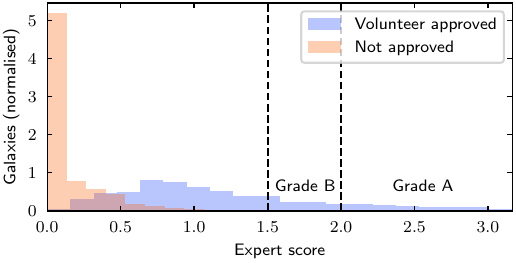}
    \caption{Distribution of expert scores for all subjects seen by experts and citizens. Citizens may have rejected (red histogram) or approved the subjects for expert vetting (based on SWAP score of $p>1\times10^{-5}$, blue histogram). Experts agree with almost all subjects rejected by volunteers. 26\% of the volunteer approved subjects were rated as Grade A or B by experts.}
    \label{fig:volunteer_vs_expert}
\end{figure}

Figure \ref{fig:volunteer_vs_expert} considers all galaxies rated by both volunteers and experts, where `Volunteer approved' means galaxies sent for expert inspection due to having a SWAP score above $p>1\times10^{-5}$. We chose this moderate threshold to include systems `worth an expert look'; not all volunteer-approved systems are expected to be highly likely candidates. 
Experts rated the galaxies approved by volunteers with a broad spread; volunteer-approved galaxies were generally considered lens-like, but only a minority (26\%) were rated as probable (grade B) to confident (grade A) lenses. This is expected given the low SWAP threshold. Moving to a higher SWAP threshold would capture most of the grade A and grade B candidates with 465 of the 497 having SWAP scores of $P>0.99$.  
Conversely, 8 of the 497 galaxies graded as A or B by experts were rejected by the citizens. While volunteer-rejected subjects would not normally be forwarded to the experts, in the initial stage of the project, experts and citizens simultaneously classified the top-scoring 1000 galaxies from each model. Within that high-ML-scoring subset of galaxies, 7 lens candidates (2 grade A and 5 grade B) were missed by the citizens. This implies a small incompleteness by Space Warps score alone for high ML score galaxies, but an ensemble analysis would help recover these (see \citealp{Q1-SP059}).

\begin{figure}
    \centering
    \includegraphics[width=\linewidth]{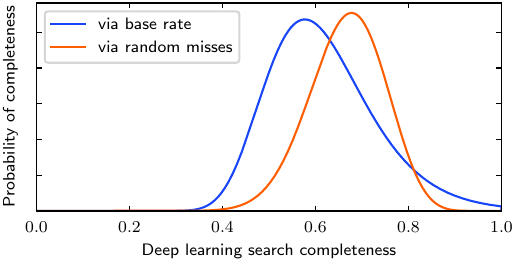}
    \caption{Posteriors for the completeness of our deep learning search. Of all the lenses that a full citizen inspection of the one million Q1 galaxies would have revealed, how many did our deep learning models correctly prioritise? Each of our two methods -- the base rate of lenses found randomly, and the fraction of random lenses correctly prioritised -- imply a completeness (compared to the number of lenses that would be found from exhaustive inspection) of around 60\%.}
    \label{fig:expected_vs_actual_lenses}
\end{figure}

How many lenses did we miss by using deep learning models to pick a subset of galaxies to inspect, rather than inspecting all one million galaxies? 
We estimate this through the forty thousand randomly-selected sources shown to citizens, of which 31 were rated as grade B candidates or better. This suggests a base rate of 0.79 lenses per thousand sources ($0.62$--$0.98$ at $10\%$--$90\%$ confidence). With that base rate, we would expect a complete visual search of all Q1 sources to find 837 candidates, while our model-prioritised search ultimately found 500, for a completeness estimate of 60\%. Alternatively, our models correctly prioritised 21 of the 31 strong lens candidates, implying a completeness of 66\%. Figure \ref{fig:expected_vs_actual_lenses} shows each posterior estimate for the completeness of our deep learning search. Understanding which lenses are missed by all models will be a major focus for DR1 and beyond.

\subsection{\label{sc:catalogue} Lens catalogue and derived data products}

Our data are publicly released on Zenodo\footnote{\url{https://doi.org/10.5281/zenodo.15003116}}. This includes our lens catalogue and our lens modelling results. We also share the underlying FITS cutouts to support further modelling work or follow-up proposals. Extensive documentation is available at that link. We summarize the presentation of the catalogue below.

The catalogue columns are ordered and grouped by theme.
The \textit{identification} group shares the sky coordinates the and \Euclid source catalogue keys (e.g., the tile index). The \textit{rating} group shares the Galaxy Judges score and suggested grade. The \textit{modelling quality} group shares our interpretation of the modelling results (e.g., if the model is consistent with a strong lens, if a counter-image was revealed, etc.). The \textit{modelling lens mass} group shares columns relating to the lens mass (e.g., the Einstein radius). The remaining \textit{modelling results} columns share additional modelling results (e.g., the signal-to-noise ratio, the lens magnification, etc.). Each modelling result is reported as a median value, plus upper and lower limits at $1\sigma$ and $3\sigma$. 

In addition to our 250 Grade A candidates and 247 Grade B candidates, we also include 585 candidates with expert scores above 1.0, and define these as Grade C candidates to indicate lens-like galaxies that are possible lenses but not persuasively so. We do not count these as lens candidates elsewhere in this work or in other papers in this series. 

\subsection{\label{sc:additional_lens_candidates} Additional lens candidates}

\subsubsection{Candidates independently identified}

Some candidates independently identified by our Discovery Engine were also found from other searches of \Euclid data. We list these candidates in our main catalogue, with metadata (e.g., the `subset' column) noting their related discovery. This ensures our selection function is not affected by other searches. 

\citet{oriordanEuclidCompleteEinstein2025} reports the discovery of a complete Einstein ring in NGC6505, found serendipitously during \Euclid Performance Verification observations aimed at measuring ice contamination. Our Discovery Engine would have found this lens; the Zoobot model ranks it 332nd of 1.08M (top 0.03\%), high enough to be sent directly for expert grading, and experts assigned it the highest grade of any candidate (eight grades of `A+' and 2 grades of `A').

\citet{Q1-SP052} used spectroscopic data to identify high-velocity-dispersion galaxies with \Euclid imaging. They then carried out an expert inspection campaign to grade those galaxies and identify strong lens candidates. Sixteen grade A and twelve grade B candidates (our scores) in this work were also identified in this targeted search. Only two candidates were rejected by our search (one was not prioritised by our machine learning models and two were rejected by Space Warps volunteers). The remaining difference in our samples (10 candidates) is due a difference in selection cuts, primarily rejections by our \textit{Gaia} cut described below.

Pearson et. al. (in prep.) asked Galaxy Zoo volunteers to flag strong lenses in HSC images, first as part of the broader process of annotating morphology and then in a targeted project on Galaxy Zoo Mobile. We were provided this list of candidates at the start of our search. All grade A and B candidates were successfully independently identified by the Discovery Engine (likely because the HSC-selected candidates were visually obvious in \Euclid imaging).

Ecker et. al (in prep.) is carrying out a dedicated search for low-redshift lenses in Q1. Comparison with this search revealed that our choice to reject \textit{Gaia} catalog sources was a convenient method to avoid stars (the vast majority of \textit{Gaia} sources) but also excludes some bright low-redshift galaxies which may be lenses. We will alter our star rejection approach for DR1. 

\subsubsection{Candidates Not Identified}

Some candidates were not identified by our Discovery Engine (e.g., due to selection cuts) but were found during the course of this project (e.g., serendipitously) and may be helpful to the community. We have graded these additional candidates through the same expert vetting (Sect. \ref{sec:expert}) process. We include them a secondary table of candidates and do not otherwise include them in the numbers reported in this work. 

Galaxy Zoo Euclid \citep{Q1-SP047} asked citizen science volunteers to label galaxy morphology (bars, spirals, etc.) in \Euclid images (\textit{outside} Q1). Volunteers had the option to flag possible lenses, either as part of a final multiple-choice question or via free text `tags' on the Galaxy Zoo forum. Between August and December 2024, volunteers flagged 814 possible candidates from 140k images. We requested expert vetting and ultimately identify 56 Grade A and 19 Grade B candidates. 

The remaining three additional lens candidates were discovered serendipitously by members of the Euclid Consortium. These are useful for revealing lenses missed by our engine.
One candidate was missed by the deep learning models; the model that ranked it highest gave a rank of seventy thousand, which compares well with one million total sources but fell well below our citizen inspection cutoff (20k). We expect that our deep learning search is around 70\% complete (Fig.~\ref{fig:expected_vs_actual_lenses}). Updating our models, most straightforwardly by retraining on the lenses found so far, will likely improve our completeness and purity, but -- given \Euclid's scale -- we consider our current models as already sufficient to find unprecedented numbers of lenses (see \citealt{Q1-SP053} and Sect. \ref{sc:conclusion_and_outlook}) 
Two candidates were rejected by citizen scientists, both `understandably'; one has a faint arc that is difficult to quickly identify, and the other was substantially off-centre. 
The off-centre example highlights how detailed modelling of a real lens search selection function involves more than just commonly-considered astrophysical parameters like Einstein radius and signal-to-noise ratio. 

\section{\label{sc:conclusion_and_outlook}  Conclusion and discussion}

\Euclid is an unparalleled instrument for finding strong gravitational lenses. 
We present a catalogue of strong lenses in Q1 -- the first core survey data from \Euclid.
We identified these lenses by combining spectroscopy, deep learning, citizen science, expert inspection, and lens modelling -- our `Discovery Engine'.  

\citet{Q1-SP052} created an initial training set for our deep learning models by using spectroscopy to identify new lenses. Our five deep learning models, described in \citet{Q1-SP053}, then ranked one million Q1 galaxies. The best-performing model, Zoobot, identified 163 lens candidates in the top 1000 galaxies. 1800 volunteers from the Space Warps citizen science project (Sect. \ref{sec:citizen_science}) searched the highest-ranked \num{78000} galaxies, and also searched \num{40000} random galaxies to robustly estimate \Euclid's lens-finding capability. 61 professional astronomers then vetted \num{7500} galaxies selected by the models and volunteers. This revealed four new double-source-plane lenses, investigated in detail in \citet{Q1-SP054}, as well as complete Einstein rings, quadruply-imaged lenses, and edge-on lensing galaxies.

We ultimately find 500 lens candidates, comparable to the most successful search campaigns (e.g., \citealt{jacobsExtendedCatalogGalaxyGalaxy2019, schuldt_holismokes_2025, gonzalez_discovering_2025})
while searching only 63 deg$^2$ in approximately six weeks. 
\citet{Q1-SP059} forecasts that even without further improvement applying our current classifiers to Euclid DR1 would yield
\num{3900} to \num{7600} grade A and B lens candidates when using Bayesian ensembling. Each of these outcomes finds more lens candidates than all previous searches combined -- we are only unsure by how much. 

The forecast uncertainty follows in part from the visual inspection budget of both citizens and experts; the conservative forecast assumes citizens inspect \num{100000} galaxies and experts inspect \num{5000} galaxies (as in this work), and the optimistic forecast assumes one million galaxies by citizens and \num{15000} galaxies by experts. While large, this is a small fraction of the roughly 36 million DR1 sources. 
Better prioritization could improve lens counts beyond the optimistic forecast, increasing towards the \num{30000} lens candidates we predict to be visually detectable in DR1.\footnote{Trivially applying our base rate estimate (Sect. \ref{sc:completeness})}

How can we prioritise better? 
Integrating known strategies could immediately help for DR1.
For the deep learning stage, we could retrain our models using our \num{900000} newly-collected human annotations of real \Euclid images.
For the citizen stage, a refinement stage asking citizens to look again at the best candidates would improve our purity \citep{Marshall2016}.
For the expert stage, applying SWAP -- as already done with citizens -- should increase efficiency by several times.

More broadly, visual searches -- whether directly by humans, or indirectly by models trained on humans -- have fundamental limits. The rapid drop in completeness for fainter lenses (Sect. \ref{sc:completeness}), combined with the faint counter-images revealed only retrospectively by lens modelling (Sect. \ref{sec:modelling}) suggests that there is an opportunity to find otherwise-undetectable lenses through automated techniques. Differentiable lens simulators \citep{guGIGALensFastBayesian2022, stoneCausticsPythonPackage2024} are one possibility. We may also see hybrid approaches such as using full-survey light subtraction (e.g., \citealt{tractor2016}) to present cleaned visual images. 

The clear visual lensing features resolved by \Euclid, the successful lens model fits, and the consistency with previous forecasts all suggest the bulk of our candidates are genuine lenses. However, 
definitive identification will require refined modelling and ultimately large-scale spectroscopy. 4MOST's planned strong lensing survey \citep{collett4MOSTStrongLensing2023} aims to measure redshifts for \num{10000} lensing pairs, which -- as this work shows -- \Euclid will comfortably provide. 

This first catalogue from \Euclid's main surveys opens a new era of strong lensing science.
Within the next two years, \Euclid's lens discoveries will expand to dwarf all previous searches combined.
We hope that the release of our catalogue will allow the community to begin using these lenses to better understand our Universe.

%
%

\begin{acknowledgements}

We gratefully acknowledge the Zooniverse team for their invaluable work on the host platform for Space Warps. 

This project would not be possible without the immense contributions of citizen scientists. Those contributing classifications for this project are listed on \href{https://www.zooniverse.org/projects/aprajita/space-warps-esa-euclid/about/team}{The Team}. We greatly appreciate their committent, enthusiasm, and time.

The Dunlap Institute is funded through an endowment established by the David Dunlap family and the University of Toronto. 

This work has received funding from the European Research Council (ERC) under the European Union's Horizon 2020 research and innovation programme (LensEra: grant agreement No 945536). TEC is funded by the Royal Society through a University Research Fellowship.

SS has received funding from the European Union’s Horizon 2022 research and innovation programme under the Marie Skłodowska-Curie grant agreement No 101105167 — FASTIDIoUS.

JP is supported by the ACME and OSCARS projects. "ACME: Astrophysics Center for Multimessenger studies in Europe" and "OSCARS: Open Science Clusters' Action for Research and Society" are funded by the European Union under grant agreement no. 101131928 and 101129751, respectively.

V.B. and C.T. acknowledge the INAF grant 2022 LEMON.

This publication uses data generated via the Zooniverse.org platform, development of which is funded by generous support, including a Global Impact Award from Google, and by a grant from the Alfred P. Sloan Foundation.

\AckEC  

\AckQone

This research makes use of ESA Datalabs (datalabs.esa.int), an initiative by
ESA’s Data Science and Archives Division in the Science and Operations
Department, Directorate of Science.

\end{acknowledgements}

%
%

\bibliography{my, Euclid, Q1}

\begin{thebibliography}{107}
\expandafter\ifx\csname natexlab\endcsname\relax\def\natexlab#1{#1}\fi

\bibitem[{Abbott {et~al.}(2018)Abbott, Abdalla, Allam, Amara, Annis, Asorey, Avila, Ballester, Banerji, Barkhouse, Baruah, Baumer, Bechtol, Becker, Benoit-Lévy, Bernstein, Bertin, Blazek, Bocquet, Brooks, Brout, Buckley-Geer, Burke, Busti, Campisano, Cardiel-Sas, Carnero~Rosell, Carrasco~Kind, Carretero, Castander, Cawthon, Chang, Chen, Conselice, Costa, Crocce, Cunha, D'Andrea, da~Costa, Das, Daues, Davis, Davis, De~Vicente, DePoy, DeRose, Desai, Diehl, Dietrich, Dodelson, Doel, Drlica-Wagner, Eifler, Elliott, Evrard, Farahi, Fausti~Neto, Fernandez, Finley, Flaugher, Foley, Fosalba, Friedel, Frieman, García-Bellido, Gaztanaga, Gerdes, Giannantonio, Gill, Glazebrook, Goldstein, Gower, Gruen, Gruendl, Gschwend, Gupta, Gutierrez, Hamilton, Hartley, Hinton, Hislop, Hollowood, Honscheid, Hoyle, Huterer, Jain, James, Jeltema, Johnson, Johnson, Kacprzak, Kent, Khullar, Klein, Kovacs, Koziol, Krause, Kremin, Kron, Kuehn, Kuhlmann, Kuropatkin, Lahav, Lasker, Li, Li, Liddle, Lima, Lin, López-Reyes, MacCrann, Maia,
  Maloney, Manera, March, Marriner, Marshall, Martini, McClintock, McKay, McMahon, Melchior, Menanteau, Miller, Miquel, Mohr, Morganson, Mould, Neilsen, Nichol, Nogueira, Nord, Nugent, Nunes, Ogando, Old, Pace, Palmese, Paz-Chinchón, Peiris, Percival, Petravick, Plazas, Poh, Pond, Porredon, Pujol, Refregier, Reil, Ricker, Rollins, Romer, Roodman, Rooney, Ross, Rykoff, Sako, Sanchez, Sanchez, Santiago, Saro, Scarpine, Scolnic, Serrano, Sevilla-Noarbe, Sheldon, Shipp, Silveira, Smith, Smith, Smith, Soares-Santos, Sobreira, Song, Stebbins, Suchyta, Sullivan, Swanson, Tarle, Thaler, Thomas, Thomas, Troxel, Tucker, Vikram, Vivas, Walker, Wechsler, Weller, Wester, Wolf, Wu, Yanny, Zenteno, Zhang, Zuntz, {DES Collaboration}, Juneau, Fitzpatrick, \& Nikutta}]{abbott_dark_2018}
Abbott, T. M.~C., Abdalla, F.~B., Allam, S., {et~al.} 2018, ApJS, 239, 18

\bibitem[{{Acevedo Barroso} {et~al.}(2025){Acevedo Barroso}, {Cl{\'e}ment}, {Courbin}, {Gavazzi}, {Lemon}, {Rojas}, {Scott}, {Gwyn}, {Hammer}, {Hudson}, \& {Magnier}}]{acevedoBarrosoEtAl25}
{Acevedo Barroso}, J.~A., {Cl{\'e}ment}, B., {Courbin}, F., {et~al.} 2025, arXiv e-prints, arXiv:2503.10610

\bibitem[{{Acevedo Barroso} {et~al.}(2024){Acevedo Barroso}, {O'Riordan}, {Cl{\'e}ment}, {et~al.}}]{AcevedoBarroso24}
{Acevedo Barroso}, J.~A., {O'Riordan}, C.~M., {Cl{\'e}ment}, B., {et~al.} 2024, A\&A, submitted, arXiv:2408.06217

\bibitem[{Agnello \& Spiniello(2019)}]{agnelloQuasarLensesSouth2019}
Agnello, A. \& Spiniello, C. 2019, MNRAS, 489, 2525

\bibitem[{Aihara {et~al.}(2019)Aihara, Alsayyad, Ando, Armstrong, Bosch, Egami, Furusawa, Furusawa, Goulding, Harikane, Hikage, Ho, Hsieh, Huang, Ikeda, Imanishi, Ito, Iwata, Jaelani, Kakuma, Kawana, Kikuta, Kobayashi, Koike, Komiyama, Li, Liang, Lin, Luo, Lupton, Lust, Macarthur, Matsuoka, Mineo, Miyatake, Miyazaki, More, Murata, Namiki, Nishizawa, Oguri, Okabe, Okamoto, Okura, Ono, Onodera, Onoue, Osato, Ouchi, Shibuya, Strauss, Sugiyama, Suto, Takada, Takagi, Takata, Takita, Tanaka, Terai, Toba, Uchiyama, Utsumi, Wang, Wang, \& Yamada}]{Aihara2019}
Aihara, H., Alsayyad, Y., Ando, M., {et~al.} 2019, PASJ, 71

\bibitem[{Andika {et~al.}(2023)Andika, Suyu, Cañameras, Melo, Schuldt, Shu, Eilers, Jaelani, \& Yue}]{andikaStreamlinedLensedQuasar2023}
Andika, I.~T., Suyu, S.~H., Cañameras, R., {et~al.} 2023, A\&A, 678, A103

\bibitem[{Barkana {et~al.}(1999)Barkana, Blandford, \& Hogg}]{barkanaPossibleGravitationalLens1999}
Barkana, R., Blandford, R., \& Hogg, D.~W. 1999, ApJ, 513, L91

\bibitem[{Bolton {et~al.}(2008)Bolton, Burles, Koopmans, Treu, Gavazzi, Moustakas, Wayth, \& Schlegel}]{boltonSloanLensACS2008}
Bolton, A.~S., Burles, S., Koopmans, L. V.~E., {et~al.} 2008, ApJ, 682, 964

\bibitem[{Bommasani {et~al.}(2021)Bommasani, Hudson, Adeli, Altman, Arora, von Arx, Bernstein, Bohg, Bosselut, Brunskill, Brynjolfsson, Buch, Card, Castellon, Chatterji, Chen, Creel, Davis, Demszky, Donahue, Doumbouya, Durmus, Ermon, Etchemendy, Ethayarajh, Fei-Fei, Finn, Gale, Gillespie, Goel, Goodman, Grossman, Guha, Hashimoto, Henderson, Hewitt, Ho, Hong, Hsu, Huang, Icard, Jain, Jurafsky, Kalluri, Karamcheti, Keeling, Khani, Khattab, Koh, Krass, Krishna, Kuditipudi, Kumar, Ladhak, Lee, Lee, Leskovec, Levent, Li, Li, Ma, Malik, Manning, Mirchandani, Mitchell, Munyikwa, Nair, Narayan, Narayanan, Newman, Nie, Niebles, Nilforoshan, Nyarko, Ogut, Orr, Papadimitriou, Park, Piech, Portelance, Potts, Raghunathan, Reich, Ren, Rong, Roohani, Ruiz, Ryan, Ré, Sadigh, Sagawa, Santhanam, Shih, Srinivasan, Tamkin, Taori, Thomas, Tramèr, Wang, Wang, Wu, Wu, Wu, Xie, Yasunaga, You, Zaharia, Zhang, Zhang, Zhang, Zhang, Zheng, Zhou, \& Liang}]{Bommasani2021}
Bommasani, R., Hudson, D.~A., Adeli, E., {et~al.} 2021, Tech. rep., Stanford, arXiv: 2108.07258

\bibitem[{Canameras {et~al.}(2024)Canameras, Schuldt, Shu, Suyu, Taubenberger, Andika, Bag, Inoue, Jaelani, Leal-Taixe, Meinhardt, Melo, \& More}]{canameras_holismokes_2024}
Canameras, R., Schuldt, S., Shu, Y., {et~al.} 2024, A\&A, 692, A72

\bibitem[{Cao {et~al.}(2022)Cao, Li, Nightingale, Massey, Robertson, Frenk, Amvrosiadis, Amorisco, He, Etherington, Cole, \& Zhu}]{cao_systematic_2022}
Cao, X., Li, R., Nightingale, J.~W., {et~al.} 2022, Research in Astronomy and Astrophysics, 22, 025014

\bibitem[{Cappellari(2002)}]{cappellari_efficient_2002}
Cappellari, M. 2002, MNRAS, 333, 400

\bibitem[{Cañameras {et~al.}(2021)Cañameras, Schuldt, Shu, Suyu, Taubenberger, Meinhardt, Leal-Taixé, Chao, Inoue, Jaelani, \& More}]{Canameras2021}
Cañameras, R., Schuldt, S., Shu, Y., {et~al.} 2021, A\&A, 653

\bibitem[{Cañameras {et~al.}(2020)Cañameras, Schuldt, Suyu, Taubenberger, Meinhardt, Leal-Taixé, Lemon, Rojas, \& Savary}]{Canameras2020}
Cañameras, R., Schuldt, S., Suyu, S.~H., {et~al.} 2020, A\&A, 644

\bibitem[{Collett(2015)}]{collettPopulationGalaxyGalaxyStrong2015}
Collett, T.~E. 2015, ApJ, 811, 20

\bibitem[{Collett \& Auger(2014)}]{collett_cosmological_2014}
Collett, T.~E. \& Auger, M.~W. 2014, MNRAS, 443, 969

\bibitem[{Collett {et~al.}(2023)Collett, Sonnenfeld, Frohmaier, Glazebrook, Sluse, Motta, Verma, Anguita, Koopmans, Tortora, Courbin, Cabanac, Frye, Smith, Diego, Alteiri, Lopez, Fassnacht, Cooray, Goobar, Ryczanowski, Serjeant, Richard, Treu, Moustakas, Li, Jacobs, Lemon, Marchetti, Hartley, Jullo, Lee, Birrer, Fritz, Nightingale, Napolitano, Plazas, Kruk, Spiniello, Grillo, Suyu, Shajib, Vernardos, Dye, Daylan, Newman, \& Schuldt}]{collett4MOSTStrongLensing2023}
Collett, T.~E., Sonnenfeld, A., Frohmaier, C., {et~al.} 2023, The Messenger, 190, 49

\bibitem[{Davies {et~al.}(2019)Davies, Serjeant, \& Bromley}]{Davies2019}
Davies, A., Serjeant, S., \& Bromley, J.~M. 2019, MNRAS, 487, 5263

\bibitem[{{ESA Datalabs}(2021)}]{Datalabscite}
{ESA Datalabs}. 2021, \url{http://dx.doi.org/10.13140/RG.2.2.36173.56807}

\bibitem[{Etherington {et~al.}(2022)Etherington, Nightingale, Massey, Cao, Robertson, Amorisco, Amvrosiadis, Cole, Frenk, He, Li, \& Tam}]{etheringtonAutomatedGalaxygalaxyStrong2022}
Etherington, A., Nightingale, J.~W., Massey, R., {et~al.} 2022, MNRAS, 517, 3275

\bibitem[{{Euclid Collaboration: Aussel} {et~al.}(2025){Euclid Collaboration: Aussel}, {Tereno}, {Schirmer}, {et~al.}}]{Q1-TP001}
{Euclid Collaboration: Aussel}, H., {Tereno}, I., {Schirmer}, M., {et~al.} 2025, \aap, submitted

\bibitem[{{Euclid Collaboration: Busillo} {et~al.}(2025){Euclid Collaboration: Busillo}, {Tortora}, {Metcalf}, {et~al.}}]{Q1-SP063}
{Euclid Collaboration: Busillo}, V., {Tortora}, C., {Metcalf}, R., {et~al.} 2025, \aap, submitted

\bibitem[{{Euclid Collaboration: Castander} {et~al.}(2024){Euclid Collaboration: Castander}, {Fosalba}, {Stadel}, {et~al.}}]{EuclidSkyFlagship}
{Euclid Collaboration: Castander}, F., {Fosalba}, P., {Stadel}, J., {et~al.} 2024, \aap, accepted, arXiv:2405.13495

\bibitem[{{Euclid Collaboration: Cropper} {et~al.}(2024){Euclid Collaboration: Cropper}, {Al Bahlawan}, {Amiaux}, {et~al.}}]{EuclidSkyVIS}
{Euclid Collaboration: Cropper}, M., {Al Bahlawan}, A., {Amiaux}, J., {et~al.} 2024, \aap, accepted, arXiv:2405.13492

\bibitem[{{Euclid Collaboration: Holloway} {et~al.}(2025){Euclid Collaboration: Holloway}, {Verma}, {Walmsley}, {et~al.}}]{Q1-SP059}
{Euclid Collaboration: Holloway}, P., {Verma}, A., {Walmsley}, M., {et~al.} 2025, \aap, submitted

\bibitem[{{Euclid Collaboration: Jahnke} {et~al.}(2024){Euclid Collaboration: Jahnke}, {Gillard}, {Schirmer}, {et~al.}}]{EuclidSkyNISP}
{Euclid Collaboration: Jahnke}, K., {Gillard}, W., {Schirmer}, M., {et~al.} 2024, \aap, accepted, arXiv:2405.13493

\bibitem[{{Euclid Collaboration: Li} {et~al.}(2025){Euclid Collaboration: Li}, {Collett}, {Walmsley}, {et~al.}}]{Q1-SP054}
{Euclid Collaboration: Li}, T., {Collett}, T., {Walmsley}, M., {et~al.} 2025, \aap, submitted

\bibitem[{{Euclid Collaboration: Lines} {et~al.}(2025){Euclid Collaboration: Lines}, {Collett}, {Walmsley}, {et~al.}}]{Q1-SP053}
{Euclid Collaboration: Lines}, N. E.~P., {Collett}, T.~E., {Walmsley}, M., {et~al.} 2025, \aap, submitted

\bibitem[{{Euclid Collaboration: McCracken} {et~al.}(2025){Euclid Collaboration: McCracken}, {Benson}, {et~al.}}]{Q1-TP002}
{Euclid Collaboration: McCracken}, H., {Benson}, K., {et~al.} 2025, \aap, submitted

\bibitem[{{Euclid Collaboration: Mellier} {et~al.}(2024){Euclid Collaboration: Mellier}, {Abdurro'uf}, {Acevedo~Barroso}, {et~al.}}]{EuclidSkyOverview}
{Euclid Collaboration: Mellier}, Y., {Abdurro'uf}, {Acevedo~Barroso}, J., {et~al.} 2024, \aap, accepted, arXiv:2405.13491

\bibitem[{{Euclid Collaboration: Polenta} {et~al.}(2025){Euclid Collaboration: Polenta}, {Frailis}, {Alavi}, {et~al.}}]{Q1-TP003}
{Euclid Collaboration: Polenta}, G., {Frailis}, M., {Alavi}, A., {et~al.} 2025, \aap, submitted

\bibitem[{{Euclid Collaboration: Rojas} {et~al.}(2025){Euclid Collaboration: Rojas}, {Collett}, {Acevedo Barroso}, {et~al.}}]{Q1-SP052}
{Euclid Collaboration: Rojas}, K., {Collett}, T., {Acevedo Barroso}, J., {et~al.} 2025, \aap, submitted

\bibitem[{{Euclid Collaboration: Romelli} {et~al.}(2025){Euclid Collaboration: Romelli}, {K\"ummel}, {Dole}, {et~al.}}]{Q1-TP004}
{Euclid Collaboration: Romelli}, E., {K\"ummel}, M., {Dole}, H., {et~al.} 2025, \aap, submitted

\bibitem[{{Euclid Collaboration: Walmsley} {et~al.}(2025){Euclid Collaboration: Walmsley}, {Huertas-Company}, {Quilley}, {et~al.}}]{Q1-SP047}
{Euclid Collaboration: Walmsley}, M., {Huertas-Company}, M., {Quilley}, L., {et~al.} 2025, \aap, submitted

\bibitem[{{Euclid Quick Release Q1}(2025)}]{Q1cite}
{Euclid Quick Release Q1}. 2025, \url{https://doi.org/10.57780/esa-2853f3b}

\bibitem[{Fassnacht {et~al.}(2004)Fassnacht, Moustakas, Casertano, Ferguson, Lucas, \& Park}]{fassnachtStrongGravitationalLens2004}
Fassnacht, C.~D., Moustakas, L.~A., Casertano, S., {et~al.} 2004, ApJ, 600, L155

\bibitem[{Faure {et~al.}(2008)Faure, Kneib, Covone, Tasca, Leauthaud, Capak, Jahnke, Smolcic, Torre, Ellis, Finoguenov, Koekemoer, Fevre, Massey, Mellier, Refregier, Rhodes, Scoville, Schinnerer, Taylor, Waerbeke, \& Walcher}]{faureFirstCatalogStrong2008}
Faure, C., Kneib, J.-P., Covone, G., {et~al.} 2008, ApJS, 176, 19

\bibitem[{Garvin {et~al.}(2022)Garvin, Kruk, Cornen, Bhatawdekar, Cañameras, \& Merín}]{garvin_hubble_2022}
Garvin, E.~O., Kruk, S., Cornen, C., {et~al.} 2022, A\&A, 667, A141

\bibitem[{Gavazzi {et~al.}(2014)Gavazzi, Marshall, Treu, \& Sonnenfeld}]{Gavazzi2014}
Gavazzi, R., Marshall, P.~J., Treu, T., \& Sonnenfeld, A. 2014, ApJ, 785

\bibitem[{Gavazzi {et~al.}(2007)Gavazzi, Treu, Rhodes, Koopmans, Bolton, Burles, Massey, \& Moustakas}]{gavazziSloanLensACS2007}
Gavazzi, R., Treu, T., Rhodes, J.~D., {et~al.} 2007, ApJ, 667, 176

\bibitem[{{Geach} {et~al.}(2015){Geach}, {More}, {Verma}, {Marshall}, {Jackson}, {Belles}, {Beswick}, {Baeten}, {Chavez}, {Cornen}, {Cox}, {Erben}, {Erickson}, {Garrington}, {Harrison}, {Harrington}, {Hughes}, {Ivison}, {Jordan}, {Lin}, {Leauthaud}, {Lintott}, {Lynn}, {Kapadia}, {Kneib}, {Macmillan}, {Makler}, {Miller}, {Monta{\~n}a}, {Mujica}, {Muxlow}, {Narayanan}, {O'Briain}, {O'Brien}, {Oguri}, {Paget}, {Parrish}, {Ross}, {Rozo}, {Rusu}, {Rykoff}, {Sanchez-Arg{\"u}elles}, {Simpson}, {Snyder}, {Schloerb}, {Tecza}, {Wang}, {Van Waerbeke}, {Wilcox}, {Viero}, {Wilson}, {Yun}, \& {Zeballos}}]{Geach2015}
{Geach}, J.~E., {More}, A., {Verma}, A., {et~al.} 2015, \mnras, 452, 502

\bibitem[{Gentile {et~al.}(2023)Gentile, Tortora, Covone, Koopmans, Li, Leuzzi, \& Napolitano}]{gentile_lemon_2023}
Gentile, F., Tortora, C., Covone, G., {et~al.} 2023, MNRAS, 522, 5442

\bibitem[{Gentile {et~al.}(2021)Gentile, Tortora, Covone, Koopmans, Spiniello, Fan, Li, Liu, Napolitano, Vaccari, \& Fu}]{Gentile+21}
Gentile, F., Tortora, C., Covone, G., {et~al.} 2021, MNRAS, 510, 500

\bibitem[{González {et~al.}(2025)González, Holloway, Collett, Verma, Bechtol, Marshall, More, Barroso, Cartwright, Martinez, Li, Rojas, Schuldt, Birrer, Diehl, Morgan, Drlica-Wagner, O'Donnell, Zaborowski, Nord, Baeten, Johnson, Macmillan, Roodman, Pieres, Walker, Malagón, Rosell, Santiago, Flaugher, Gruen, Brooks, Burke, James, Cid, Hollowood, Tucker, Buckley-Geer, Gaztanaga, Suchyta, Sanchez, Gutierrez, Giannini, Tarle, Sevilla-Noarbe, Marshall, Carretero, Frieman, Vicente, García-Bellido, Mena-Fernández, Myles, Honscheid, Kuehn, Lima, Pereira, Smith, Aguena, Weaverdyck, Lahav, Doel, Miquel, Gruendl, Cawthon, Hinton, Allam, Desai, Samuroff, Everett, Lee, Davis, Abbott, \& Vikram}]{gonzalez_discovering_2025}
González, J., Holloway, P., Collett, T., {et~al.} 2025, arXiv:2501.15679

\bibitem[{Gu {et~al.}(2022)Gu, Huang, Sheu, Aldering, Bolton, Boone, Dey, Filipp, Jullo, Perlmutter, Rubin, Schlafly, Schlegel, Shu, \& Suyu}]{guGIGALensFastBayesian2022}
Gu, A., Huang, X., Sheu, W., {et~al.} 2022, ApJ, 935, 49

\bibitem[{He {et~al.}(2023)He, Nightingale, Robertson, Amvrosiadis, Cole, Frenk, Massey, Li, Amorisco, Metcalf, Cao, \& Etherington}]{heTestingStrongLensing2023}
He, Q., Nightingale, J., Robertson, A., {et~al.} 2023, MNRAS, 518, 220

\bibitem[{He {et~al.}(2024)He, Nightingale, Amvrosiadis, Robertson, Cole, Frenk, Massey, Li, Cao, Lange, \& França}]{he_unveiling_2024}
He, Q., Nightingale, J.~W., Amvrosiadis, A., {et~al.} 2024, arXiv:2403.16253

\bibitem[{Holloway {et~al.}(2024)Holloway, Marshall, Verma, More, Cañameras, Jaelani, Ishida, \& Wong}]{holloway_bayesian_2024}
Holloway, P., Marshall, P.~J., Verma, A., {et~al.} 2024, MNRAS, 530, 1297

\bibitem[{Huang {et~al.}(2025)Huang, Baltasar, Ratier-Werbin, Storfer, Sheu, Agarwal, Tamargo-Arizmendi, Schlegel, Aguilar, Ahlen, Aldering, Banka, BenZvi, Bianchi, Bolton, Brooks, Cikota, Claybaugh, Macorra, Dey, Doel, Edelstein, Filipp, Forero-Romero, Gaztanaga, Gontcho, Gu, Gutierrez, Honscheid, Jullo, Juneau, Kehoe, Kirkby, Kisner, Kremin, Kwon, Lambert, Landriau, Lang, Guillou, Liu, Meisner, Miquel, Moustakas, Myers, Perlmutter, Perez-Rafols, Prada, Rossi, Rubin, Sanchez, Schubnell, Shu, Silver, Sprayberry, Suzuki, Tarle, Weaver, \& Zou}]{huang_desi_2025}
Huang, X., Baltasar, S., Ratier-Werbin, N., {et~al.} 2025, arXiv:2502.03455

\bibitem[{Huang {et~al.}(2021)Huang, Storfer, Gu, Ravi, Pilon, Sheu, Venguswamy, Banka, Dey, Landriau, Lang, Meisner, Moustakas, Myers, Sajith, Schlafly, \& Schlegel}]{huang_discovering_2021}
Huang, X., Storfer, C., Gu, A., {et~al.} 2021, ApJ, 909, 27

\bibitem[{Huang {et~al.}(2020)Huang, Storfer, Ravi, Pilon, Domingo, Schlegel, Bailey, Dey, Gupta, Herrera, Juneau, Landriau, Lang, Meisner, Moustakas, Myers, Schlafly, Valdes, Weaver, Yang, \& Yèche}]{huang_finding_2020}
Huang, X., Storfer, C., Ravi, V., {et~al.} 2020, ApJ, 894, 78

\bibitem[{Ishida {et~al.}(2025)Ishida, Wong, Jaelani, \& More}]{ishida_combining_2025}
Ishida, Y., Wong, K.~C., Jaelani, A.~T., \& More, A. 2025, PASJ, 77, 105

\bibitem[{Jacobs {et~al.}(2019)Jacobs, Collett, Glazebrook, Buckley-Geer, Diehl, Lin, McCarthy, Qin, Odden, Caso~Escudero, Dial, Yung, Gaitsch, Pellico, Lindgren, Abbott, Annis, Avila, Brooks, Burke, Carnero~Rosell, Carrasco~Kind, Carretero, da~Costa, De~Vicente, Fosalba, Frieman, García-Bellido, Gaztanaga, Goldstein, Gruen, Gruendl, Gschwend, Hollowood, Honscheid, Hoyle, James, Krause, Kuropatkin, Lahav, Lima, Maia, Marshall, Miquel, Plazas, Roodman, Sanchez, Scarpine, Serrano, Sevilla-Noarbe, Smith, Sobreira, Suchyta, Swanson, Tarle, Vikram, Walker, Zhang, \& Collaboration}]{jacobsExtendedCatalogGalaxyGalaxy2019}
Jacobs, C., Collett, T., Glazebrook, K., {et~al.} 2019, ApJS, 243, 17

\bibitem[{Jacobs {et~al.}(2017)Jacobs, Glazebrook, Collett, More, \& McCarthy}]{jacobs_finding_2017}
Jacobs, C., Glazebrook, K., Collett, T., More, A., \& McCarthy, C. 2017, MNRAS, 471, 167

\bibitem[{{Jaelani} {et~al.}(2024){Jaelani}, {More}, {Wong}, {Inoue}, {Chao}, {Premadi}, \& {Ca{\~n}ameras}}]{jaelani24}
{Jaelani}, A.~T., {More}, A., {Wong}, K.~C., {et~al.} 2024, \mnras, 535, 1625

\bibitem[{Kormann {et~al.}(1994)Kormann, Schneider, \& Bartelmann}]{kormann_isothermal_1994}
Kormann, R., Schneider, P., \& Bartelmann, M. 1994, A\&A, 284, 285

\bibitem[{Lang {et~al.}(2016)Lang, Hogg, \& Mykytyn}]{tractor2016}
Lang, D., Hogg, D.~W., \& Mykytyn, D. 2016, ascl:1604.008

\bibitem[{Lange(2023)}]{langeNautilusBoostingBayesian2023}
Lange, J.~U. 2023, MNRAS, 525, 3181

\bibitem[{Lanusse {et~al.}(2018)Lanusse, Ma, Li, Collett, Li, Ravanbakhsh, Mandelbaum, \& Póczos}]{Lanusse2017}
Lanusse, F., Ma, Q., Li, N., {et~al.} 2018, MNRAS, 473, 3895

\bibitem[{Lemon {et~al.}(2024)Lemon, Courbin, More, Schechter, Cañameras, Delchambre, Leung, Shu, Spiniello, Hezaveh, Klüter, \& McMahon}]{lemon_searching_2024}
Lemon, C., Courbin, F., More, A., {et~al.} 2024, Space Science Reviews, 220, 23

\bibitem[{Li {et~al.}(2021)Li, Napolitano, Spiniello, Tortora, Kuijken, Koopmans, Schneider, Getman, Xie, Long, Shu, Vernardos, Huang, Covone, Dvornik, Heymans, Hildebrandt, Radovich, \& Wright}]{liHighqualityStrongLens2021}
Li, R., Napolitano, N.~R., Spiniello, C., {et~al.} 2021, ApJ, 923, 16

\bibitem[{Li {et~al.}(2020)Li, Napolitano, Tortora, Spiniello, Koopmans, Huang, Roy, Vernardos, Chatterjee, Giblin, \& et~al.}]{Li+20}
Li, R., Napolitano, N.~R., Tortora, C., {et~al.} 2020, ApJ, 899, 30

\bibitem[{Li {et~al.}(2024)Li, Collett, Krawczyk, \& Enzi}]{liCosmologyLargePopulations2024}
Li, T., Collett, T.~E., Krawczyk, C.~M., \& Enzi, W. 2024, MNRAS, 527, 5311

\bibitem[{Lintott {et~al.}(2009)Lintott, Schawinski, Keel, Arkel, Bennert, Edmondson, Thomas, Smith, Herbert, Jarvis, Virani, Andreescu, Bamford, Land, Murray, Nichol, Raddick, Slosar, Szalay, \& Vandenberg}]{lintottGalaxyZooHannys2009}
Lintott, C., Schawinski, K., Keel, W., {et~al.} 2009, MNRAS, 399, 129

\bibitem[{Lupton {et~al.}(2004)Lupton, Blanton, Fekete, Hogg, O’Mullane, Szalay, \& Wherry}]{Lupton2004}
Lupton, R., Blanton, M., Fekete, G., {et~al.} 2004, \pasp, 116, 133

\bibitem[{Margalef-Bentabol {et~al.}(2024)Margalef-Bentabol, Wang, Marca, Blanco-Prieto, Chudy, Domínguez-Sánchez, Goulding, Guzmán-Ortega, Huertas-Company, Martin, Pearson, Rodriguez-Gomez, Walmsley, Bickley, Bottrell, Conselice, \& O'Ryan}]{margalef-bentabolGalaxyMergerChallenge2024}
Margalef-Bentabol, B., Wang, L., Marca, A.~L., {et~al.} 2024, A\&A, 687, A24

\bibitem[{Marshall {et~al.}(2016)Marshall, Verma, More, Davis, More, Kapadia, Parrish, Snyder, Wilcox, Baeten, Macmillan, Cornen, Baumer, Simpson, Lintott, Miller, Paget, Simpson, Smith, K??ng, Saha, \& Collett}]{Marshall2016}
Marshall, P.~J., Verma, A., More, A., {et~al.} 2016, MNRAS, 455, 1171

\bibitem[{Melo {et~al.}(2024)Melo, Cañameras, Schuldt, Suyu, Andika, Bag, \& Taubenberger}]{melo_holismokes_2024}
Melo, A., Cañameras, R., Schuldt, S., {et~al.} 2024, arXiv:2411.18694

\bibitem[{Metcalf {et~al.}(2019)Metcalf, Meneghetti, Avestruz, Bellagamba, Bom, Bertin, Cabanac, Courbin, Davies, Decencière, Flamary, Gavazzi, Geiger, Hartley, Huertas-Company, Jackson, Jacobs, Jullo, Kneib, Koopmans, Lanusse, Li, Ma, Makler, Li, Lightman, Petrillo, Serjeant, Schäfer, Sonnenfeld, Tagore, Tortora, Tuccillo, Valentín, Velasco-Forero, Kleijn, \& Vernardos}]{metcalf_strong_2019}
Metcalf, R.~B., Meneghetti, M., Avestruz, C., {et~al.} 2019, A\&A, 625, A119

\bibitem[{{More} {et~al.}(2016){More}, {Verma}, {Marshall}, {More}, {Baeten}, {Wilcox}, {Macmillan}, {Cornen}, {Kapadia}, {Parrish}, {Snyder}, {Davis}, {Gavazzi}, {Lintott}, {Simpson}, {Miller}, {Smith}, {Paget}, {Saha}, {K{\"u}ng}, \& {Collett}}]{moreSpaceWarpsII2016}
{More}, A., {Verma}, A., {Marshall}, P.~J., {et~al.} 2016, \mnras, 455, 1191

\bibitem[{Myers {et~al.}(2003)Myers, Jackson, Browne, de~Bruyn, Pearson, Readhead, Wilkinson, Biggs, Blandford, Fassnacht, Koopmans, Marlow, McKean, Norbury, Phillips, Rusin, Shepherd, \& Sykes}]{myersCosmicLensAllSky2003}
Myers, S.~T., Jackson, N.~J., Browne, I. W.~A., {et~al.} 2003, MNRAS, 341, 1

\bibitem[{Nagam {et~al.}(2025)Nagam, Barroso, Wilde, Andika, Manjón-García, Pearce-Casey, Stern, Nightingale, Moustakas, McCarthy, Moravec, Leuzzi, Rojas, Serjeant, Collett, Matavulj, Walmsley, Clément, Tortora, Gavazzi, Metcalf, O'Riordan, Kleijn, Koopmans, Valentijn, Busillo, Schuldt, Courbin, Vernardos, Meneghetti, Díaz-Sánchez, Diego, Ecker, Thai, Cooray, Courtois, Delchambre, Despali, Sluse, Ulivi, Melo, Corcho-Caballero, Altieri, Amara, Andreon, Auricchio, Aussel, Baccigalupi, Baldi, Balestra, Bardelli, Battaglia, Bonino, Branchini, Brescia, Brinchmann, Caillat, Camera, Capobianco, Carbone, Carretero, Casas, Castellano, Castignani, Cavuoti, Cimatti, Colodro-Conde, Congedo, Conselice, Conversi, Copin, Cropper, Silva, Degaudenzi, Lucia, Giorgio, Dinis, Dubath, Duncan, Dupac, Dusini, Fabricius, Farina, Farrens, Ferriol, Frailis, Franceschi, Fumana, George, Gillard, Gillis, Giocoli, Gómez-Alvarez, Grazian, Grupp, Guzzo, Haugan, Hoar, Holmes, Hook, Hormuth, Hornstrup, Hudelot, Jahnke, Jhabvala,
  Joachimi, Keihänen, Kermiche, Kubik, Kuijken, Kümmel, Kunz, Kurki-Suonio, Laureijs, Mignant, Ligori, Lilje, Lindholm, Lloro, Mainetti, Maiorano, Mansutti, Marggraf, Markovic, Martinelli, Martinet, Marulli, Massey, Medinaceli, Melchior, Mellier, Merlin, Meylan, Moresco, Moscardini, Nakajima, Neissner, Nichol, Niemi, Padilla, Paltani, Pasian, Pedersen, Percival, Pettorino, Pires, Polenta, Poncet, Popa, Pozzetti, Raison, Rebolo, Renzi, Rhodes, Riccio, Romelli, Roncarelli, Rossetti, Saglia, Sakr, Sánchez, Sapone, Sartoris, Schirmer, Schneider, Schrabback, Secroun, Seidel, Serrano, Sirignano, Sirri, Skottfelt, Stanco, Starck, Steinwagner, Tallada-Crespí, Tavagnacco, Taylor, Teplitz, Tereno, Toledo-Moreo, Torradeflot, Tsyganov, Tutusaus, Valenziano, Vassallo, Veropalumbo, Wang, Weller, Zacchei, Zucca, Burigana, Mora, Pöntinen, \& Scottez}]{nagam_euclid_2025}
Nagam, B.~C., Barroso, J. A.~A., Wilde, J., {et~al.} 2025, arXiv:2502.09802

\bibitem[{Nagam {et~al.}(2023)Nagam, Koopmans, Valentijn, Kleijn, de~Jong, Napolitano, Li, \& Tortora}]{nagamDenseLensUsingDenseNet2023}
Nagam, B.~C., Koopmans, L. V.~E., Valentijn, E.~A., {et~al.} 2023, MNRAS, 523, 4188

\bibitem[{Nagam {et~al.}(2024)Nagam, Koopmans, Valentijn, Kleijn, de Jong, Napolitano, Li, Tortora, Busillo, \& Dong}]{nagam+24}
Nagam, B.~C., Koopmans, L. V.~E., Valentijn, E.~A., {et~al.} 2024, MNRAS, 533, 1426

\bibitem[{Nightingale {et~al.}(2021)Nightingale, Hayes, Kelly, Amvrosiadis, Etherington, He, Li, Cao, Frawley, Cole, Enia, Frenk, Harvey, Li, Massey, Negrello, \& Robertson}]{nightingalePyAutoLensOpenSourceStrong2021}
Nightingale, J.~W., Hayes, R.~G., Kelly, A., {et~al.} 2021, Journal of Open Source Software, 6, 2825

\bibitem[{Nightingale {et~al.}(2024)Nightingale, He, Cao, Amvrosiadis, Etherington, Frenk, Hayes, Robertson, Cole, Lange, Li, \& Massey}]{nightingaleScanningDarkMatter2024}
Nightingale, J.~W., He, Q., Cao, X., {et~al.} 2024, MNRAS, 527, 10480

\bibitem[{Nightingale {et~al.}(2023)Nightingale, Smith, He, O’Riordan, Kegerreis, Amvrosiadis, Edge, Etherington, Hayes, Kelly, Lucey, \& Massey}]{nightingaleAbell1201Detection2023}
Nightingale, J.~W., Smith, R.~J., He, Q., {et~al.} 2023, MNRAS, 521, 3298

\bibitem[{O’Riordan {et~al.}(2023)O’Riordan, Despali, Vegetti, Lovell, \& Moline}]{oriordanSensitivityStrongLensing2023}
O’Riordan, C.~M., Despali, G., Vegetti, S., Lovell, M.~R., \& Moline, A. 2023, MNRAS, 521, 2342

\bibitem[{O’Riordan {et~al.}(2025)O’Riordan, Oldham, Nersesian, Li, Collett, Sluse, Altieri, Clément, Vasan, Rhoades, Chen, Jones, Adami, Gavazzi, Vegetti, Powell, Barroso, Andika, Bhatawdekar, Cooray, Despali, Diego, Ecker, Galan, Gómez-Alvarez, Leuzzi, Meneghetti, Metcalf, Schirmer, Serjeant, Tortora, Vaccari, Vernardos, Walmsley, Amara, Andreon, Auricchio, Aussel, Baccigalupi, Baldi, Balestra, Bardelli, Basset, Battaglia, Bender, Bonino, Branchini, Brescia, Brinchmann, Caillat, Camera, Capobianco, Carbone, Carretero, Casas, Castander, Castellano, Castignani, Cavuoti, Cimatti, Colodro-Conde, Congedo, Conselice, Conversi, Copin, Corcione, Courbin, Courtois, Cropper, Silva, Degaudenzi, Lucia, Giorgio, Dinis, Dubath, Duncan, Dupac, Dusini, Farina, Farrens, Faustini, Ferriol, Fourmanoit, Frailis, Franceschi, Fumana, Galeotta, Gillard, Gillis, Giocoli, Granett, Grazian, Grupp, Guzzo, Haugan, Hoar, Hoekstra, Holmes, Hook, Hormuth, Hornstrup, Hudelot, Jahnke, Jhabvala, Joachimi, Keihänen, Kermiche,
  Kiessling, Kilbinger, Kohley, Kubik, Kümmel, Kunz, Kurki-Suonio, Lahav, Laureijs, Mignant, Ligori, Lilje, Lindholm, Lloro, Mainetti, Maiorano, Mansutti, Marggraf, Markovic, Martinelli, Martinet, Marulli, Massey, Medinaceli, Mei, Melchior, Mellier, Merlin, Meylan, Moresco, Moscardini, Nakajima, Nichol, Niemi, Nightingale, Padilla, Paltani, Pasian, Pedersen, Percival, Pettorino, Pires, Polenta, Poncet, Popa, Pozzetti, Raison, Rebolo, Renzi, Rhodes, Riccio, Rix, Romelli, Roncarelli, Rossetti, Rusholme, Saglia, Sakr, Sánchez, Sapone, Sartoris, Schneider, Schrabback, Secroun, Seidel, Serrano, Sirignano, Sirri, Stanco, Steinwagner, Tallada-Crespí, Tereno, Toledo-Moreo, Torradeflot, Tutusaus, Valenziano, Vassallo, Kleijn, Veropalumbo, Wang, Weller, Zacchei, Zamorani, Zucca, Burigana, Casenove, Mora, Scottez, Viel, Jauzac, \& Dannerbauer}]{oriordanEuclidCompleteEinstein2025}
O’Riordan, C.~M., Oldham, L.~J., Nersesian, A., {et~al.} 2025, A\&A, 694, A145

\bibitem[{{Pearce-Casey} {et~al.}(2024){Pearce-Casey}, {Nagam}, {Wilde}, {et~al.}}]{Pearce-Casey24}
{Pearce-Casey}, R., {Nagam}, B.~C., {Wilde}, J., {et~al.} 2024, A\&A, submitted, arXiv:2411.16808

\bibitem[{Petrillo {et~al.}(2017)Petrillo, Tortora, Chatterjee, Vernardos, Koopmans, Kleijn, Napolitano, Covone, Schneider, Grado, \& McFarland}]{Petrillo2017}
Petrillo, C.~E., Tortora, C., Chatterjee, S., {et~al.} 2017, MNRAS, 472, 1129

\bibitem[{Petrillo {et~al.}(2019{\natexlab{a}})Petrillo, Tortora, Chatterjee, Vernardos, Koopmans, Verdoes Kleijn, Napolitano, Covone, Kelvin, \& Hopkins}]{Petrillo2019a}
Petrillo, C.~E., Tortora, C., Chatterjee, S., {et~al.} 2019{\natexlab{a}}, MNRAS, 482, 807

\bibitem[{Petrillo {et~al.}(2019{\natexlab{b}})Petrillo, Tortora, Vernardos, Koopmans, Verdoes~Kleijn, Bilicki, Napolitano, Chatterjee, Covone, Dvornik, Erben, Getman, Giblin, Heymans, De~Jong, Kuijken, Schneider, Shan, Spiniello, \& Wright}]{Petrillo2019b}
Petrillo, C.~E., Tortora, C., Vernardos, G., {et~al.} 2019{\natexlab{b}}, MNRAS, 484, 3879

\bibitem[{Pourrahmani {et~al.}(2018)Pourrahmani, Nayyeri, \& Cooray}]{Pourrahmani+18}
Pourrahmani, M., Nayyeri, H., \& Cooray, A. 2018, ApJ, 856, 68

\bibitem[{Rezaei {et~al.}(2022)Rezaei, McKean, Biehl, de Roo, \& Lafontaine}]{rezaei_machine_2022}
Rezaei, S., McKean, J.~P., Biehl, M., de Roo, W., \& Lafontaine, A. 2022, MNRAS, 517, 1156

\bibitem[{Rojas {et~al.}(2023)Rojas, Collett, Ballard, Magee, Birrer, Buckley-Geer, Chan, Clément, Diego, Gentile, González, Joseph, Mastache, Schuldt, Tortora, Verdugo, Verma, Daylan, Millon, Jackson, Dye, Melo, Mahler, Ogando, Courbin, Fritz, Herle, Acevedo~Barroso, Cañameras, Cornen, Dhanasingham, Glazebrook, Martinez, Ryczanowski, Savary, Góis-Silva, Arturo Ureña-López, Wiesner, Wilde, Valim~Calçada, Cabanac, Pan, Sierra, Despali, Cavalcante-Gomes, Macmillan, Maresca, Grudskaia, O'Donnell, Paic, Niemiec, de~la Bella, Bromley, Williams, More, \& Levine}]{rojasImpactHumanExpert2023}
Rojas, K., Collett, T.~E., Ballard, D., {et~al.} 2023, MNRAS, 523, 4413

\bibitem[{Rojas {et~al.}(2022)Rojas, Savary, Clément, Maus, Courbin, Lemon, Chan, Vernardos, Joseph, Cañameras, \& Galan}]{rojas_search_2022}
Rojas, K., Savary, E., Clément, B., {et~al.} 2022, A\&A, 668, A73

\bibitem[{Savary {et~al.}(2022)Savary, Rojas, Maus, Clément, Courbin, Gavazzi, Chan, Lemon, Vernardos, Cañameras, Schuldt, Suyu, Cuillandre, Fabbro, Gwyn, Hudson, Kilbinger, Scott, \& Stone}]{savaryStrongLensingUNIONS2022}
Savary, E., Rojas, K., Maus, M., {et~al.} 2022, A\&A, 666, A1

\bibitem[{Schaefer {et~al.}(2018)Schaefer, Geiger, Kuntzer, \& Kneib}]{Schaefer2017}
Schaefer, C., Geiger, M., Kuntzer, T., \& Kneib, J.~P. 2018, A\&A, 611

\bibitem[{{Schuldt} {et~al.}(2025{\natexlab{a}}){Schuldt}, {Ca{\~n}ameras}, {Andika}, {Bag}, {Melo}, {Shu}, {Suyu}, {Taubenberger}, \& {Grillo}}]{schuldt_holismokes_2025}
{Schuldt}, S., {Ca{\~n}ameras}, R., {Andika}, I.~T., {et~al.} 2025{\natexlab{a}}, \aap, 693, A291

\bibitem[{{Schuldt} {et~al.}(2025{\natexlab{b}}){Schuldt}, {Ca{\~n}ameras}, {Shu}, {Andika}, {Bag}, {Grillo}, {Melo}, {Suyu}, \& {Taubenberger}}]{schuldtEtAl25b}
{Schuldt}, S., {Ca{\~n}ameras}, R., {Shu}, Y., {et~al.} 2025{\natexlab{b}}, arXiv e-prints, arXiv:2503.07733

\bibitem[{Shajib {et~al.}(2024)Shajib, Smith, Birrer, Verma, Arendse, \& Collett}]{shajib_strong_2024}
Shajib, A.~J., Smith, G.~P., Birrer, S., {et~al.} 2024, arXiv:2406.08919

\bibitem[{Sharma {et~al.}(2023)Sharma, Collett, \& Linder}]{sharmaTestingCosmologyDouble2023}
Sharma, D., Collett, T.~E., \& Linder, E.~V. 2023, JCAP, 04, 001

\bibitem[{{Shu} {et~al.}(2022){Shu}, {Ca{\~n}ameras}, {Schuldt}, {Suyu}, {Taubenberger}, {Inoue}, \& {Jaelani}}]{shu22_HOLISMOKES8}
{Shu}, Y., {Ca{\~n}ameras}, R., {Schuldt}, S., {et~al.} 2022, \aap, 662, A4

\bibitem[{Sonnenfeld {et~al.}(2018)Sonnenfeld, Chan, Shu, More, Oguri, Suyu, Wong, Lee, Coupon, Yonehara, Bolton, Jaelani, Tanaka, Miyazaki, \& Komiyama}]{sonnenfeldSurveyGravitationallylensedObjects2018}
Sonnenfeld, A., Chan, J. H.~H., Shu, Y., {et~al.} 2018, PASJ, 70, S29

\bibitem[{Sonnenfeld {et~al.}(2023)Sonnenfeld, Li, Despali, Gavazzi, Shajib, \& Taylor}]{sonnenfeldStrongLensingSelection2023}
Sonnenfeld, A., Li, S.-S., Despali, G., {et~al.} 2023, A\&A, 678, A4

\bibitem[{Sonnenfeld {et~al.}(2020)Sonnenfeld, Verma, More, Baeten, Macmillan, Wong, Chan, Jaelani, Lee, Oguri, Rusu, Veldthuis, Trouille, Marshall, Hutchings, Allen, Donnell, Cornen, Davis, McMaster, Lintott, \& Miller}]{sonnenfeldSurveyGravitationallylensedObjects2020a}
Sonnenfeld, A., Verma, A., More, A., {et~al.} 2020, A\&A, 642, A148

\bibitem[{Speagle(2020)}]{speagleDYNESTYDynamicNested2020}
Speagle, J.~S. 2020, MNRAS, 493, 3132

\bibitem[{Stein {et~al.}(2022)Stein, Blaum, Harrington, Medan, \& Lukić}]{Stein2022}
Stein, G., Blaum, J., Harrington, P., Medan, T., \& Lukić, Z. 2022, ApJ, 932, 107

\bibitem[{Stone {et~al.}(2024)Stone, Adam, Coogan, Yantovski-Barth, Filipp, Setiawan, Core, Legin, Wilson, Barco, Hezaveh, \& Perreault-Levasseur}]{stoneCausticsPythonPackage2024}
Stone, C., Adam, A., Coogan, A., {et~al.} 2024, Journal of Open Source Software, 9, 7081

\bibitem[{Storfer {et~al.}(2024)Storfer, Huang, Gu, Sheu, Banka, Dey, Inchausti~Reyes, Jain, Kwon, Lang, Lee, Meisner, Moustakas, Myers, Tabares-Tarquinio, Schlafly, \& Schlegel}]{storferNewStrongGravitational2024}
Storfer, C., Huang, X., Gu, A., {et~al.} 2024, ApJS, 274, 16

\bibitem[{Treu {et~al.}(2011)Treu, Dutton, Auger, Marshall, Bolton, Brewer, Koo, \& Koopmans}]{treuSWELLSSurveyLarge2011}
Treu, T., Dutton, A.~A., Auger, M.~W., {et~al.} 2011, MNRAS, 417, 1601

\bibitem[{Walmsley {et~al.}(2023)Walmsley, Allen, Aussel, Bowles, Gregorowicz, Slijepcevic, Lintott, Scaife, Jabłońska, Karchev, Lanzieri, Mohan, O’Ryan, Saiguhan, Suárez, Guerra-Varas, \& Velu}]{Walmsley2023zoobot}
Walmsley, M., Allen, C., Aussel, B., {et~al.} 2023, Journal of Open Source Software, 8, 5312

\bibitem[{Walmsley {et~al.}(2024)Walmsley, Bowles, Scaife, Makechemu, Gordon, Ferguson, Mann, Pearson, Popp, Bovy, Speagle, Dickinson, Fortson, Géron, Kruk, Lintott, Mantha, Mohan, O'Ryan, \& Slijepevic}]{walmsley_scaling_2024}
Walmsley, M., Bowles, M., Scaife, A. M.~M., {et~al.} 2024, arXiv:2404.02973

\bibitem[{Warren \& Dye(2003)}]{warrenSemilinearGravitationalLens2003}
Warren, S.~J. \& Dye, S. 2003, ApJ, 590, 673

\bibitem[{Wong {et~al.}(2022)Wong, Chan, Chao, Jaelani, Kayo, Lee, More, \& Oguri}]{wongSurveyGravitationallyLensed2022}
Wong, K.~C., Chan, J. H.~H., Chao, D. C.-Y., {et~al.} 2022, PASJ, 74, 1209

\bibitem[{Zhao {et~al.}(2020)Zhao, Queralta, \& Westerlund}]{zhaoSimtoRealTransferDeep2020}
Zhao, W., Queralta, J.~P., \& Westerlund, T. 2020, in 2020 {IEEE} {Symposium} {Series} on {Computational} {Intelligence} ({SSCI}), 737--744

\end{thebibliography}

%

\begin{appendix}
  \onecolumn 

\section{Strong lens gallery}
\label{app:strong_lens_gallery}

\begin{figure*}
    \centering
    \includegraphics[width=0.96\textwidth]{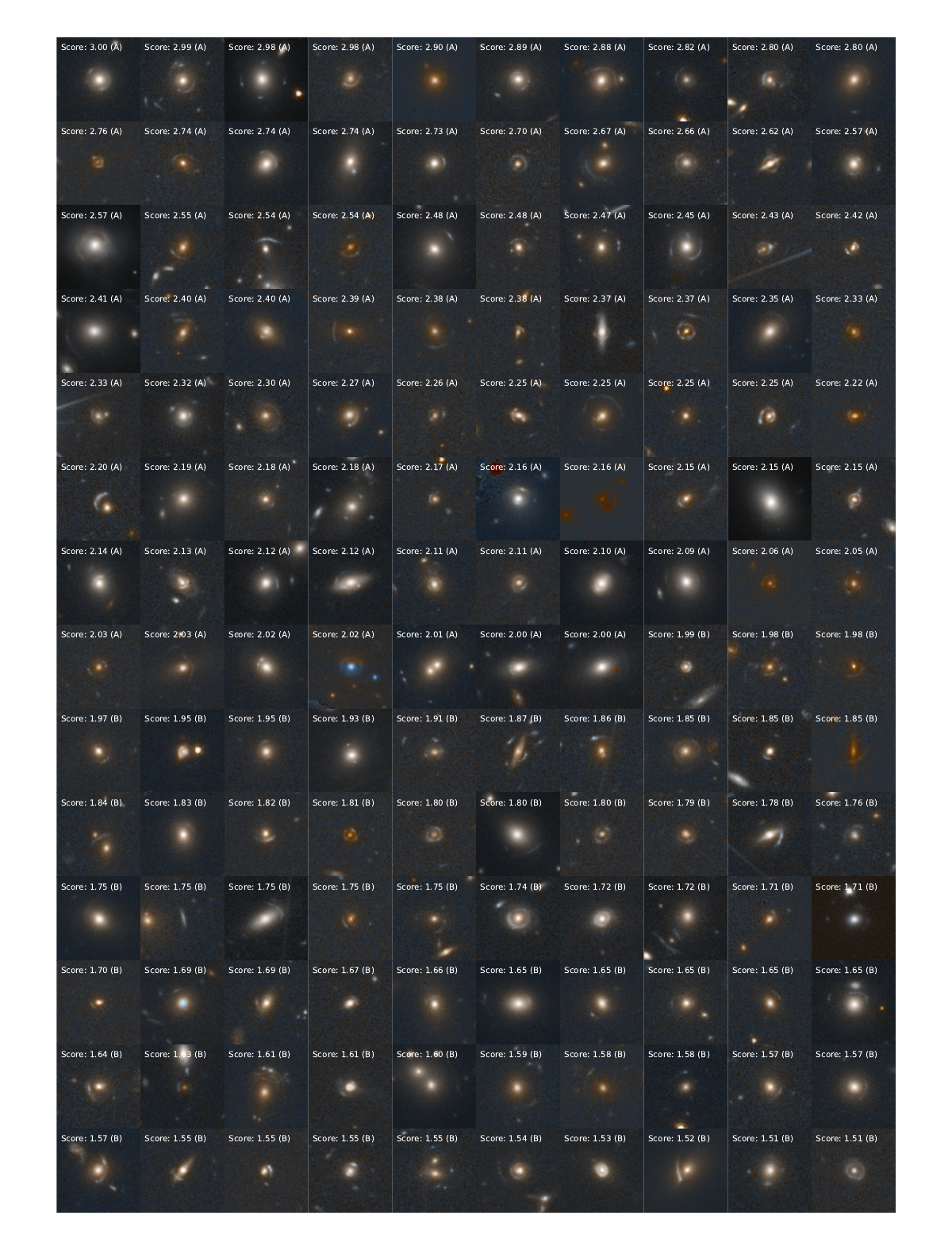}
    \caption{Strong lens candidates as a function of expert visual inspection score. Random subset shown for conciseness (140 of 500). We refer to a score above 2.0 as Grade A and above 1.5 as Grade B in this work, but suggest that readers choose their own selection cut according to their science goals.}
    \label{fig:lens_subset_by_score}
\end{figure*}

Figure \ref{fig:lens_subset_by_score} shows our lens candidates (random subset for conciseness) ordered by expert visual inspection score. Use this to choose your own score cuts.

\section{Calibrating For individual expert optimism}
\label{app:calibrating_experts}

Individual experts can disagree on strong lens grades. A common explanation for part of this disagreement is that experts can be, on average, more optimistic or more conservative \citep{rojasImpactHumanExpert2023}. We introduce a simple re-weighting technique to account for any systematic offset in optimism between experts. In practice, we find this has a minimal effect on our aggregate grades, likely because we ask ten experts for every galaxy and so any offsets are largely averaged out. We report our rebalancing here for completeness.

\begin{figure}
    \centering
    \includegraphics[width=0.5\linewidth]{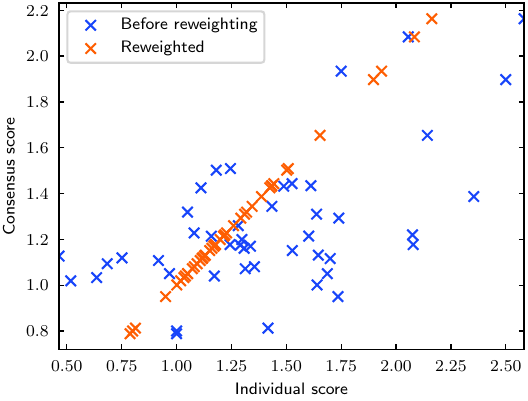}
    \caption{Removing systematic offsets in grading optimism between experts. Each mark is an expert. The $x$-axis shows the mean score of an expert for all the galaxies they graded. The $y$-axis shows the mean score of all other experts over the same set of galaxies. In blue is the original scores, and in orange is the same calculation after reweighting (matching the individual mean with the group mean). The range in mean scores (both axes, before and after reweighting) is due to experts joining the grading project at different times and seeing a different balance of galaxies.}
    \label{fig:debiasing}
    \label{LastPage}
\end{figure}

We calculate our calibration by aggregating over all galaxies seen by one expert. The intuition is that for the set of galaxies seen by each expert, their mean grade should be the same as the mean of the grades from all other experts for that same set of galaxies. For example, if I see three galaxies and grade them all as A (score of 3.0), but the mean grade for those galaxies from all other experts is B (score of 2.0), one should discount my scores by 1.0 to account for my overall optimism. Repeating this calculation individually for all experts provides a calibration such that the optimism of every expert is accounted for.
Figure \ref{fig:debiasing} shows the result.

\section{Lens modelling pipeline}
\label{app:modelling}

\subsection{Data preprocessing}

Before lens modelling, several preprocessing steps are required, streamlined to enable scalability for large lens samples:

\begin{itemize}

\item \textbf{Mask:} A $\ang{;;5.0}$ circular mask is applied to every imaging dataset,  limiting the fit to pixels within this radius. Using a fixed radius for all candidates eliminates human input, enabling scalability to large lens samples.

\item \textbf{Contaminant removal:} Emission from nearby line-of-sight galaxies within the mask can interfere with the model. A GUI is used to manually `spray-paint' these regions, replacing them with random Gaussian noise and increasing the RMS noise map to ensure the model ignores them. This method, commonly used in lens modelling (e.g., \citealt{etheringtonAutomatedGalaxygalaxyStrong2022, nightingaleScanningDarkMatter2024}), takes approximately ten seconds per lens and will require future automation for scalability to thousands of candidates.

\item \textbf{Multiple image positions:} Previous studies input $(x,y)$ coordinates of multiple lensed images, enforcing that the mass model maps them within a threshold (e.g., $\ang{;;0.1}$) in the source plane. This manual step (about ten seconds per lens) is omitted in the \Euclid pipeline. Instead, after the first lens model fit, a lens equation solver automatically computes these positions for use in subsequent fits, enabling full automation.
\end{itemize}

\subsection{Fitting}

The initial stage of the pipeline uses an MGE source model, which is efficient and flexible with only one set of 30 Gaussians spanning $\sigma$ values from $\ang{;;0.001}$ to $\ang{;;1.0}$. As discussed in H24, the MGE source model is effective for automated lens modeling because it can fit the data is a highly flexible way whilst retaining a relatively low number of non-linear free parameters. For the source, we use only one set of 30 Gaussians, whose $\sigma$ values span $\log10$ increments from $\ang{;;0.001}$ to $\ang{;;1.0}$. The MGE's assumption of symmetry limits its ability to model complex high-redshift source morphologies. Later stages use an adaptive Delaunay mesh for irregular sources, employing bilinear interpolation and cross-regularization from H24. Details of the linear algebra, interpolation, and regularization are in \citet{nightingaleScanningDarkMatter2024} and H24. 

The pipeline performs five chained fits, the first two and final two use the nested sampler \texttt{nautilus}\footnote{\url{https://github.com/johannesulf/nautilus}} \citep{langeNautilusBoostingBayesian2023} and the third stage uses the nested sampler \texttt{dynesty} \citep{speagleDYNESTYDynamicNested2020}. The pipeline is the `Source' and `Mass' pipelines of the \texttt{PyAutoLens} SLaM (Source, Light, and Mass) pipelines used by various other studies (e.g., \citealt{etheringtonAutomatedGalaxygalaxyStrong2022, cao_systematic_2022, heTestingStrongLensing2023, nightingaleAbell1201Detection2023, nightingaleScanningDarkMatter2024}). 

All pipeline stages decompose the lens light into 2D elliptical Gaussians using \citet{cappellari_efficient_2002}'s MGE framework, implemented in the semi-linear inversion method \citep{warrenSemilinearGravitationalLens2003} using a fast non-negative least-square (fnnls) algorithm which enforces positivity on the solution. Gaussians are grouped into two sets of $30$ Gaussians which share the same centres, position angles, axis ratios, and their $\sigma$ values are fixed to preset values which evenly increase in $\log10$-spaced intervals between $\ang{;;0.02}$ and $\ang{;;5.0}$. The implementation is described fully in \citet{he_unveiling_2024}.

\section{\label{app:data_availability} Data availability}

All data underlying this article is available on Zenodo at \url{https://doi.org/10.5281/zenodo.15003116}. This article builds on data released during Euclid Quick Release 1, available from \cite{Q1-TP001}.

\end{appendix}

\end{document}